\documentclass[journal=aelccp,manuscript=letter]{achemso}

\usepackage[version=3]{mhchem} 
\usepackage[utf8]{inputenc}
\usepackage{graphicx}
\usepackage{listings}
\usepackage{xcolor}
\usepackage{natbib}
\usepackage{graphicx}
\usepackage{multirow}
\usepackage{caption}
\usepackage{subcaption}
\usepackage{enumitem}
\usepackage{hyperref}
\usepackage{matlab-prettifier}
\usepackage[english]{babel}
\usepackage[capitalise]{cleveref}
\usepackage{algorithm}
\usepackage{algpseudocode}
\usepackage{xpatch}
\usepackage{subcaption}
\usepackage{breqn}

\newcommand{\lion}{Li$^+$}
\newcommand{\papertitle}{Differentiable Solvation Shell Model for Rational Electrolyte Design}

\author{Hancheng Zhao}
\author{Hongyi Lin}
\author{Celia Kelly}
\affiliation[University of Michigan]
{Department of Mechanical Engineering, University of Michigan, Ann Arbor, Michigan 48103, USA}
\author{Venkatasubramanian Viswanathan}
\altaffiliation{Department of Aerospace Engineering, University of Michigan, Ann Arbor, Michigan 48103, USA}

\email{venkvis@umich.edu}
\affiliation[University of Michigan]
{Department of Mechanical Engineering, University of Michigan, Ann Arbor, Michigan 48103, USA}

\title{\papertitle}

\abbreviations{IR,NMR,UV}
\keywords{American Chemical Society, \LaTeX}

\usepackage{tabularx}
\usepackage{booktabs}

\makeatletter
\@ifpackageloaded{mciteplus}{\mciteErrorOnUnknownfalse}{}
\makeatother

\makeatletter
\let\acs@saved@maketitle\maketitle
\let\acs@saved@@maketitle\@maketitle
\let\acs@saved@title\title
\let\acs@saved@SectionsOn\SectionsOn
\let\acs@saved@SectionNumbersOn\SectionNumbersOn
\makeatother

\begin{document}

\begin{tocentry}




  
\includegraphics[width=1.0\linewidth]{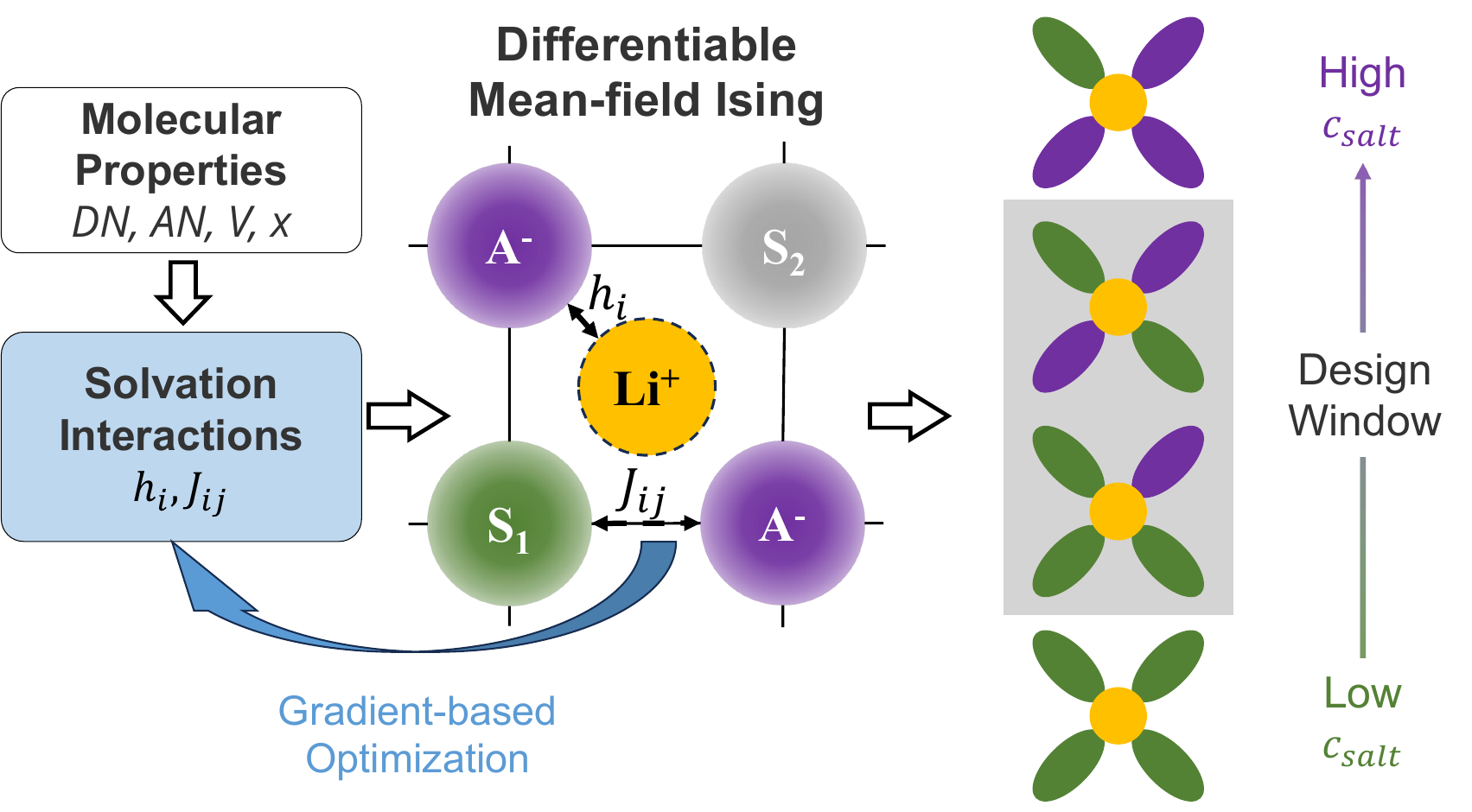}
\end{tocentry}

\begin{abstract}
Tailoring \lion \space solvation structures has emerged as a promising design principle across multiple Li metal battery (LMB) electrolytes, with localized high concentration electrolytes (LHCE) emerging as a leading candidate. However, rational design of these electrolytes remains limited by a poor quantitative understanding of how molecular properties govern \lion \space solvation shell composition. Here we introduce a mean field modeling framework, built on an Ising model, that predicts solvation shell composition from donor number (DN), acceptor number (AN), molar ratio, and molecular size. This framework is end-to-end differentiable, which enables parameterization directly from molecular dynamics (MD) solvation structures via gradient-based optimization. Focusing on LHCE, the model achieves 10.7\% RMSE and $R^2=0.87$ for shell composition, 2.3\% RMSE on free solvent ratio, and further reproduces solvation trends in real LHCE electrolytes, at a fraction of MD's cost. Our model also shows good generalizability on other electrolyte systems in addition to LHCE. Analysis of its interaction terms shows that DN dominates \lion \space solvation energetics, providing a thermodynamic basis for empirical DN-based design rules. We further demonstrate the model's design utility on the LiTFSI/tetraglyme (G4) system absent from training. The model identifies fluorobenzene (FB) as a promising diluent and predicts a salt-concentration window with anion-rich solvation shells and low free solvent, favorable for stable anion-derived SEIs and high oxidative stability, respectively. This prediction is validated by higher-fidelity MD. This work establishes an interpretable differentiable framework for predicting \lion \space solvation shell composition, with potential to extend to other electrolyte classes.
\end{abstract}

Lithium-ion batteries (LIBs) power many modern electronic devices, yet their energy density remains insufficient for emerging applications such as heavy-duty transportation and electric aviation \cite{choi_promise_2016,sripad_performance_2017,viswanathan_challenges_2022}. Meeting these demands requires moving beyond the graphite anode to high-capacity anodes such as Li metal, i.e., Li metal batteries (LMB) \cite{yang_reviewchallenges_2024}. However, commercial carbonate-based liquid electrolytes, optimized for graphite compatibility and transport properties \cite{xu_electrolytes_2014,liu_molecular_2026}, fail to passivate Li metal surfaces \cite{cao_reviewlocalized_2021}, causing electrolyte depletion, active material loss, and dendrite formation \cite{chang_correlating_2015, lu_failure_2015, ding_dendrite-free_2013}. 

This challenge has motivated the exploration of several new classes of non-aqueous liquid electrolytes, including high-concentration electrolytes (HCEs) \cite{qian_high_2015, yamada_advances_2019}, localized high concentration electrolytes (LHCEs) \cite{cao_reviewlocalized_2021,chen_high-voltage_2018}, weakly solvating electrolytes (WSEs) \cite{karbak_weakly_2025}, and high-entropy electrolytes (HEEs) \cite{kim_high-entropy_2023, xia_rational_2026}, each of which has demonstrated improved cycling stability against lithium metal. Despite their compositional diversity, these electrolytes share one common underlying design principle: engineering the thermodynamics of the \lion \space solvation shell to promote the formation of solid electrolyte interphases (SEIs) compatible with Li metal. 
For HCE, LHCE, and WSE, enthalpically favorable \lion-anion coordination results in anion-rich inner solvation shells \cite{yamada_advances_2019,qian_high_2015}; while for HEE, high configurational entropy can facilitate multi-anion coordination with \lion \cite{wang_high_2023}. 
These anion-rich solvation structures enable preferential reduction of anions on electrode surfaces and the formation of Li-metal-friendly anion-derived stable SEI \cite{yamada_advances_2019,xia_rational_2026}.
Beyond interfacial stability, solvation shell composition also governs desolvation kinetics, high-voltage stability, and transport in practical LMBs \cite{liu_molecular_2026,yamada_advances_2019,zhang_predicting_2022}.
Electrolyte solvation shell engineering at the molecular level has thus attracted increasing attention and is emerging as a promising direction for future LMB electrolyte exploration \cite{liu_molecular_2026,weintz_nanoengineering_2026}.

\begin{figure}
  \centering
  \begin{subfigure}[b]{0.6\textwidth}
    \centering
    \caption{}
    \includegraphics[width=\linewidth]{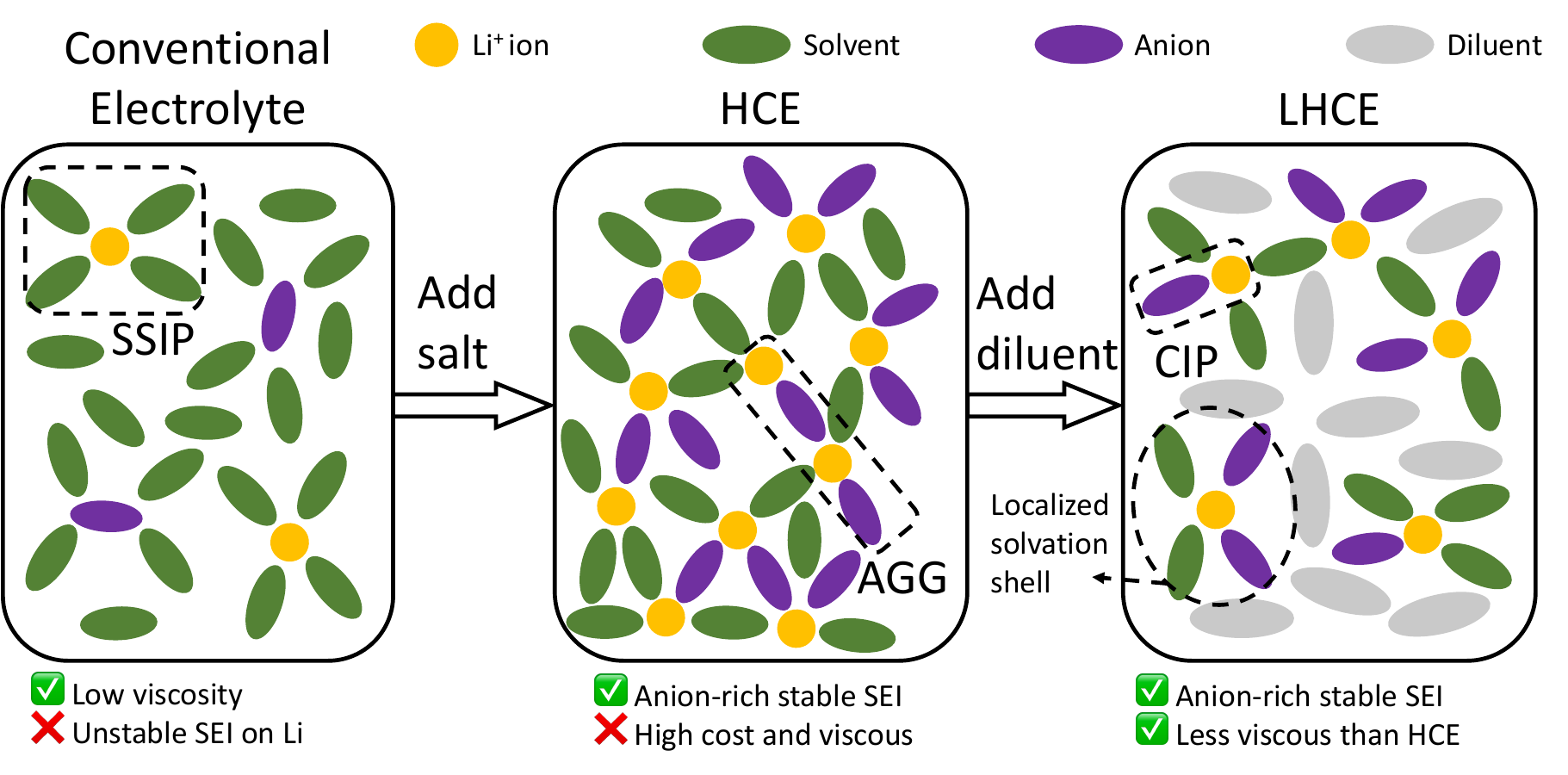}
    \label{fig:lhce-schematic}
  \end{subfigure}
  \begin{subfigure}[b]{0.6\textwidth}
    \centering
    \caption{}
    \includegraphics[width=\linewidth]{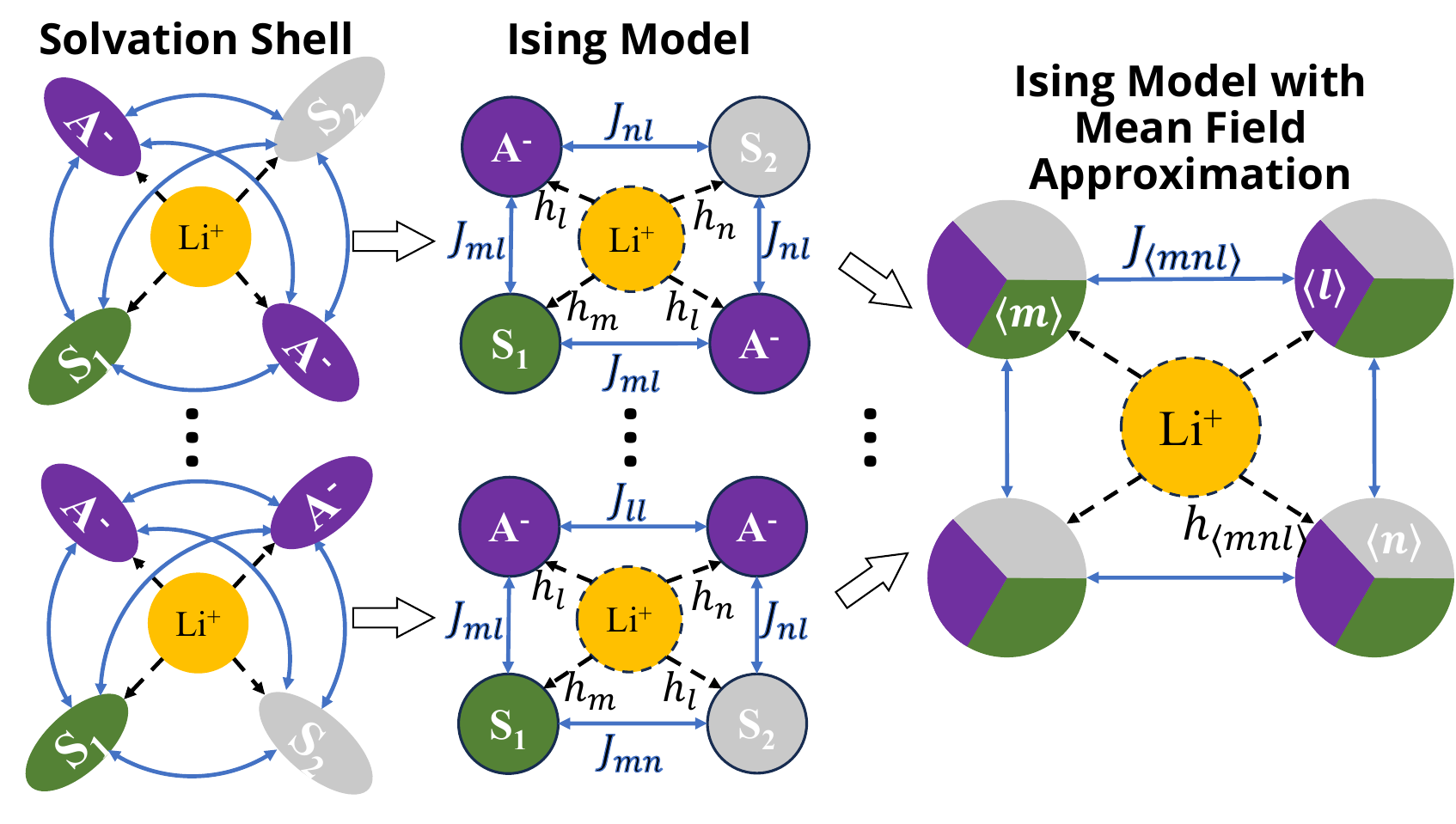}
    \centering
    \label{fig:ising-abstraction}
  \end{subfigure}
  \begin{subfigure}[b]{0.6\textwidth}
    \centering
    \caption{}
    \includegraphics[width=\linewidth]{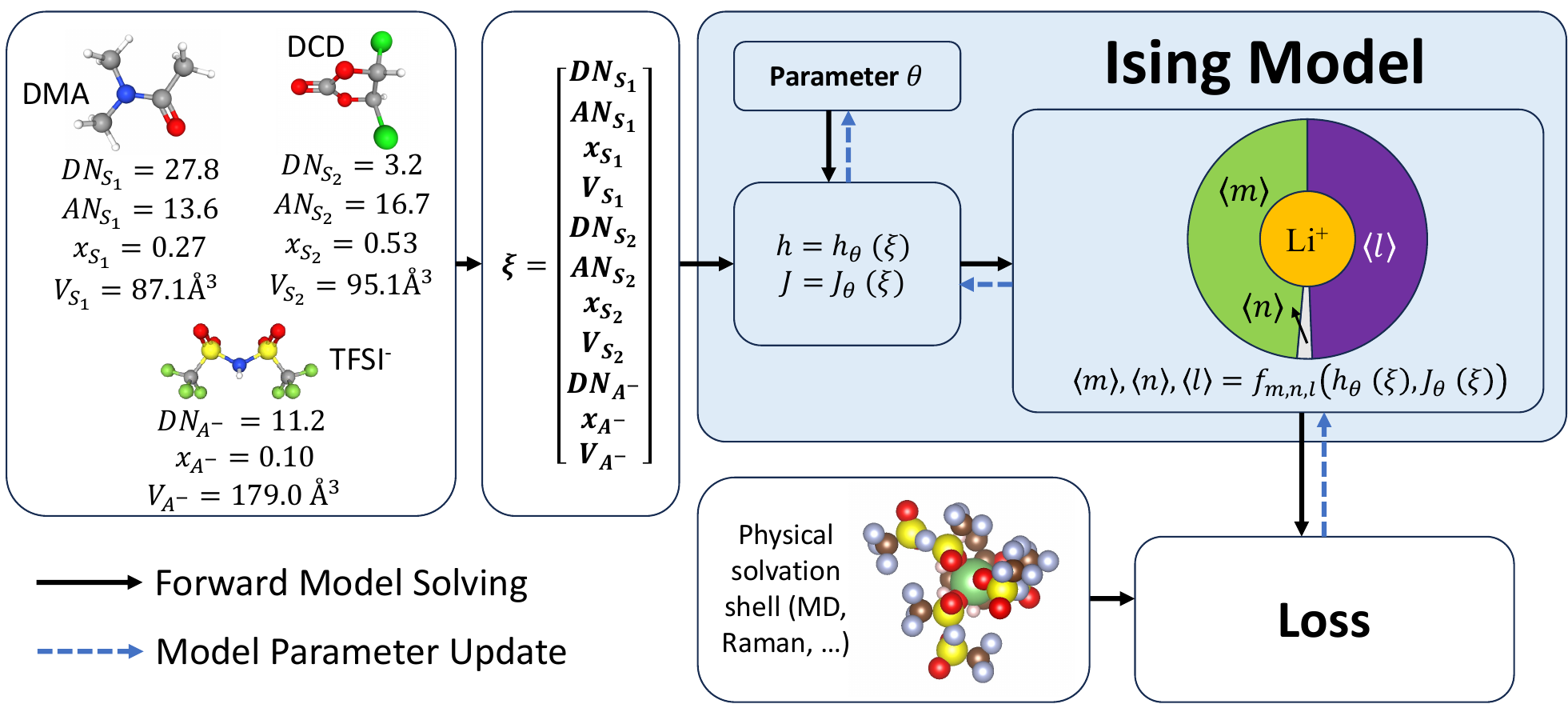}
    \centering
    \label{fig:differentiable-framework}
  \end{subfigure}
  \caption{(a) Schematic solvation structures of conventional electrolytes, HCE, and LHCE. (b) Mapping from a physical solvation shell to an Ising model. \lion-solvent and \lion-anion interactions correspond to external field energy $h_i$ (black dashed arrows), while interactions between coordinated molecules correspond to site-site interactions $J_{ij}$ (blue solid arrows). We invoke the mean field approximation to obtain mean solvation shell compositions efficiently. $m$, $n$ and $l$ correspond to solvent 1 ($S_1$), solvent 2 ($S_2$) and anion ($A^-$). (c) Model parametrization via an end-to-end differentiable framework. The black solid arrows denote forward solvation shell composition calculation. The loss is computed against MD results, and the blue dashed arrows show parameter learning through auto differentiation.}
\end{figure}

Among the electrolyte classes introduced above, LHCEs are among the most studied systems \cite{cao_reviewlocalized_2021, ni_research_2025} and have shown some of the highest Li plating/stripping coulombic efficiencies (CE) \cite{weintz_nanoengineering_2026,yang_advancing_2025}. As shown in \cref{fig:lhce-schematic}, introducing a non-coordinating, low-viscosity diluent into an HCE preserves the anion-rich solvation shells, thereby retaining their beneficial interphase compositions and outstanding electrochemical stability, while greatly reducing viscosity for practical operation \cite{chen_design_2023, chen_high-voltage_2018,cao_reviewlocalized_2021}. 
Many such formulations, spanning diverse salt-solvent-diluent combinations, have shown strong cycling stability and electrochemical performance in LMBs \cite{chen_high-voltage_2018, efaw_localized_2023, ko_omics-enabled_2024}, establishing LHCE among the most promising next-generation electrolytes.
This promise, however, comes with substantial design complexity. With three independent major components -- salt, solvent, and diluent -- each from a vast candidate space, the LHCE formulation space is far too large to explore exhaustively, and design has so far relied largely on expert knowledge and trial-and-error \cite{chen_design_2023}.

To rationalize LHCE design, prior works have explored molecular-property-based descriptors for solvent and diluent selection. The dielectric constant ($\varepsilon$) has long served as a descriptor of the solvation structure \cite{xu_electrolytes_2014, yao_atomic_2021}, with diluents typically chosen for low $\varepsilon$ \cite{takada_optimized_2019}.
Another molecular property, the Gutmann donor number (DN), which quantifies a solvent's electron-donating ability toward Lewis acids \cite{gutmann_empirical_1976}, has recently gained attention as a descriptor for \lion-solvent binding energy \cite{zhou_understanding_2024}, with qualitative design rules selecting high DN solvents and low DN diluents \cite{zhou_rationally_2022,xu_electrolyte_2023}.
\citeauthor{chen_design_2023} showed that solvents (DN $\geq$10) and diluents (DN $<$10) fall into separate DN ranges, a trend unseen in more widely used parameters such as dielectric constant and dipole moment \cite{chen_design_2023}, suggesting that DN is a leading descriptor for LHCE design.
These rules help narrow the candidate space but remain qualitative: formulations are classified only as forming ``anion-rich" or ``solvent-rich" solvation shells, without quantifying the shell composition. The role of anion-solvent interactions inside the solvation shell also remains underexplored \cite{wang_designing_2025}.

Atomistic simulations such as Density Functional Theory (DFT) and MD can in principle close this gap, providing quantitative predictions of solvation energetics and solvation shell compositions, and have been pivotal in characterizing specific LHCE structures \cite{jiao_stable_2018, cao_effects_2021, efaw_localized_2023, yang_advancing_2025}, but their high computational cost makes screening across the vast LHCE design space impractical. Chemical-physics models such as the Advanced Electrolyte Model (AEM) instead build on solvation thermodynamics and agree well with experiment across diverse formulations \cite{mceldrew_theory_2020, goodwin_theory_2023, gering_prediction_2006, gering_prediction_2017}, yet depend on extensive empirical fitting and do not transfer easily to novel formulations \cite{zhang_predicting_2022}. Machine learning models offer a complementary screening route \cite{wadell_foundation_2025, yang_unified_2026}, but their accuracy depends on training-data availability and their interpretability is limited relative to physics-based models.

Given the increasing interest in tailoring solvation structures, there is a clear need for a framework combining the accuracy of atomistic simulation with the speed and interpretability of chemical-physics models. The model should predict solvation shell composition directly from accessible molecular properties, without proprietary parameterization or extensive empirical fitting.  
In an earlier work, our team showed initial viability of this approach for low-concentration Li-O$_2$ electrolytes, using a mean-field Ising model with DN- and AN-based interactions parameterized from half-wave potentials, which explained the capacity improvement with LiNO$_3$ concentration in the LiNO$_3$/LiTFSI/DME electrolyte \cite{burke_enhancing_2015}.
However, this framework cannot be directly applied in LHCE or other electrolyte systems.

In this work, we develop an end-to-end differentiable mean-field Ising model that quantitatively predicts \lion \space solvation shell composition in LHCE systems from accessible molecular properties. Extending our earlier framework \cite{burke_enhancing_2015}, we introduce three key advances: (1) physically motivated interaction terms that respect monotonicity against DN and AN; (2) effective DN and AN values with concentration and molecular-size corrections; and (3) direct parameterization from MD solvation shell compositions via gradient-based optimization. Using only DN, AN, molar ratios, and molecular sizes as inputs, the model reaches remarkable quantitative agreement with MD for both solvation shell composition and free solvent population at a fraction of MD's cost. The model also agrees with MD on HCE, despite not trained on this class, showing promising generalizability as a universal electrolyte design tool. Analyzing its interaction terms, we find that DN dominates \lion \space solvation energetics over AN, providing a thermodynamic basis for empirical DN-based design rules. We then demonstrate the model's design utility on LiTFSI in tetraglyme (G4), absent from training: it identifies fluorobenzene as a promising diluent and predicts a concentration window with anion-rich solvation shells and low free solvent population, validated by higher-fidelity MD. 
This establishes a general, interpretable differentiable framework for designing electrolytes with target solvation shell compositions, with LHCE as a testbed but potentially extendable to other next-generation electrolyte classes.

Following the setup in our previous work \cite{burke_enhancing_2015}, we schematically illustrate in \cref{fig:ising-abstraction} how a \lion-centered solvation shell maps to an Ising model lattice. $S_1$ ($m$) and $S_2$ ($n$) represent the two different types of organic solvents in the electrolyte, which in LHCE correspond to the solvent and the diluent. $A^- \space (l)$ stands for the anion.
The composition of a \lion-centered solvation shell is intrinsically determined by energetics of interactions between shell components, including interactions between \lion and coordinating molecules and interactions among non-\lion molecules.
The former interactions involving \lion \space are mathematically equivalent to $h_\alpha$ ($\alpha=m,n,l$), or the external field energy, in the Ising model; the latter interactions are equivalent to $J_{\alpha\beta}$ ($\alpha, \beta=m,n,l$), or the neighboring site interaction energies. Each lattice site $i$ is occupied by either the solvent, diluent, or anion, denoted $m_i$, $n_i$, and $l_i$. We assume that solvation shell coordination numbers are invariant, so $m_i,n_i,l_i=0,1$ and $m_i+n_i+l_i=1$. The site occupations and interactions can be used to construct the system's Hamiltonian (Eq.~\ref{eq:full_hamiltonian}), which governs the system's equilibrium composition.

In an actual electrolyte, the \lion \space solvation shell composition fluctuates around a mean value due to microscopic thermal motion. Capturing this via conventional Monte Carlo (MC) sampling is accurate but computationally demanding. We instead invoke the mean field approximation to solve directly for a \textit{representative} mean solvation shell composition (\cref{fig:ising-abstraction}). Under this approximation, neighboring site interaction terms are replaced by the energy of a site under a mean field exerted by averaged occupations. For example, the total solvent-anion interactions $J_{ml}\sum_{i,j}m_il_j$ in the full Hamiltonian (Eq.~\ref{eq:full_hamiltonian}) becomes $J_{ml}\frac{z}{2}\langle l \rangle \sum_i m_i$ in the mean-field Hamiltonian (Eq.~\ref{eq:hamiltonian_mean_field}), using coordination number $z=N/2=1.86$, with $N=3.72$ being the average \lion \space coordination number from MD. This reduces the problem to solving for average occupations $\langle m \rangle$, $\langle n \rangle$, $\langle l \rangle$ within a single lattice as a root-finding problem, eliminating the need for lengthy MC simulations (full derivation in Supporting Information (SI)).

To compute fractional occupations, the interaction terms $h_\alpha$ and $J_{\alpha\beta}$ need to be parameterized. \lion \space is a Lewis acid, so the \lion-solvent interaction free energy depends on the solvent's donating ability. Building on our group's prior finding that Li/\lion \space half-wave potential correlates with solvent DN \cite{khetan_trade-offs_2015}, we use a sigmoid-based function for the enthalpic contribution to $h$. This enforces the physical constraint of monotonic saturation at high DN, a constraint the previous quadratic fit did not guarantee. The entropic contribution is dependent on molar ratio $x$, and the full $h$ term is:
\begin{equation}
  h(DN, x) = \left[ a_0 + \frac{a_1} {1 + a_2 \exp(-a_3 DN)} \right]+ a_4 \log(x)
  \label{eq:h-term}
\end{equation}
where $a_t (t=0,1,2,3,4)$ are model parameters. We use the same functional form with a separate parameter set for \lion-anion interactions.

The anion-solvent interaction depends on the solvent's accepting ability and the anion's donating ability, i.e., DN$_{A^-}$ and AN$_{sol}$, described using a similar functional form:
\begin{equation}
  J_{\alpha\beta}(DN, AN, x_\alpha, x_\beta) = \left[b_0 + \frac{b_1} {1 + \exp(b_2 + b_3 DN + b_4 AN)} \right] + b_5 \left[\log (x_\alpha) + \log(x_\beta)\right]
  \label{eq:an_j}
\end{equation}
Solvent-solvent and anion-anion ($J_{ll}$, modeled as functions of anion DN) interactions are parameterized analogously, with parameters constrained such that enthalpic terms decrease monotonically with increasing DN/AN. We use \textit{effective} DN$_{eff}$ and AN$_{eff}$ scaled by molar ratio and molecular size, rather than experimental values, to correct for concentration and volume effects (details in SI).

In total, the model contains 45 learnable parameters (Table~\ref{tab:parameter_count}). This number of parameters is too large to optimize efficiently with traditional methods such as grid search, Bayesian optimization, or Covariance Matrix Adaptation Evolution Strategy (CMA-ES) \cite{chen_differentiable_2026}. We therefore implement the full model in JAX \cite{bradbury_jax_2018}, an end-to-end differentiable programming framework enabling gradient-based parameter optimization (\cref{fig:differentiable-framework}). Model inputs $\xi$ (DN, AN, molar ratio, molecular size) determine $h$ and $J$ via parameters $\theta$. Occupations are solved via a Broyden root-finding solver, and the loss is the RMSE between predicted and MD average occupations (Eq.~\ref{eq:train-loss}). Gradients $\partial L/\partial\theta$ are obtained via automatic differentiation and used to update $\theta$ with the ADAM optimizer \cite{kingma_adam_2015} (training schedule in SI).

\begin{figure}
  \centering
  \begin{subfigure}[b]{0.8\textwidth}
    \centering
    \caption{}
    \includegraphics[width=\linewidth]{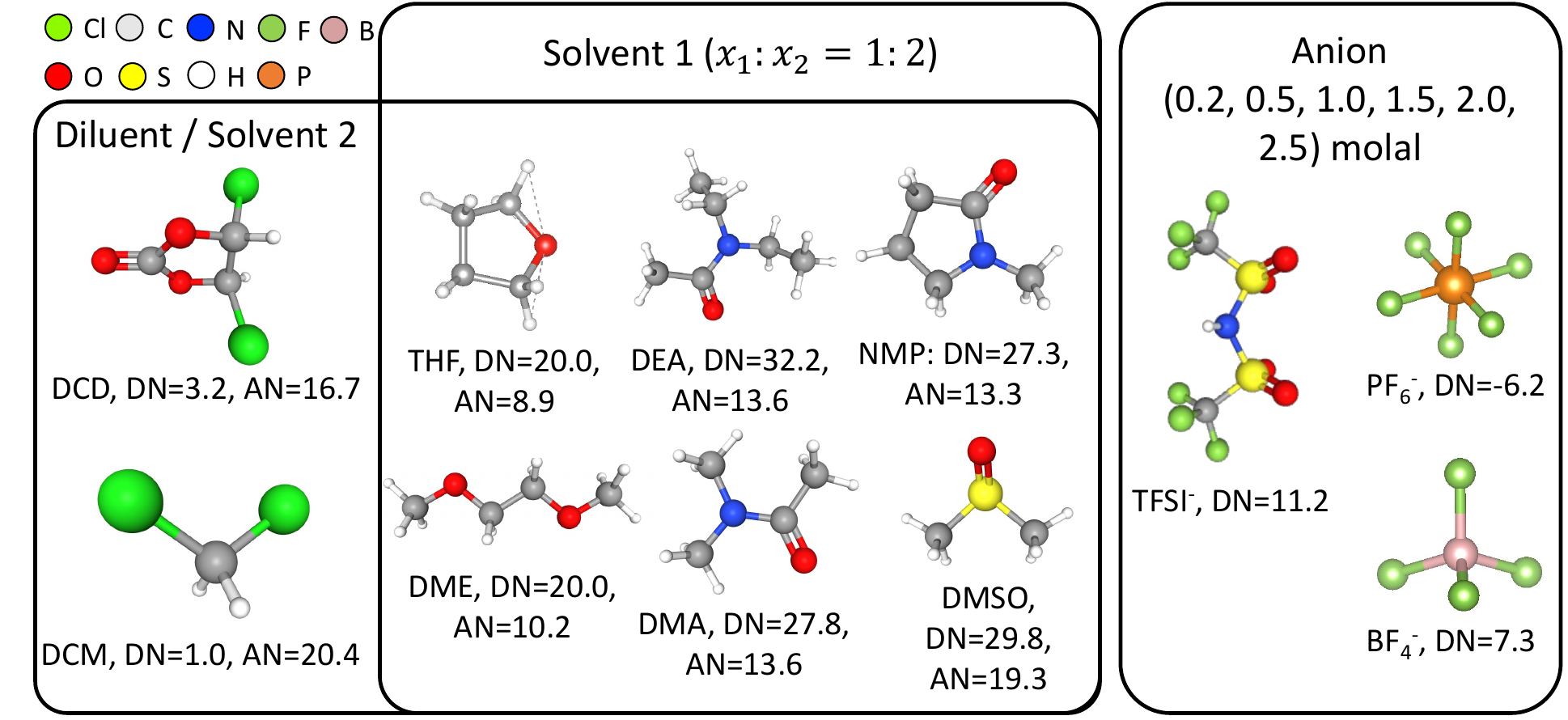}
    \centering
    \label{fig:modeled-system}
  \end{subfigure}
  \begin{subfigure}[b]{0.4\textwidth}
    \centering
    \caption{}
    \includegraphics[width=\linewidth]{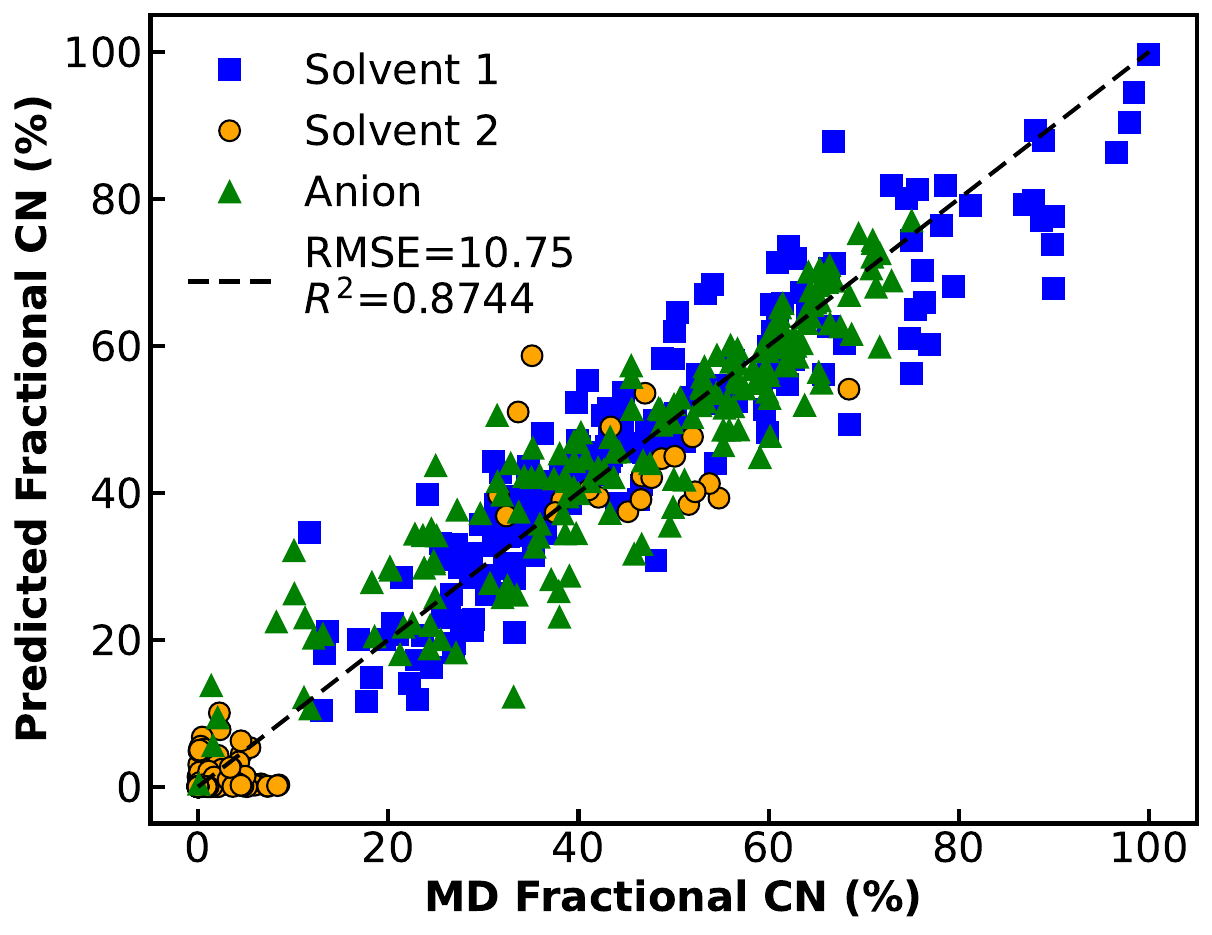}
    \centering
    \label{fig:kfold_parity}
  \end{subfigure}
  \begin{subfigure}[b]{0.39\textwidth}
    \centering
    \caption{}
    \includegraphics[width=\linewidth]{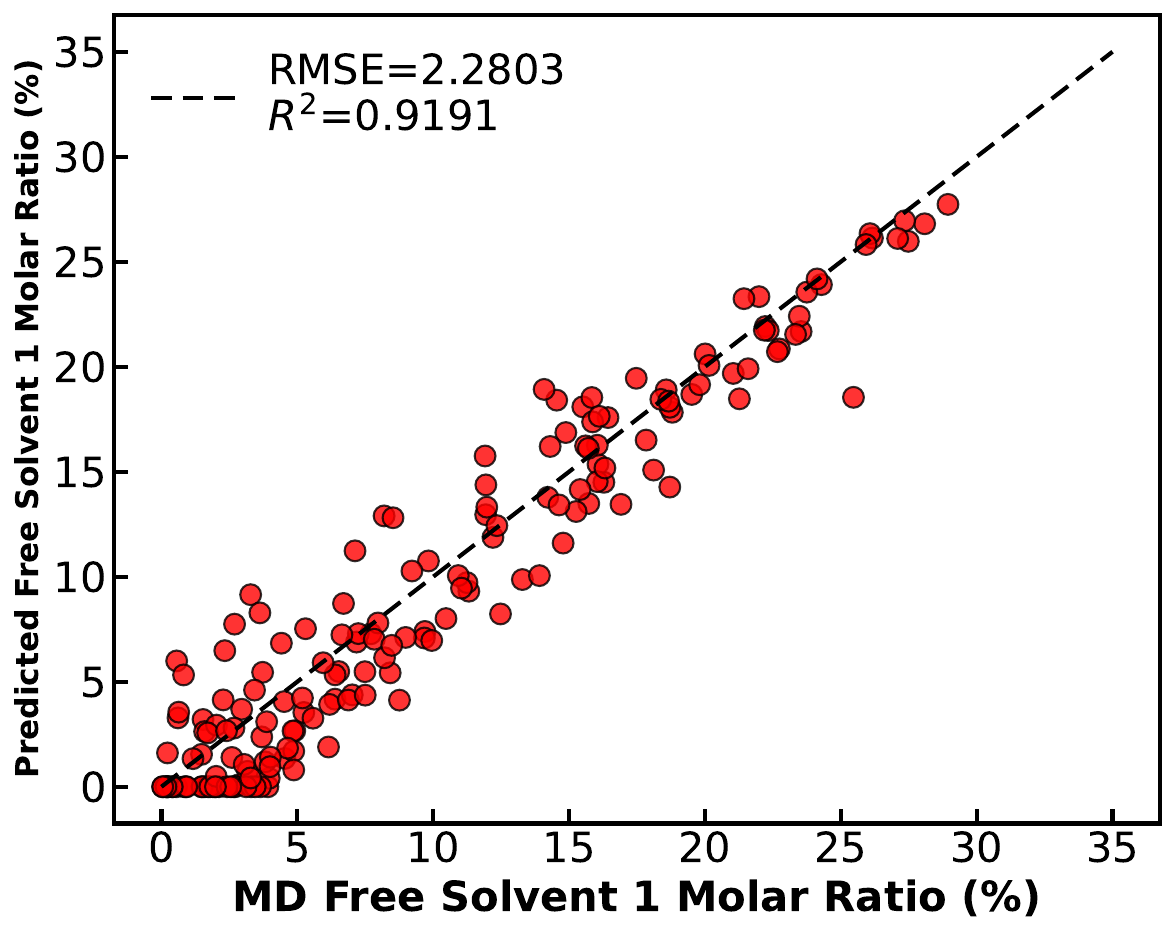}
    \centering
    \label{fig:free_solvent_test}
  \end{subfigure}
  \begin{subfigure}[b]{0.4\textwidth}
    \centering
    \caption{}
    \includegraphics[width=\linewidth]{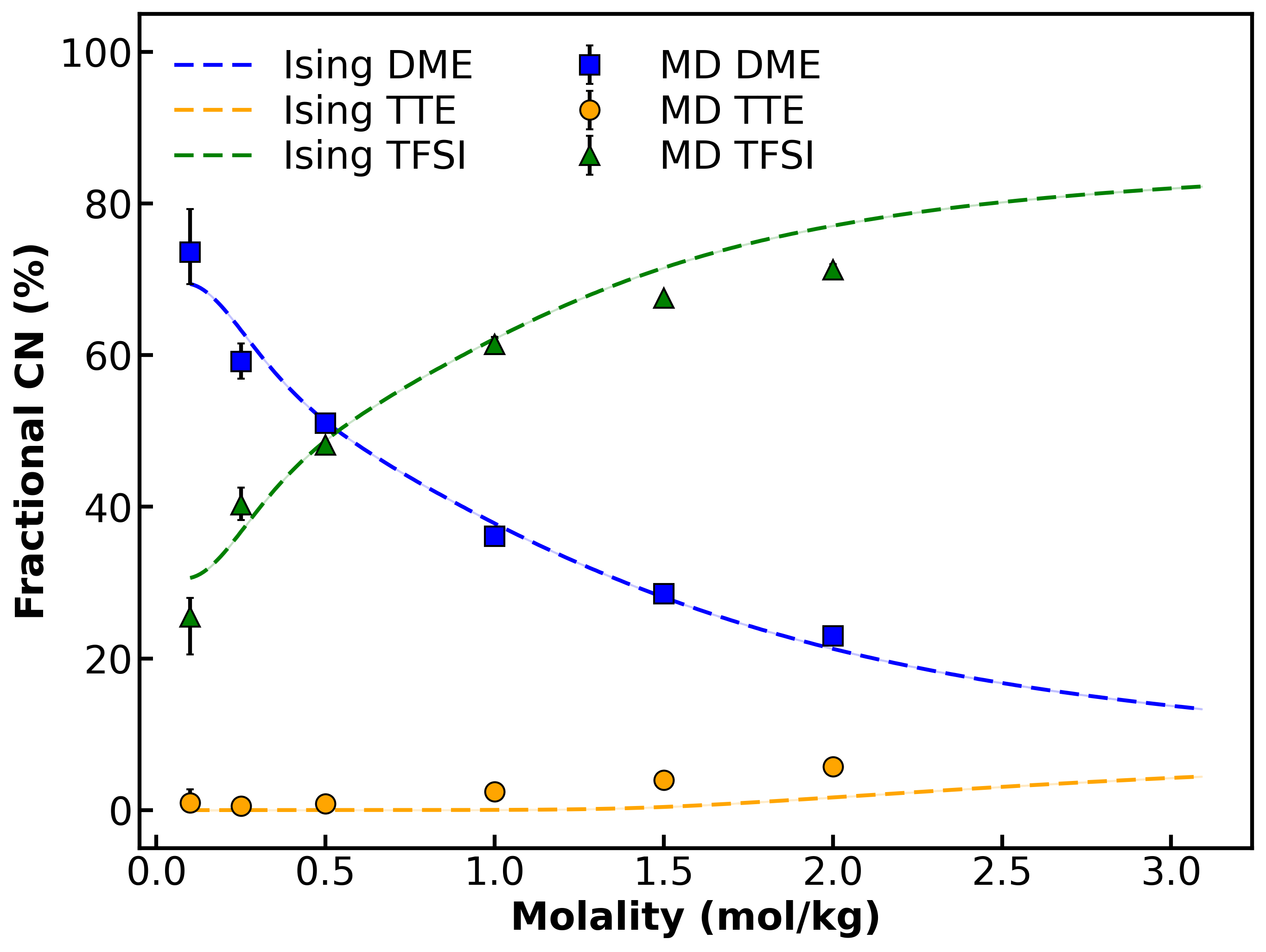}
    \label{fig:dme_tte_litfsi}
  \end{subfigure}
  \begin{subfigure}[b]{0.4\textwidth}
    \centering
    \caption{}
    \includegraphics[width=\linewidth]{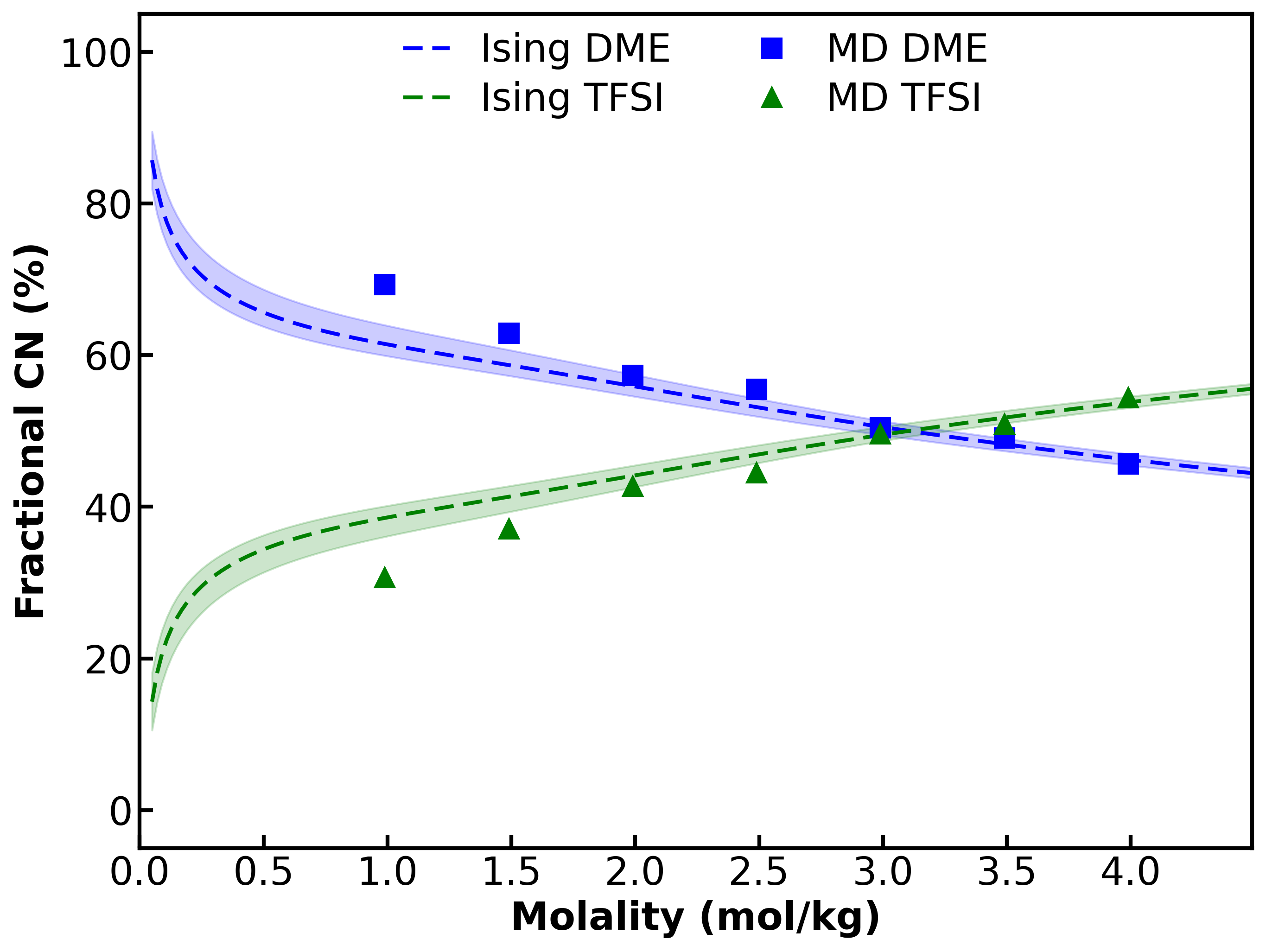}
    \label{fig:dme_litfsi}
  \end{subfigure}
  \caption{(a) MD-modeled electrolyte systems in this work, in total 182 formulations in the training dataset. (b), (c) 5-fold cross validation parity plot between our model and MD on: (b) fractional \lion \space coordination number (CN), and (c) free solvent molar ratio in the electrolyte. Both results show good model accuracy. (d) Comparison between model-predicted and MD-predicted \lion \space fractional CN on 0.5-3.0m LiTFSI in DME:TTE=1:2 (mol:mol). (e) Despite not trained explicitly on HCE data, our model shows promising generalizability on LiTFSI/DME HCE system.}
  \label{fig:accuracy}
\end{figure}

We use MD to obtain electrolyte solvation structures for model training, and DFT to compute molecular sizes (full simulation details are provided in SI).
The training dataset contains 182 electrolyte formulations, each with 2 solvents and 1 salt. The whole dataset covers 8 solvents/diluents selected from the Stenutz Gutmann DN/AN dataset \cite{stenutz_gutmann_2026}, and 3 salts (LiPF$_6$, LiBF$_4$, LiTFSI) under multiple concentrations (\cref{fig:modeled-system}). The solvents/diluents are selected based on experimental DN reliability, electrochemical stability, and likelihood of LHCE formation. The last criterion follows the design principle proposed by \citeauthor{chen_design_2023}, where molecules with DN$<$10 are likely diluents and those with DN$\geq$10 are likely solvents\cite{chen_design_2023} (full selection criteria and molecule list in Table S3). Many chosen solvents have been previously explored in the HCE regime \cite{ko_omics-enabled_2024,yang_stable_2024,kim_tailored_2023}. Of the 182 formulations, 22 have $DN_2>10$, where solvent 2 acts as a co-solvent rather than a diluent, to improve model generalizability.

\cref{fig:kfold_parity} shows the 5-fold cross-validation parity plot comparing model-predicted \lion \space fractional coordination numbers (CN) against MD, with blue squares, orange circles, and green triangles representing solvent 1, solvent 2 (diluent/co-solvent), and anions, respectively. The model reaches quantitative agreement with MD, with a summed RMSE of 10.75\% and $R^2=0.87$, and component-wise RMSEs reaching 7.43\%/4.40\%/7.17\% for solvent 1/solvent 2/anion (Fig.~\ref{fig:parity_sep_mnl}). Our model requires only a small fraction of MD's $\sim$100 CPU-hour cost ($<1$s per formulation), enabling high-throughput screening without HPC access. The largest errors ($>20\%$ RMSE) occur for a low-concentration LiTFSI in DME/DMA electrolyte with $>75\%$ uncoordinated free solvent, and a DMSO/DCM system where DMSO's high DN, small size, and high \lion \space coordination numbers contribute to the large error (Fig.~\ref{fig:rmse_vs_cn}). Since DMSO-containing electrolytes typically do not form anion-rich shells, these errors are not critical for LHCE design.

In addition to anion-rich \lion \space coordination, low free solvent population is critical in LHCE for suppressing electrode corrosion and extending the electrochemical stability window \cite{yamada_advances_2019,ren_localized_2018,li_concentrated_2023}. We calculate this property via $x_{free-solvent} = x_{solvent} - x_{Li^+}\langle m \rangle N$, which matches MD quantitatively (RMSE 2.29\%, $R^2=0.92$; \cref{fig:free_solvent_test}), with major discrepancies at low free-solvent fractions where triple-ion-pairs and aggregates become more dominant \cite{zhang_predicting_2022}. Since first-pass screening should avoid missing viable low-free-solvent candidates, we also evaluate classification accuracy at a $<10\%$ free-solvent threshold. The model achieves a false negative rate of 5.5\% and recall of 0.94 (Fig.~\ref{fig:conf_matrix_10}). Together, these results confirm the model's ability to predict both \lion \space coordination for anion-rich shell design, and free solvent ratio for low-free-solvent design.

Since low test-set RMSE does not guarantee practical usefulness \cite{phuthi_accurate_2024, zhao_effect_2025}, we validate against common LHCE formulations absent from training -- LiTFSI \& LiPF$_6$ in DME/DMC solvents with TTE/BTFE as diluents. \cref{fig:dme_tte_litfsi} shows that our model faithfully reproduces the trend of increasing anion and decreasing solvent occupations with salt concentration, correctly predicting that LiTFSI anion occupation exceeds solvent occupation above 0.5m. Our model also predicted that LiPF$_6$ dissociates more completely than LiTFSI at the same molality, consistent with MD and common observations \cite{zhang_predicting_2022} (Fig.~\ref{fig:real-system-result-si}). Larger errors are seen below 0.5m, likely because solvation at low concentrations is also governed by dielectric constant, which our model does not consider. Despite this limitation, the model accuracy remains high above 0.5m, which is the concentration range relevant for LHCE design. In addition, as a preliminary test of model generalizability beyond LHCE, we also apply the same LHCE-trained parameter set directly to HCE and HEE formulations without re-parameterization. \cref{fig:dme_litfsi} shows that the model achieves good agreement on LiTFSI/DME HCE, with a mean absolute error of $\sim2.5\%$. For other HCE systems, our model reproduces anion occupation increase with salt concentration (Fig. \ref{fig:hce-generalizability}); while for HEE the model correctly predicts higher BF$_4^-$ coordination tendency with \lion \space than TFSI$^-$ resulting from molecular size effects (Figs.~\ref{fig:hee-generalizability},~\ref{fig:hcee-generalizability}). These results indicate that the model has learned the underlying solvation physics rather than memorized its training data, and suggests that the framework can be extended to other electrolyte classes and used as a general solvation-shell design tool if targeted training data is available.

\begin{figure}[h!]
  \centering
  \begin{subfigure}[b]{0.32\textwidth}
    \centering
    \caption{}
    \includegraphics[width=\linewidth]{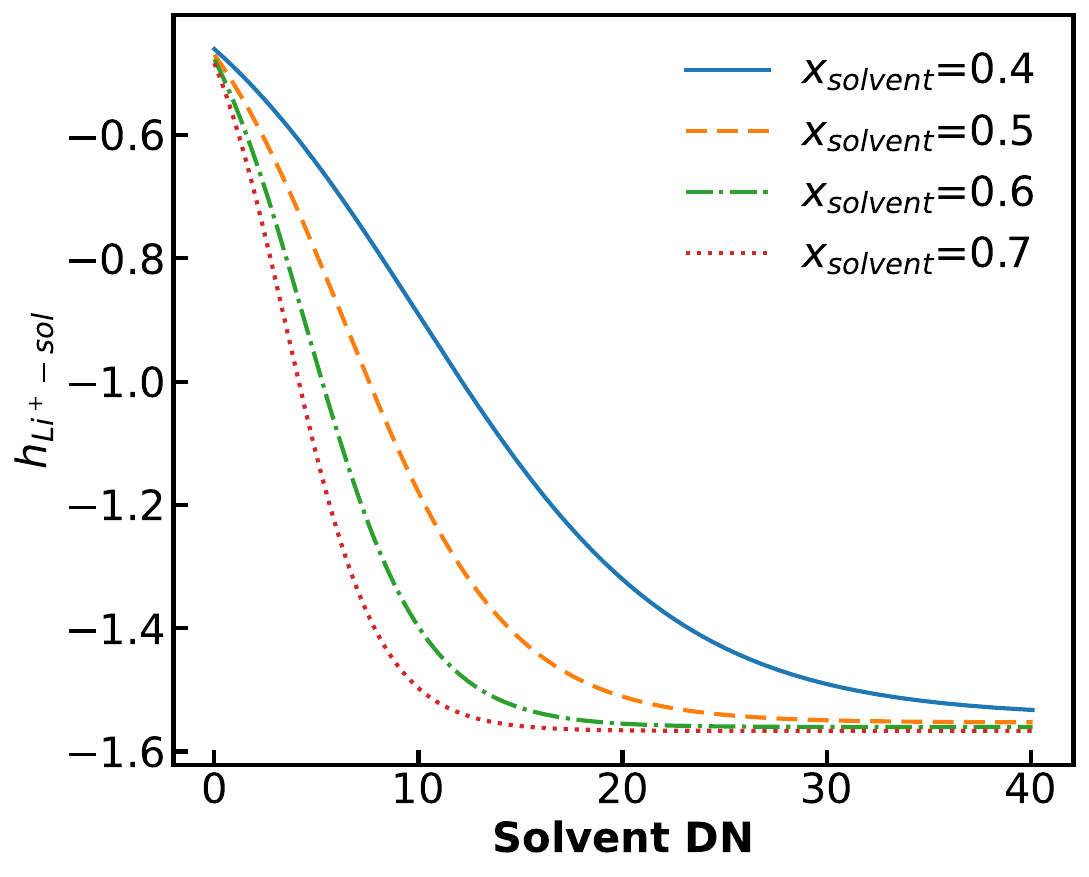}
    \centering
    \label{fig:sol_dn_func}
  \end{subfigure}
  \begin{subfigure}[b]{0.32\textwidth}
    \centering
    \caption{}
    \includegraphics[width=\linewidth]{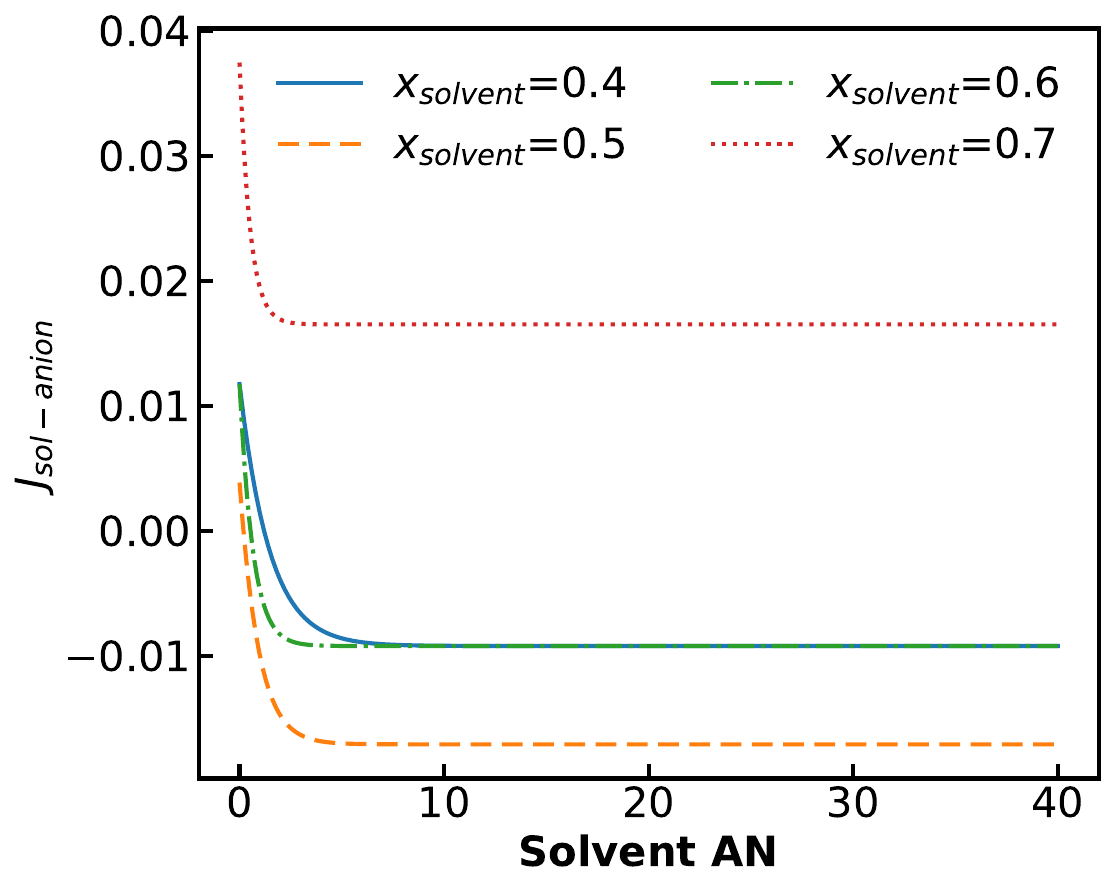}
    \centering
    \label{fig:an_m_func}
  \end{subfigure}
  \begin{subfigure}[b]{0.33\textwidth}
    \centering
    \caption{}
    \includegraphics[width=\linewidth]{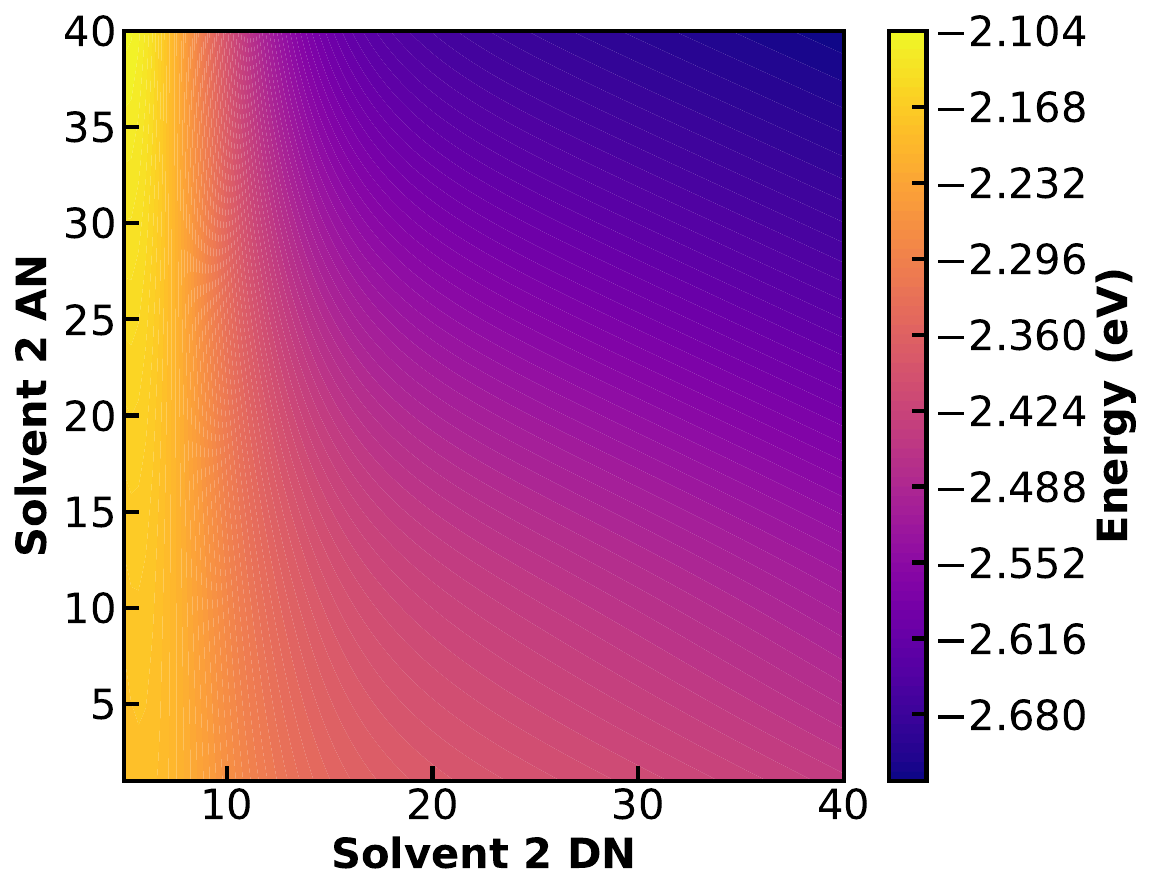}
    \centering
    \label{fig:li_free_energy}
  \end{subfigure}
  \caption{Interactions within our model as a function of solvent DN/AN and molar ratio. (a) $h_{Li^+-solvent}$ as a function of solvent DN and molar ratio. The interaction strengthens with increasing DN and molar ratio. (b) Anion-solvent interaction energy as a function of solvent AN and molar ratio. Interaction energy changes much more significantly with varying DN than AN. Both figures assume $x_{A^-}+x_{Li^+}+x_{sol}=1.0$ and $x_{sol}:x_{dil}=1:2$. (c) \lion \space solvation free energy contour as a function of solvent 2 DN and AN with fixed solvent 1 DN (20.0) and anion molar ratio (0.14). Compared to DN, AN has a much weaker effect on \lion \space solvation energy, implying that DN plays a dominant role in controlling \lion \space solvation environment.}
  \label{fig:physical-meaning}
\end{figure}

One benefit of the Ising model lies in its physical interpretability. We visualize the key energetic terms to provide insight for LHCE design. \cref{fig:sol_dn_func} shows $h_{Li^+-sol}$ as a function of solvent DN and molar ratio: the interaction strengthens by $\sim1.2$eV from $DN=0$ to $DN=40$, increasing almost linearly below $DN=20$ before plateauing, matching the trend in Li/\lion \space half-wave potentials \cite{khetan_trade-offs_2015}. This saturation is not an artifact of our sigmoid form but is learned directly from solvation data: no constraints are placed on its range or slope during parameterization, and no analogous plateau appears at low DN (full discussion in SI). Solvent concentration also modulates $h_{Li^+-sol}$: increasing $x_{solvent}$ from 0.2 to 0.5 lowers the plateau by $0.044$eV and shifts its onset from $DN\sim30$ to $DN\sim15$, implying weaker DN-sensitivity at low salt concentration (high solvent fraction).

\cref{fig:an_m_func} shows the solvent-anion interaction $J_{sol-anion}$ as a function of solvent AN and molar ratio. In contrast to DN, AN has minimal impact: $J_{sol-anion}$ saturates by $AN\sim4$ with only $\leq0.021$eV variation below this value, likely because anions are larger and less electron-dense than \lion, weakening their sensitivity to solvent properties. Molar fraction affects $J_{sol-anion}$ non-monotonically via entropic stabilization (maximized at $x=0.5$), contributing to SSIP-dominated structures at low salt concentration and AGG-dominated structures at high concentration.

The absolute magnitude of $J_{sol-anion}$ is $\sim$1.5eV smaller than $h_{Li^+-sol}$, with similar behavior for $h_{Li^+-A^-}$ (Fig.~\ref{fig:salt_dn_func}), indicating that \lion \space solvation energetics are dominated by DN with only a marginal AN contribution. To quantify this, we compute the \lion \space solvation free energy $G_{Li^+} = \langle m \rangle h_{Li^+-sol_1} + \langle n \rangle h_{Li^+-sol_2} + \langle l \rangle h_{Li^+-A^-}$ for LiTFSI in a fixed solvent 1 ($DN_{DME}=20.0$kcal/mol) plus a varying solvent 2 ($x_{A^-}=0.14$; \cref{fig:li_free_energy}). At fixed solvent 2 DN=10, varying AN changes $G_{Li^+}$ by only 0.13eV, whereas at fixed AN=10, increasing DN from 10 to 40 drops $G_{Li^+}$ by 0.3eV. This confirms that DN, not AN, is the dominant lever for tuning solvation shell composition, explaining its established role as the leading-order LHCE design descriptor \cite{chen_design_2023}.

\begin{figure}[h!]
  \centering
  \includegraphics[width=0.9\linewidth]{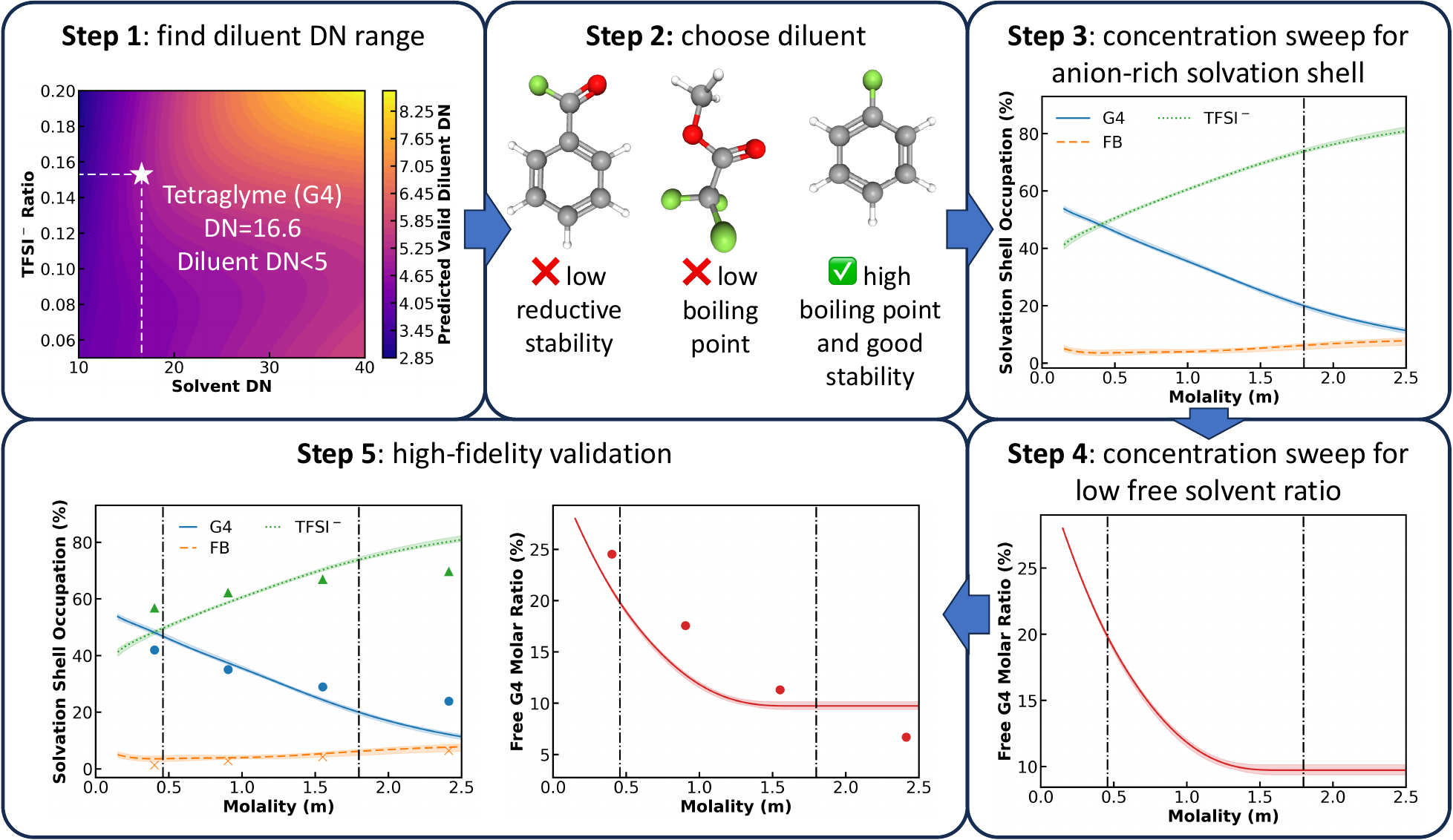}
  \caption{Representative LHCE design boundary prediction with our model, using LiTFSI in tetraglyme (G4, DN=16.6 kcal/mol) as an example. Step 1: screening anion molar ratios identifies molecules with DN$<$5 as valid diluents (fractional \lion \space CN $<$10\%). Step 2: among fluorinated candidates in this range, fluorobenzene (FB, DN=3.0 kcal/mol) is chosen for its electrochemical stability and boiling point. Step 3: sweeping salt concentration for anion-rich solvation shells. Although anion-rich solvation shells form even at low concentrations, we discard formulations above 1.8m, where solvent shell occupation $<20\%$, risking salt precipitation. Step 4: sweeping for low free solvent ratio ($<20\%$) sets a lower concentration bound of 0.46m. Step 5: higher-fidelity MD validates that 0.46-1.8m LiTFSI is a promising LHCE range, with our model agreeing well with MD in both solvation shell occupations and free solvent ratios. This electrolyte formulation promotes both a stable anion-derived SEI via anion-rich \lion \space coordination and good oxidative stability via low free-solvent population.}
  \label{fig:design-workflow}
\end{figure}

Having established DN as the primary LHCE design descriptor, we demonstrate our model's design utility (\cref{fig:design-workflow}). Since molecules with DN$<$10 cannot adequately dissolve Li salts \cite{chen_design_2023}, solvent choice is restricted to the DN=10-40 range. This choice can be made via expert knowledge or database lookup \cite{stenutz_gutmann_2026}, or facilitated by ML property prediction \cite{wadell_foundation_2025}. We choose tetraglyme (G4, DN=16.6 kcal/mol) from the ether family, which is unseen in training, as our base solvent. In step 1, we sweep diluent DN and anion molar ratio to identify the diluent DN boundary (diluent shell occupation $<10\%$). This boundary shifts to higher DN as solvent DN increases, since stronger solvents require stronger co-solvents to displace them from the solvation shell (full discussion in SI). For G4/LiTFSI, our model predicts diluents with DN$<$5, a sufficient but not necessary criterion (Fig.~\ref{fig:diluent-choice-md-comparison}). In step 2, we select fluorinated candidates in this DN range for their favorable LMB electrochemical stability \cite{ni_research_2025}. Among the three candidates in the Stenutz database \cite{stenutz_gutmann_2026}, fluorobenzene (FB, DN=3.0 kcal/mol) offers the best combination of reductive stability, boiling point, and viscosity (Fig.~\ref{fig:3-diluent-candidate-comparison}), and is chosen as our diluent.

In steps 3-4, we sweep salt concentration to identify the range yielding both an anion-rich solvation shell and low free solvent ratio. We set criteria for anion and solvent shell occupations to be above 30\% and 20\% respectively based on model accuracy and literature knowledge \cite{chen_high-voltage_2018,yang_advancing_2025,yang_unified_2026} (full justification in SI), resulting in concentrations below 1.8m, while a separate $<20\%$ free-solvent threshold requires salt concentrations above 0.46m. Thus, the target design window is 0.46-1.8m LiTFSI for the G4/FB system. In step 5, higher-fidelity MD validates this range: predicted and MD solvation shell occupations and free solvent ratios agree well across concentrations, with larger discrepancies concentrated outside the design window (more discussion in SI). This agreement confirms that our model identifies promising formulations ahead of costly MD or experimental validation. We further demonstrate generalizability of this design workflow across additional systems (Figs.~\ref{fig:pc-fb-design}--\ref{fig:g4-fb-no3-design}). 

In summary, we develop an end-to-end differentiable mean-field Ising model for rational LHCE design, extending our group's previous framework with physically monotonic interaction terms and direct parameterization from MD-predicted solvation shell compositions. The model reaches quantitative agreement with MD in both solvation shell prediction (RMSE $<11\%$, $R^2=0.87$) and free solvent prediction (RMSE 2.3\%) at a fraction of MD's cost. Analysis of the model's interaction terms reveals that DN dominates \lion \space solvation energetics over AN, establishing DN as the leading-order LHCE design descriptor. We demonstrate the model's design utility on LiTFSI in tetraglyme (G4), a system absent from training: the model identifies fluorobenzene as a promising diluent and predicts a 0.46-1.8m LiTFSI design window with anion-rich, low-free-solvent solvation structures useful for LMB electrolytes, confirmed by higher-fidelity MD.

Beyond LHCE, this work establishes a general, interpretable thermodynamic framework connecting molecular descriptors to solvation shell composition, complementing qualitative design rules and data-driven ML approaches without requiring costly atomistic simulations. We additionally find that the current parameter set captures expected solvation behavior for HCE and HEE without re-parameterization, suggesting straightforward extensibility as more training data becomes available for these and other classes (e.g., WSEs) or cations (Na$^+$, Zn$^{2+}$). The model's end-to-end differentiability allows systematic re-parameterization rather than redesign in these cases.

\begin{acknowledgement}

This work was supported by ARPA-E DOE JOULES-1K award DE-AR0001883 and ARPA-E DOE JOULES award DE-AR0001728. The views and opinions of authors expressed herein do not necessarily state or reflect those of the United States Government or any agency thereof. 
This work used Bridges-2 at Pittsburgh Supercomputing Center through allocation CTS180061 from the Advanced Cyberinfrastructure Coordination Ecosystem: Services \& Support (ACCESS) program. 
H.Z. acknowledges help from Victor Azumah, Hongshuo Huang, and Dr. Haotian Chen in idea formulation and paper writing.

\end{acknowledgement}

\section{Data Availability}

Data will be made available upon publication.

\begin{suppinfo}
Details of model setup, additional information about interaction terms, training dataset molecule selection, MD benchmark results, additional model accuracy and interaction term visualizations, additional discussions on interaction physical monotonicity, diluent DN boundary, LHCE design workflow, and model scope and future extensions.
\end{suppinfo}

\newpage
\bibliography{ref}

\clearpage

\makeatletter
\let\SectionsOn\acs@saved@SectionsOn
\let\SectionNumbersOn\acs@saved@SectionNumbersOn
\makeatother
\SectionsOn
\SectionNumbersOn
\setcounter{section}{0}
\setcounter{figure}{0}
\setcounter{table}{0}
\setcounter{equation}{0}
\renewcommand{\thesection}{S\arabic{section}}
\renewcommand{\thefigure}{S\arabic{figure}}
\renewcommand{\thetable}{S\arabic{table}}
\renewcommand{\theequation}{S\arabic{equation}}

\makeatletter
\let\maketitle\acs@saved@maketitle
\let\@maketitle\acs@saved@@maketitle
\let\title\acs@saved@title
\makeatother
\title{Supporting Information for \\
Differentiable Solvation Shell Model for Rational Electrolyte Design}
\maketitle
\section{Details of model setup}
As stated in the main text, the full Hamiltonian of the system is formulated as:
\begin{dmath}
    H = h_m\sum_{i=1}^N m_i + h_n\sum_{i=1}^N n_i + h_l\sum_{i=1}^N l_i + J_{mm} \sum_{i,j} m_i m_j + J_{nn}\sum_{i,j} n_i n_j + J_{ll}\sum_{i,j} l_i l_j + J_{mn} \sum_{i,j} m_i n_j + J_{nm} \sum_{i,j} n_i m_j + J_{ml} \sum_{i,j} m_i l_j + J_{lm} \sum_{i,j} l_i m_j + J_{nl} \sum_{i,j} n_i l_j + J_{ln} \sum_{i,j} l_i n_j
    \label{eq:full_hamiltonian}
\end{dmath}
One potential risk of directly solving Eq.\ref{eq:full_hamiltonian} lies in the fact that each site is fully occupied by one species only, which may lead to failure in capturing the generalized solvation shell composition. Full occupation of one species on one site essentially describes only one possible solvation shell structure, but molecules constantly exchange between the solvation shell and the bulk electrolyte driven by thermal motion, so multiple different shell configurations may exist throughout space and time.
Furthermore, using one snapshot with singular site occupancy may leave out possible interactions between molecules within the solvation shell that are not described as nearest neighbors in the Ising model lattice in this snapshot.

One possible solution is to run Monte Carlo (MC) simulations to generate an ensemble of shell compositions and calculate the ensemble average, but this is quite computationally demanding. Another possible solution is to invoke the mean field approximation to describe site occupancies. In the mean field approximation, rather than summing over all $\sigma_i\sigma_j(\sigma=m,n,l)$ pairs when calculating pairwise interactions, the interaction between the two molecules is modeled as molecule $\sigma_i$ on site $i$ in the mean field produced by all molecules of the same species as $\sigma_j$. In other words, the $J_{\alpha\beta}\sum_{i,j}\sigma_i\sigma_j$ term in Eq.\ref{eq:full_hamiltonian} is replaced by $J_{\alpha\beta}z\langle \sigma \rangle\sum_i \sigma_i$, where the averaged occupation $\langle \sigma \rangle = \frac{1}{N}\sum_{j=1}^N \langle \sigma_j\rangle$ describes the mean field, and is allowed to take any real value from 0 to 1 (as opposed to 0 or 1 in a single snapshot). This treatment ensures that the modeled solvation shell is representative, as the mean occupation is essentially equivalent to the ensemble-averaged occupation. It also eliminates the issue of overlooking possible interactions in one snapshot through the use of partial occupations. An important property of the average occupation $\langle \sigma \rangle$ is its spatial invariance $\langle \sigma \rangle = \langle \sigma_j \rangle$, which corresponds to the homogeneity of the bulk electrolyte. $z=N/2$ is the lattice coordination number. The Ising model lattice is chosen to be a modified two-dimensional square lattice with a nearest neighbor count of $N=3.72$, based on our MD mean \lion \space coordination numbers.

We also note that in addition to interactions between molecules within the first solvation shell, there are interactions between different solvation shells. In these inter-solvation-shell interactions, each solvation shell acts as a whole and interacts with other \lion-centered solvation shells. These interactions are non-negligible when two solvation shells are spatially very close to each other, such as in the case of triple-ion-pairs (TIP) \cite{gering_prediction_2017}. However, these interactions are not well-captured in the original Ising model, since the pairwise interactions only involve nearest neighbors. In our model, we do not explicitly apply a hierarchical Ising model approach, but correct for these long-range interactions through an additional implementation of the mean field approximation. For a modeled solvation shell, the interaction energy between itself and all the solvation shells in the vicinity can be treated as the effect of a mean field exerted by all the components $\langle \sigma \rangle$ in those shells. This means that we can apply the mean field treatment once more and rewrite the pairwise interaction terms as $J_{\alpha\beta}\frac{z}{2}\sum_i \sigma_i$, where the $z/2$ accounts for the double counting in the inter-solvation-shell mean field. In this treatment, the $J_{\alpha\beta}$ terms not only account for the intra-shell interaction within the first solvation shell, but also account for the inter-shell interaction between different solvation shells. We need to note that when $\alpha=\beta$, the pairwise interaction term should be written as $J_{\alpha\alpha}z\sum_i \sigma_i$, using $z$ instead of $z/2$ as the coordination number, because component $\alpha$ in the modeled solvation shell acts not only as a receptor of the mean field but also as a contributor.

Overall, the full Hamiltonian of the system after applying the mean field approximation could be written as:
\begin{dmath}
    H = h_{m}\sum_{i=1}^{N}m_{i} + h_{n}\sum_{i=1}^{N}n_{i} + h_{l}\sum_{i=1}^{N}l_{i} + J_{mm}z\langle m\rangle\sum_{i=1}^{N}m_{i} + J_{nn}z\langle n\rangle\sum_{i=1}^{N}n_{i} + J_{ll}z\langle l\rangle\sum_{i=1}^{N}l_{i} + J_{mn}\frac{z}{2}\langle n\rangle\sum_{i=1}^{N}m_{i} + J_{nm}\frac{z}{2}\langle m\rangle\sum_{i=1}^{N}n_{i} + J_{ml}\frac{z}{2}\langle l\rangle\sum_{i=1}^{N}m_{i} + J_{lm}\frac{z}{2}\langle m\rangle\sum_{i=1}^{N}l_{i} + J_{nl}\frac{z}{2}\langle l\rangle\sum_{i=1}^{N}n_{i} + J_{ln}\frac{z}{2}\langle n\rangle\sum_{i=1}^{N}l_{i}
    \label{eq:hamiltonian_mean_field}
\end{dmath}

All $h$ and $J$ terms are treated as site-independent values to capture the averaged behavior of all solvation shells in the electrolyte. In a physical sense, these interaction terms, whether describing short-range van der Waals interactions or long-range Coulombic interactions, should be permutationally invariant, as the interaction itself holds the same invariance. There should be no difference between a solvent-anion interaction and an anion-solvent interaction, as long as the same anion and solvent are involved. This leads to the symmetry in Eq.\ref{eq:full_hamiltonian} that $J_{mn}=J_{nm}$, $J_{ml}=J_{lm}$ and $J_{nl}=J_{ln}$.

Upon applying the mean field approximation, the full Hamiltonian of the system simply becomes a sum of $i$ non-interacting lattice sites, with each site obeying $m_i+n_i+l_i=1 (m_i,n_i,l_i=0,1)$. Using the spatial invariance property, we can calculate the average occupation of one site to represent the average occupation of the full system to reduce computation cost. The partition function of one site is given by
\begin{dmath}
    q = \sum_{m,n,l=0,1} \exp\left(-\frac{H_i}{kT}\right) = \exp\left(-\frac{h_m + J_{mm}z\langle m\rangle + J_{mn}\frac{z}{2}\langle n \rangle + J_{ml}\frac{z}{2} \langle l \rangle}{kT}\right) + \exp\left(-\frac{h_n + J_{nn}z\langle n\rangle + J_{nm}\frac{z}{2}\langle m \rangle + J_{nl}\frac{z}{2} \langle l \rangle}{kT}\right) + \exp\left(-\frac{h_l + J_{ll}z\langle l\rangle + J_{lm}\frac{z}{2}\langle m \rangle + J_{ln}\frac{z}{2} \langle n \rangle}{kT}\right)
    \label{eq:partition}
\end{dmath}
Note that there are 3 terms in the partition function, corresponding to each of the three possible site occupancy: $(m=1, n,l=0)$, $(n=1, m,l=0)$ and $(l=1, m,n=0)$. The average occupancy of each component can be computed from the following equations:
\begin{subequations}\label{eq:average_occupancy}
    \begin{align}
    \langle m \rangle = \langle m_i \rangle &= \frac{1}{q}
    \sum_{m_i,n_i,l_i=0,1}
    m_i\exp\left(-\tfrac{H_i}{kT}\right) = \frac{1}{q} \exp\left(-\frac{h_m + J_{mm}z\langle m\rangle + J_{mn}\frac{z}{2}\langle n \rangle + J_{ml}\frac{z}{2} \langle l \rangle}{kT}\right)\\
    \langle n \rangle = \langle n_i \rangle &= \frac{1}{q}
    \sum_{m_i,n_i,l_i=0,1}
    n_i\exp\left(-\tfrac{H_i}{kT}\right) = \frac{1}{q} \exp\left(-\frac{h_n + J_{nn}z\langle n\rangle + J_{nm}\frac{z}{2}\langle m \rangle + J_{nl}\frac{z}{2} \langle l \rangle}{kT}\right)\\
    \langle l \rangle = \langle l_i \rangle &= \frac{1}{q} \sum_{m_i,n_i,l_i=0,1}l_i\exp\left(-\tfrac{H_i}{kT}\right) = \frac{1}{q}\exp\left(-\frac{h_l + J_{ll}z\langle l\rangle + J_{lm}\frac{z}{2}\langle m \rangle + J_{ln}\frac{z}{2} \langle n \rangle}{kT}\right)
    \end{align}
\end{subequations}
where q is given by Eq. \ref{eq:partition}. Eq.\ref{eq:average_occupancy} shows that $\langle m \rangle$ is a function of all three averaged occupations, and so are $\langle n \rangle$ and $\langle l \rangle$. We can rearrange the equation and formulate it as
\begin{subequations}\label{eq:occupancy_root_finding}
    \begin{align}
    \langle m \rangle - f_1 (\langle m \rangle, \langle n \rangle, \langle l \rangle) &= 0\\
    \langle n \rangle - f_2 (\langle m \rangle, \langle n \rangle, \langle l \rangle) &= 0\\
    \langle l \rangle - f_3 (\langle m \rangle, \langle n \rangle, \langle l \rangle) &= 0
    \end{align}
\end{subequations}
From Eq.\ref{eq:occupancy_root_finding} it becomes clear that the solution to the average occupancy is the root of the system of functions $[f_1, f_2, f_3]^T$, therefore we can use any root finding algorithm to solve for the average occupations in an efficient manner.

\section{Details of interaction terms}
\begin{figure}[h!]
    \centering
    \begin{subfigure}[b]{0.4\textwidth}
        \centering 
        \caption{}
        \includegraphics[width=\linewidth]{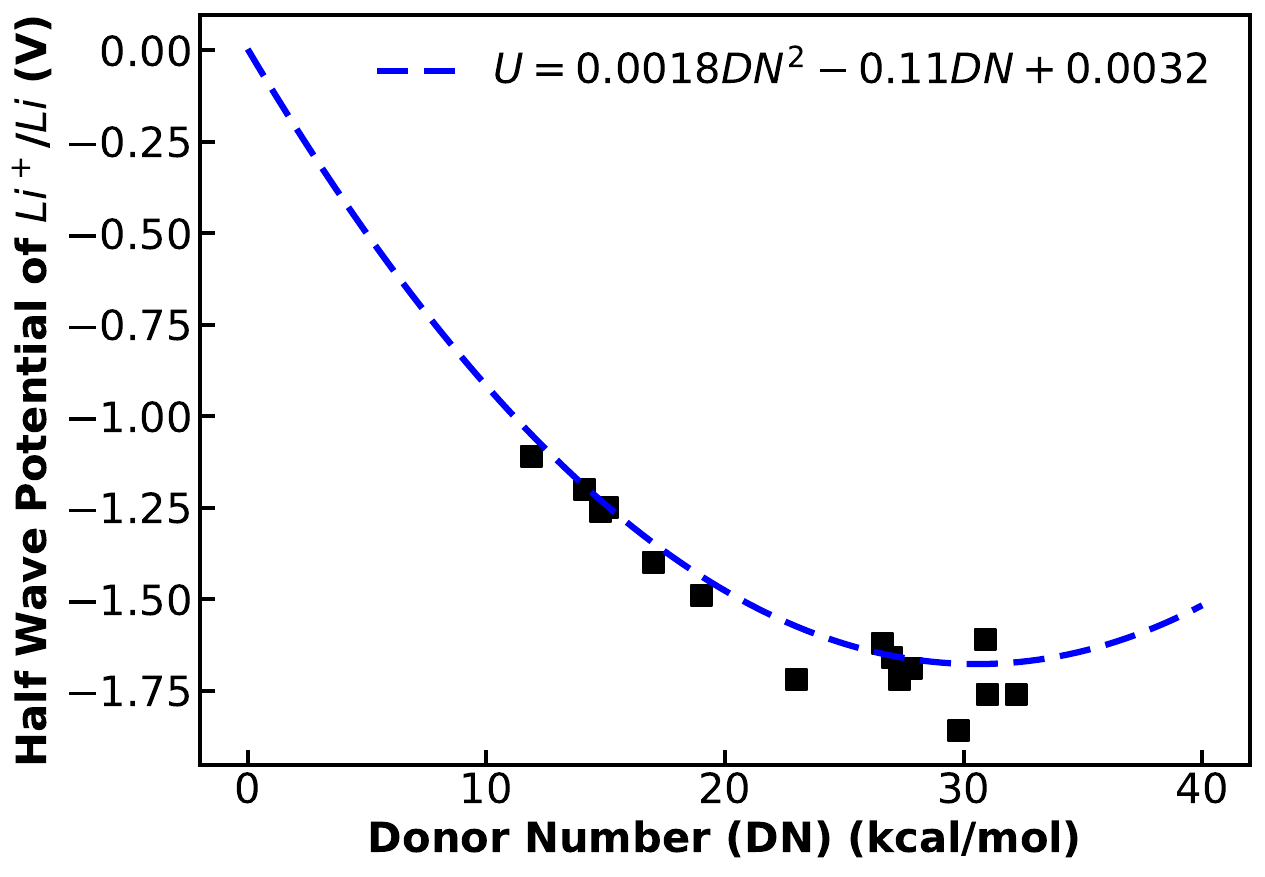}
         \centering
         \label{fig:khetan_fit}
    \end{subfigure}
    \begin{subfigure}[b]{0.4\textwidth}
        \centering 
        \caption{}
        \includegraphics[width=\linewidth]{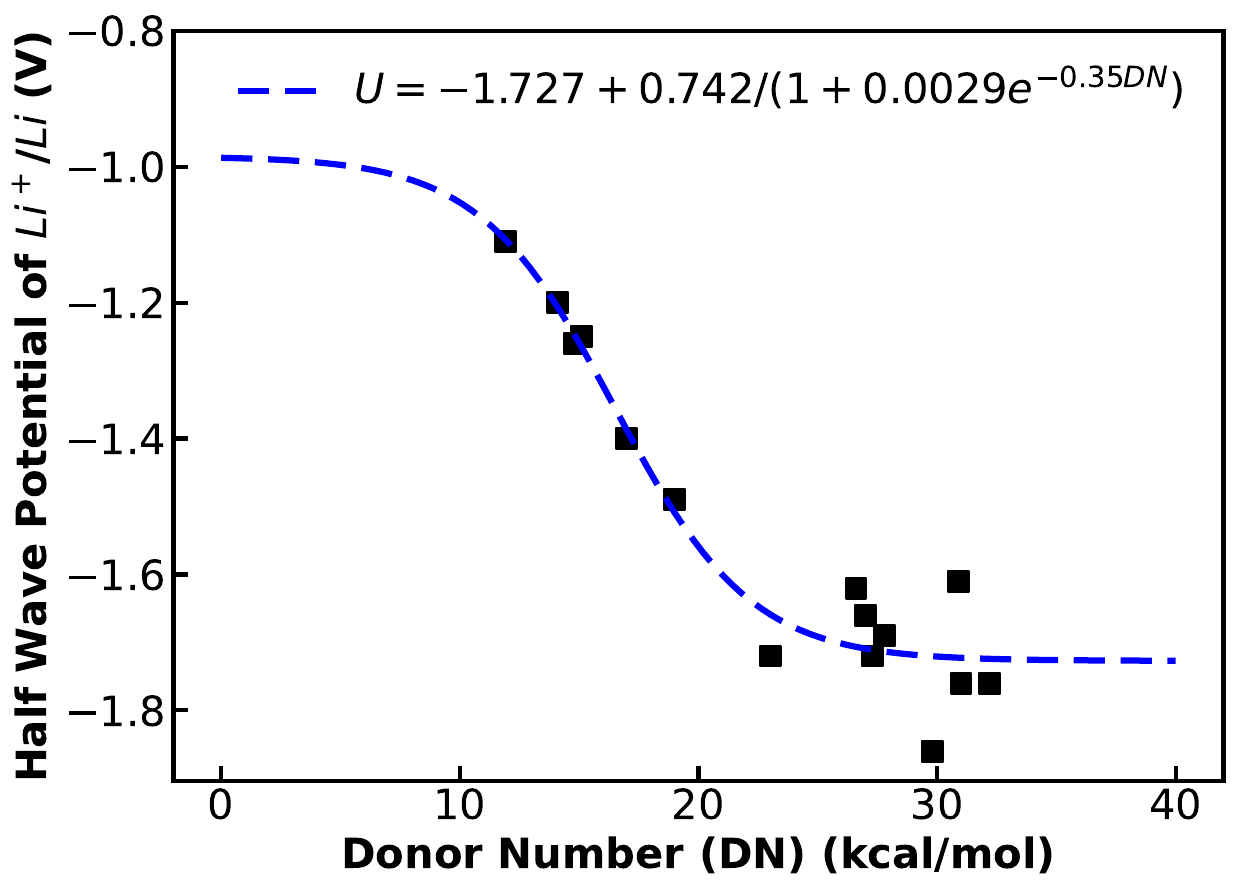}
         \centering
         \label{fig:sigmoid_fit_half_wave}
    \end{subfigure}    
    \caption{Fitted functions for Li/\lion \space half-wave potential vs DN: (a) 2$^{nd}$-order polynomial by \citeauthor{khetan_trade-offs_2015}, (b) our approach using sigmoid-like functional form. Our fitting ensures that half-wave potential monotonically decreases as DN increases.}
\end{figure}

As mentioned previously, the $h_u$ term correlates with the \lion-solvent or \lion-anion interaction strength. 
Our group's previous work has shown that \lion \space solvation free energy in a solvent has a linear relation with Li/\lion \space half-wave potential, and the latter correlates with the solvent's DN \cite{khetan_trade-offs_2015}. The fitted correlation is shown in \cref{fig:khetan_fit}. The half-wave potential first decreases almost linearly with increase of solvent DN, and then reaches a plateau as DN further increases. We expect a similar behavior for our Ising model's $h$ terms, which also describes \lion \space interactions with solvents. In our previous work, \citeauthor{khetan_trade-offs_2015} fitted the correlation as a second order polynomial, which could not capture the saturation behavior at very high DN (\cref{fig:khetan_fit}). Due to lack of data at very low DN, it is currently not clear whether such saturation behavior would also occur at low DN regime, but we nevertheless hypothesize that a similar saturation should also occur as \lion-solvent interaction free energy cannot increase indefinitely.
Therefore, we use a sigmoid-type function to describe the enthalpic part of \lion-solvent interactions:
\begin{equation}
  h_H(DN) = \left[ a_0 + \frac{a_1} {1 + a_2 \exp(-a_3 DN)} \right]
  \label{eq:sigmoid}
\end{equation}
The correlation between the Li/\lion \space half-wave potential and DN fitted under this functional form is shown in \cref{fig:sigmoid_fit_half_wave}.

We note that concentration effects are not explicitly considered by \citeauthor{khetan_trade-offs_2015} since the half-wave potential usually does not show concentration dependence. 
In the case of LHCE, there is an additional contribution from configurational entropy changes when solvents and anions enter the solvation shell. We capture this effect by adding an entropic term to $h$, which is given by
\begin{equation}\label{eq:entropy}
    h_{TS} (x) = a_4 \log (x)
\end{equation}
The entropy term treatment is similar to that of \citeauthor{burke_enhancing_2015}, despite that in this work we do not use explicit $k_BT$ as the coefficient but treat it as a learnable parameter. Eq.\ref{eq:sigmoid} and Eq.\ref{eq:entropy} constitute the full $h$ interaction term:
\begin{equation}
  h(DN, x) = \left[ a_0 + \frac{a_1} {1 + a_2 \exp(-a_3 DN)} \right]+ a_4 \log(x)
  \label{eq:h-term-si}
\end{equation}

We expect the same trends with \lion-anion interactions. The problem for anions is that the DN is not measured in the same experimental setup as organic solvents \cite{schmeisser_gutmann_2012}, so we apply a different set of parameters in Eq.\ref{eq:h-term-si} for anions in $h_l$.

For solvent-solvent, solvent-anion and anion-anion interactions, we use functional forms similar to those of Eq.\ref{eq:h-term-si}. Since the electron acceptor is not fixed for these interactions (as opposed to \lion-solvent and \lion-anion interactions, where the electron acceptor is consistently \lion), the interaction strength should depend on both donating and accepting ability of the two molecules involved. The solvent-anion interaction is described as:
\begin{equation}
    J_{ml, nl} = b_0 +  \frac{b_1}{1 + \exp\left[b_2 + b_3 DN_{m,n} + b_4 AN_{l}\right]} + b_5 \left[\log (x_{m,n}) + \log(x_l)\right]
    \label{eq:sol-anion}
\end{equation}
Note that in the entropic contribution, we consider both the anion molar ratio and solvent molar ratio to ensure permutation invariance of $(m,n)$ and $l$.

In solvent-solvent interactions, both solvents could act as electron donor and electron acceptor, so we take into account both possibilities. In addition, there could be repulsions between electron-donating functional groups as well as between electron-accepting functional groups. With these considerations, we model the solvent-solvent interactions as:
\begin{dmath}
    J_{\alpha\beta} = \left[c_0 + \frac{c_1}{1 + \exp\left[c_2 + c_3 DN_{\alpha} + c_4 AN_{\beta}\right]}\right] + \left[c_0+\frac{c_1}{1 + \exp\left[c_2 + c_3 DN_{\beta} + c_4 AN_{\alpha}\right]}\right] + \left[c_5+\frac{c_6}{1 + \exp\left[c_7 + c_8 DN_{\alpha} + c_9 DN_{\beta}\right]}\right] + \left[c_{10}+\frac{c_{11}}{1 + \exp\left[c_{12} + c_{13} AN_{\alpha} + c_{14} AN_{\beta}\right]}\right] + c_{15} \left[\log(x_{\alpha})+ \log(x_{\beta})\right]
\end{dmath}
where $\alpha=m,n$ and $\beta=m,n$. The first two terms correspond to interactions resulting from electron donating and accepting, the third and fourth terms correspond to lone-pair electron and empty orbital repulsions, and the last term corresponds to the entropic contributions. 

Finally, the anion-anion repulsion depends on the charge density of the anions. We can use anion DN to quantify this repulsion, without needing to perform Density Functional Theory (DFT) calculations to obtain charge densities. In our model, the anion-anion repulsion is modeled as:
\begin{equation}
  J_{ll} = \left[ d_0 + \frac{d_1} {1 + d_2 \exp(-d_3 DN_l)} \right]+ d_4 \log(x_l)
  \label{eq:anion-anion}
\end{equation}

We would like to note that in our model, we use effective DN and AN that are scaled by concentrations and molecular sizes to account for these effects. The scaling functions are given by:
\begin{subequations}
    \begin{align}
        f_x (x, x_l)= e_0+\frac{e_1}{1+e_2 \exp[-e_3({x}/{x_{l}})]}\\
        f_V (V, V_l) = e_4+\frac{e_5}{1+e_6 \exp[-e_7({V_l}/{V})]}
    \end{align}
\end{subequations}
subsequently $DN_{eff}$ and $AN_{eff}$ can be calculated:
\begin{subequations}
    \begin{align}
        DN_{eff} &= DN \times f_x(x,x_l) \times f_V(V,V_l)  \\
        AN_{eff} &= AN \times f_x(x,x_l) \times f_V(V,V_l) 
    \end{align}
    \label{eq:dn_eff}
\end{subequations}
where $x_l$ and $V_l$ correspond to the molar ratio  and the molecular size of the anion. The $({x}/{x_{l}})$ term is introduced such that solvent is more likely to dissociate \lion from anions when the system approaches low concentrations. The $(V_l/V)$ term accounts for the solvent molecular size effect, because we observe in our MD simulations that smaller molecules are generally more likely to coordinate with \lion in the first solvation shell.

To obtain physically consistent monotonic interaction terms, some model parameters are constrained in sign via a softplus transformation. For example, $a_1$ in $h_{m}$ is calculated via: 
\begin{equation}
    a_1 = \log (1+\exp(a_1'))
\end{equation}
and we update $a_1'$ in the model parameterization process. The physical constraints that we pose on different interaction terms are summarized in \cref{tab:sign-param}. 
\begin{table}[h]
    \centering
    \caption{Sign constraint for parameters}
    \label{tab:sign-param}
    \begin{tabular}{cc}
        \hline
        \textbf{Term} &
        \textbf{Constraint} \\
        \hline
        $h_m, h_n, h_l$ & $a_1,a_2\geq 0, a_3,a_4\leq 0$  \\
        $J_{ml}, J_{nl}$ & $b_1,b_3,b_4\geq 0, b_5\leq 0$  \\        
        $J_{mm}, J_{mn}, J_{nn}$ & $c_1,c_3,c_4,c_6,c_{11}\geq 0, c_8, c_9, c_{13},c_{14},c_{15}\leq 0$   \\
        $J_{ll}$  & $d_1, d_2, d_3\geq0, d_4\leq 0$\\
        $f_x$ & $e_1,e_3\leq 0, e_2\geq 0$ \\
        $f_V$ & $e_5,e_6,e_7\geq 0$ \\
        \hline
    \end{tabular}
\end{table}

In total 45 parameters are used, and we show parameter number breakdown in each term in \cref{tab:parameter_count}.
\begin{table}[h]
    \centering
    \caption{Parameter breakdown table of the model}
    \label{tab:parameter_count}
    \begin{tabular}{cc}
        \hline
        \textbf{Term} &
        \textbf{Total Number of parameters} \\
        \hline
        $h_{\mathrm{Li-sol}}$ & 5  \\
        $h_{\mathrm{Li-anion}}$& 5  \\
        $J_{\mathrm{sol-sol}}$  & 16 \\
        $J_{\mathrm{sol-anion}}$   & 6  \\
        $J_{\mathrm{anion-anion}}$  & 5  \\
        $f_x$ & 4 \\
        $f_V$ & 4 \\
        \textbf{Total} & \textbf{45} \\
        \hline
    \end{tabular}
\end{table}

Model parameters $\theta$ are updated with the ADAM optimizer \cite{kingma_adam_2015}. The learning rate is initialized at 0.002 and controlled by an exponential decay scheduler with a decay rate of 0.95. The loss is the summed RMSE between model prediction and MD data:
\begin{equation}\label{eq:train-loss}
  L = \sqrt{(\langle m_{model}\rangle - \langle m_{MD}\rangle)^2+(\langle n_{model}\rangle - \langle n_{MD}\rangle)^2+ (\langle l_{model}\rangle - \langle l_{MD}\rangle)^2}
\end{equation}

\section{Training dataset molecule selection criteria}
We apply 3 screening criteria when choosing formulations for MD simulation: (1) the DN measurement should be reliable; (2) the molecule should possess good electrochemical stability and compatible with LIB/LMB anodes; and (3) the formulation should potentially form LHCE.
For the first criterion, previous work has shown that amines are polymerized by SbCl$_5$ which is used as the reference Lewis acid in DN measurement \cite{vitale_synthesis_2013}. Consistent with this, we observe that there are only 3 molecules with DN higher than 40, and all of them belong to the amine family. This exceptionally high DN may be inaccurate due to the aforementioned side reaction, so we neglect these molecules.
To quantify the second criterion, we use SolventNet, a machine learning model developed to predict molecular properties \cite{annevelink_automat_2022}, to predict highest occupied molecular orbital (HOMO) and lowest unoccupied molecular orbital (LUMO) levels. To a first approximation, the HOMO level governs the solvent molecule's oxidative stability, and the LUMO level governs the reductive stability \cite{cao_reviewlocalized_2021}. We set a cutoff at HOMO$\leq$-6eV and LUMO$\geq$0eV to ensure that the electrolyte system we model is not electrochemically unstable. We also remove all the molecules containing hydroxyl (-OH) groups because their active protons are unstable against Li metal \cite{xu_nonaqueous_2004}. These downselections lead to a shortened stenutz dataset containing 30 organic molecules. The third criterion builds on the work of \citeauthor{chen_design_2023}, in which they proposed that LHCE diluents should possess DN$<$10 and solvents should possess DN$\geq$10. We choose 2 potential diluents with $DN<10$ and 6 solvents with $DN>10$. The chosen solvent/diluent molecules are listed in \cref{tab:molecule_table}.

\begin{table}[h]
\centering
\caption{Full list of molecules used in training dataset}
\label{tab:molecule_table}
    \begin{tabularx}{\linewidth}{X p{2.5cm} p{2.0cm} p{1.2cm} p{2.2cm} p{2.2cm}}
        \toprule
        Molecule & PubChem CID & DN (kcal$/$mol) & AN & Molar Mass (g$/$mol) & Molecular Size ($\AA^3$) \\
        \midrule
        Dichloromethane & 6344 & 1.0 & 20.4 & 84.93 & 88.23\\
        4,5-dichloro-1,3-dioxolan-2-one & 19869 & 3.2 & 16.7 & 156.95 & 143.48\\
        1,2-Dimethoxyethane & 8071 & 20.0 & 10.0 & 90.12 & 135.53 \\
        Tetrahydrofuran & 8028 & 20.0 & 8.0 & 72.11 & 108.85 \\
        N-methyl-2-pyrrolidone & 13387 & 27.3 & 13.3 & 99.13 & 139.66 \\
        Dimethylacetamide & 31374 & 27.8 & 13.6 & 87.12 & 130.99 \\
        Dimethyl sulfoxide & 679 & 29.8 & 19.3 & 78.14 & 104.63 \\
        N,N-diethylacetamide & 12703 & 32.2 & 13.6 & 115.17 & 177.36\\
        \bottomrule
    \end{tabularx}
\end{table}

\section{MD and DFT simulation details}
We use MD to obtain electrolyte solvation structures as the ground truth for parameterizing our model. MD simulations are performed with GROMACS version 2022.5\cite{abraham_gromacs_2015} using the OPLS-AA force field \cite{jorgensen_development_1996}. 
OPLS-AA-based classical MD has been widely used to resolve \lion \space solvation structure in nonaqueous battery electrolytes, with predicted coordination environments validated against spectroscopic and transport measurements \cite{shi_molecular_2025}.
Force field parameters for dimethoxyethane and TFSI$^-$ anion are taken from our previous works \cite{khetan_understanding_2018}, while parameters for BF$_4^-$ and PF$_6^-$ anions are obtained from \cite{doherty_revisiting_2017}. All other parameters are obtained from the LigParGen server \cite{jorgensen_potential_2005,dodda_114cm1a-lbcc_2017,dolgov_parallel_2020}. Charge-scaling is performed to ensure agreement between simulated densities of the pure substances and corresponding experimental densities (\cref{tab:md-benchmark}). Molecules are randomly initialized in the simulation box with solvent:diluent = 1:2 (mol:mol). Energy minimization is performed for a maximum of 50,000 steps. The system is then equilibrated to 300K using the Nose-Hoover thermostat for 0.5 ns, followed by equilibration to 300K and 1 bar with the Nose-Hoover thermostat and Parrinello-Rahman barostat ensemble for 1 ns. A production run using the Nose-Hoover thermostat set to 300K is then performed for 20 ns. We use MDAnalysis \cite{michaud-agrawal_mdanalysis_2011, gowers_mdanalysis_2016} and solvation-analysis \cite{cohen_solvationanalysis_2023} Python packages to analyze solvation structures and calculate coordination numbers. Molecular sizes are calculated by Density Functional Theory (DFT), carried out using Gaussian 16 \cite{frisch_gaussian16_2016}. We use the B3LYP exchange correlation functional with the 6-311+G(2d,p) basis set, with van der Waals interactions implemented through D3 dispersion correction \cite{grimme_consistent_2010}. The molecule size is computed as the volume in which the electron density $\rho>0.001$e/Bohr$^3$.

\section{MD benchmark results}
\cref{tab:md-benchmark} shows MD density benchmark results for the pure substance systems composed of molecules in the training set. The pure substance densities are obtained from the Stenutz database \cite{stenutz_collection_nodate}.
\begin{table}[h]
\centering
\caption{MD benchmark results}
\label{tab:md-benchmark}
    \begin{tabularx}{\linewidth}{X p{2.7cm} p{2.0cm} p{2.0cm} p{2.2cm}}
        \toprule
        Molecule & Charge scaling (\%) & $\rho_{experiment}$ (g/mL) & $\rho_{MD}$ (g/mL) & Percentage error (\%)\\
        \midrule
        Dichloromethane & 110 & 1.325 & 1.298 & 2.04\\
        4,5-dichloro-1,3-dioxolan-2-one & 90 & 1.670 & 1.665 & 0.30 \\
        1,2-Dimethoxyethane & 110 & 0.869 & 0.875 & 0.70\\
        Tetrahydrofuran & 110 & 0.889 & 0.880 & 0.97 \\
        N-methyl-2-pyrrolidone & 100 & 1.028 & 1.027 & 0.10\\
        Dimethylacetamide & 90 & 0.937 & 0.932 & 0.55\\
        Dimethyl sulfoxide & 90 & 1.100 & 1.102 & 0.22\\
        N,N-diethylacetamide & 110 & 0.925 & 0.928 & 0.37\\
        \bottomrule
    \end{tabularx}
\end{table}


\section{Model accuracy}
\begin{figure}[h]
    \centering
    \includegraphics[width=0.9\linewidth]{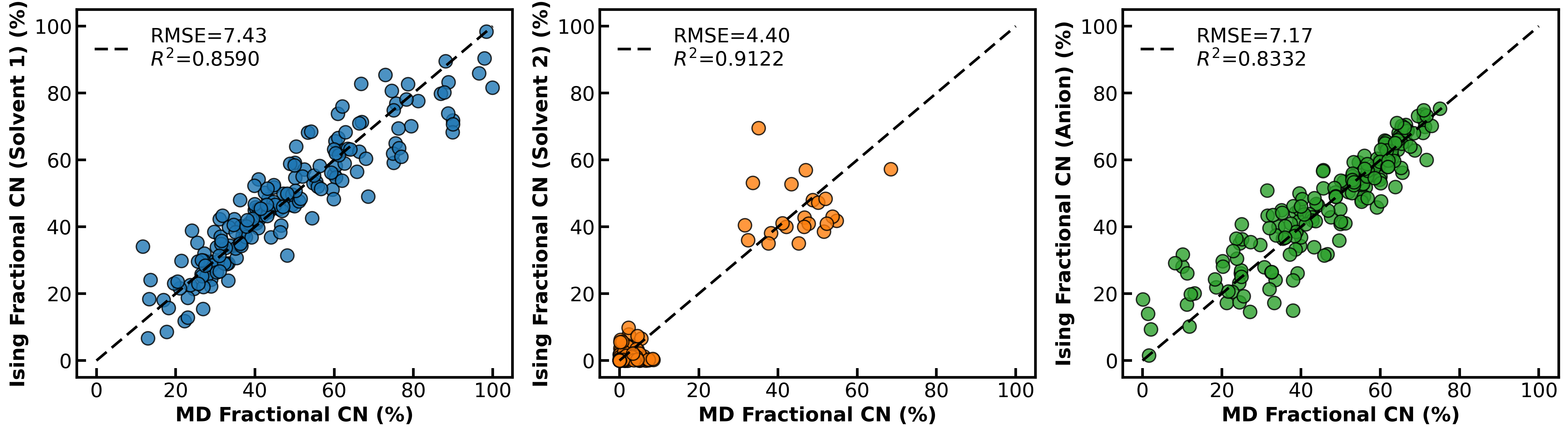}
    \caption{5-fold cross validation parity plots, separating solvent 1, solvent 2 and anion.}
    \label{fig:parity_sep_mnl}
\end{figure}

\begin{figure}[h]
    \centering
    \includegraphics[width=0.9\linewidth]{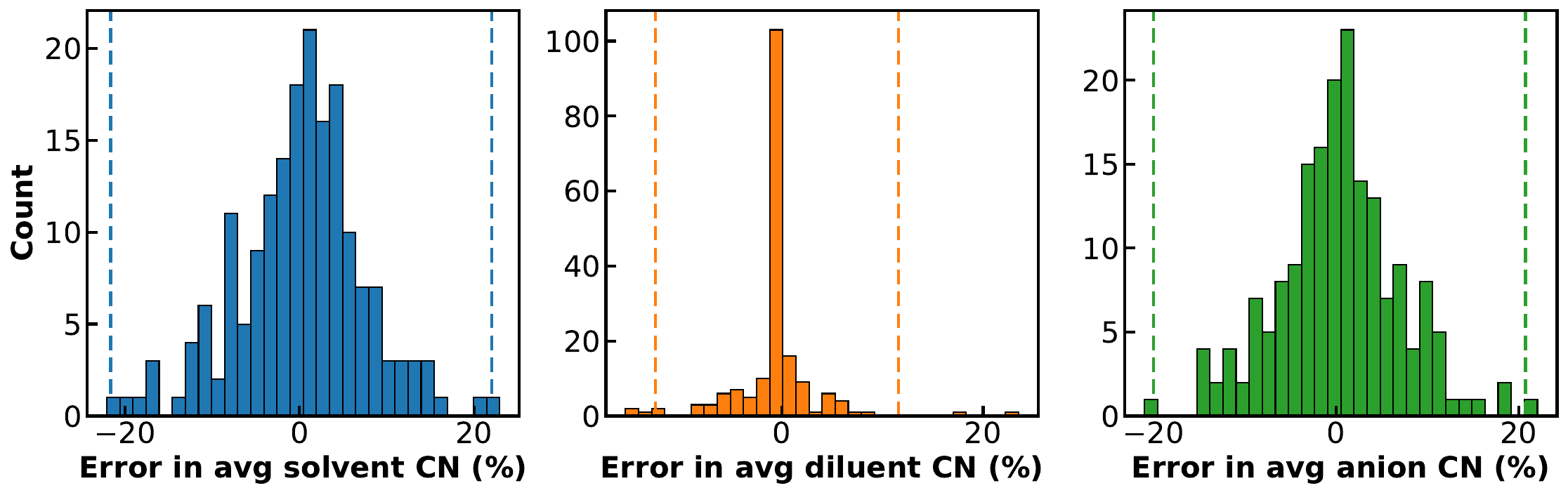}
    \caption{Histogram of model error on each component, the dashed lines enclose $[\mu-3\sigma, \mu + 3\sigma]$}
    \label{fig:kfold_error_hist}
\end{figure}

In \cref{fig:parity_sep_mnl,fig:kfold_error_hist} we show the errors on separate components. We would like to point out that our model solves for the 3 occupations simultaneously via root finding, so plotting solvent 1, solvent 2 and anion occupation predictions separately is just for visualization purpose and not indicating that our model has 3 independent prediction heads. We see that for single-component prediction, model RMSE falls in $4\%-8\%$. Assuming model residuals follow a normal distribution, the standard deviation for solvent, diluent and anion fractional CN prediction is $7.28\%$, $4.03\%$ and $6.78\%$ respectively. 

\begin{figure}[h]
    \centering
    \includegraphics[width=0.5\linewidth]{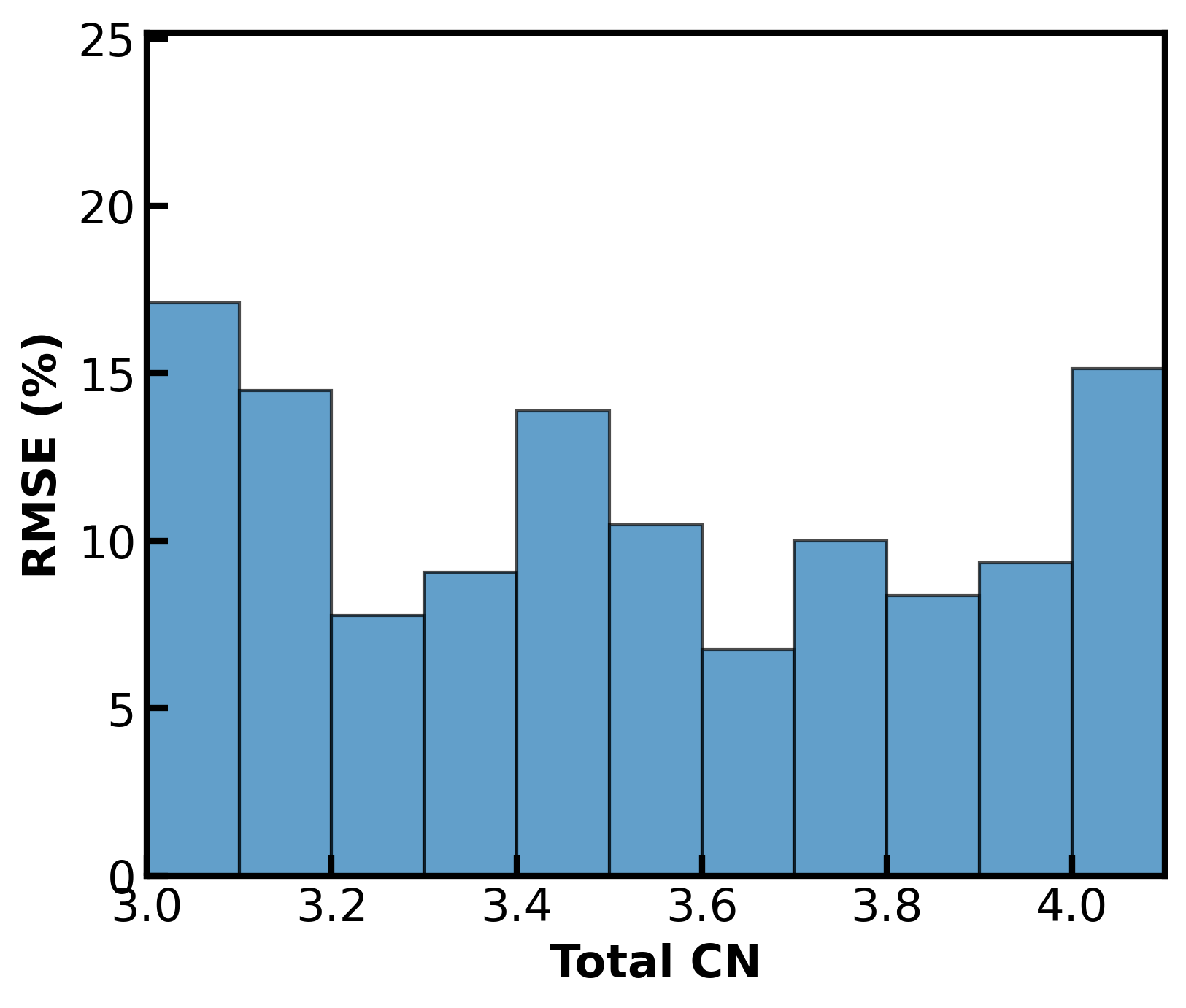}
    \caption{RMSE with different Li coordination numbers.}
    \label{fig:rmse_vs_cn}
\end{figure}

\begin{figure}[h!]
    \centering
    \begin{subfigure}[b]{0.6\textwidth}
        \centering 
        \caption{}
        \includegraphics[width=\linewidth]{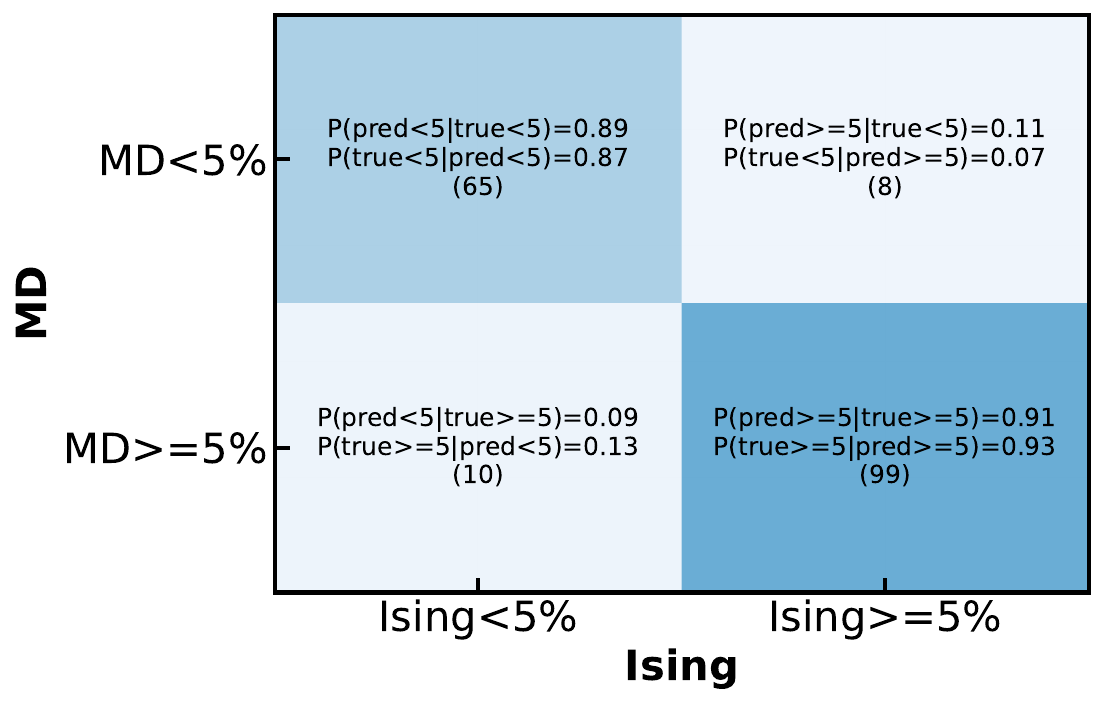}
         \centering
         \label{fig:conf_matrix_5}
    \end{subfigure}
    \begin{subfigure}[b]{0.6\textwidth}
        \centering 
        \caption{}
        \includegraphics[width=\linewidth]{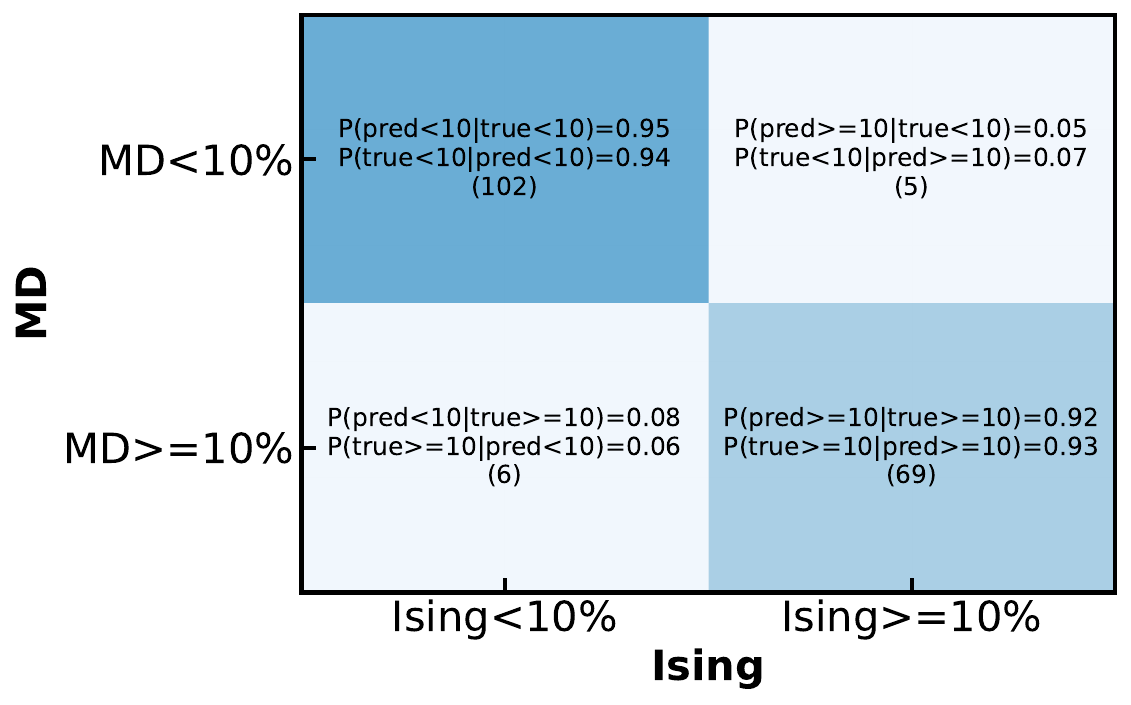}
         \centering
         \label{fig:conf_matrix_10}
    \end{subfigure}    
    \caption{Confusion matrix of free solvent molar ratio prediction under different criteria: (a) 5\%, and (b) 10\%}
    \label{fig:conf_matrix}
\end{figure}

Two formulations exceed 20\% RMSE: 1.0m LiTFSI in DME/DMA and in DMSO/DCM. The former is a low-concentration electrolyte ($x_{A^-}\sim0.07$) containing two high-DN solvents ($DN_{DME}=20.0$ kcal/mol and $DN_{DMA}=27.8$ kcal/mol) with $>75\%$ uncoordinated free solvent. The latter contains DMSO, whose small molecular size and high DN ($DN_{DMSO}=29.8$ kcal/mol) contribute to its unusual solvation behavior: DMSO shows an excessively negative Li/\lion \space half-wave potential relative to our fitted trend \cite{khetan_trade-offs_2015}, which is consistent with our model underestimating its coordination ($\langle m\rangle_{Ising}=0.683$ vs. $\langle m\rangle_{MD}=0.898$). DMSO-containing systems also generally exhibit higher total \lion \space coordination numbers (\cref{fig:rmse_vs_cn}), which further contributes to error given our invariant-CN assumption. Despite these errors, electrolytes containing DMSO typically do not form anion-rich solvation shells, so this is not a critical concern for LHCE design.

We also visualize model RMSE as a function of \lion \space total coordination number to investigate the validity of the model assumption that all formulations share the same invariant total CN. In \cref{fig:rmse_vs_cn} we observe that RMSE are typically lower where \lion \space CN falls within the vicinity of our model assumption 3.72, while at very low CN and high CN model RMSE rises slightly. The reason for abnormally high RMSE around CN=3.5 is due to the DME/DMA formulation discussed above. Based on these results, we argue that using an invariant CN=3.72 across all formulations does not significantly hamper the model accuracy. Nevertheless, developing tools/models to accurately predict total CN just through formulations and solvent/salt properties would be an interesting direction moving forward.

\section{Demonstration of model generalizability}
\begin{figure}[h!]
    \begin{subfigure}[b]{0.4\textwidth}
        \centering
        \caption{}
        \includegraphics[width=\linewidth]{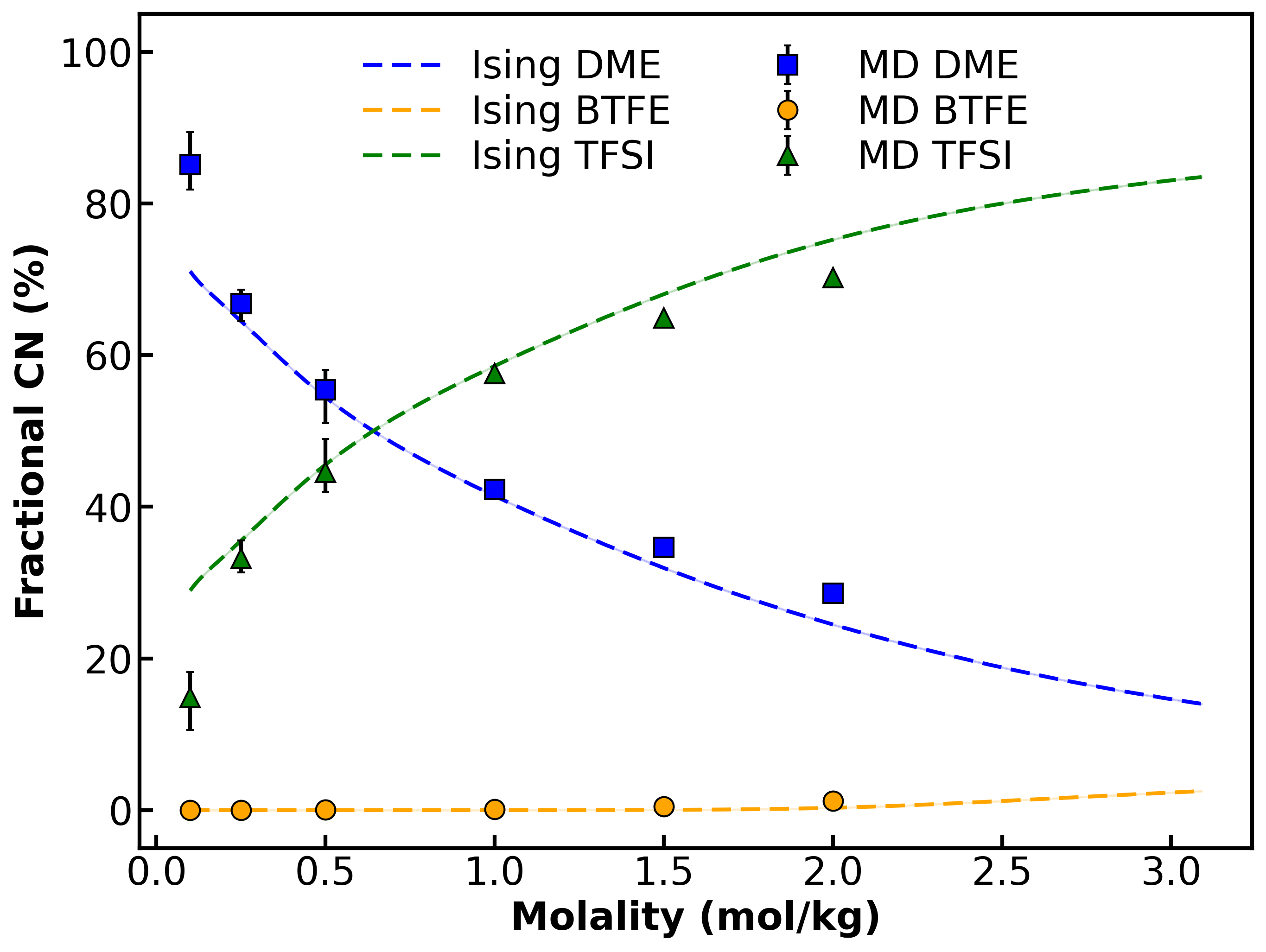}
        \label{fig:dme_btfe_litfsi}
    \end{subfigure}    
    \begin{subfigure}[b]{0.4\textwidth}
        \centering
        \caption{}
        \includegraphics[width=\linewidth]{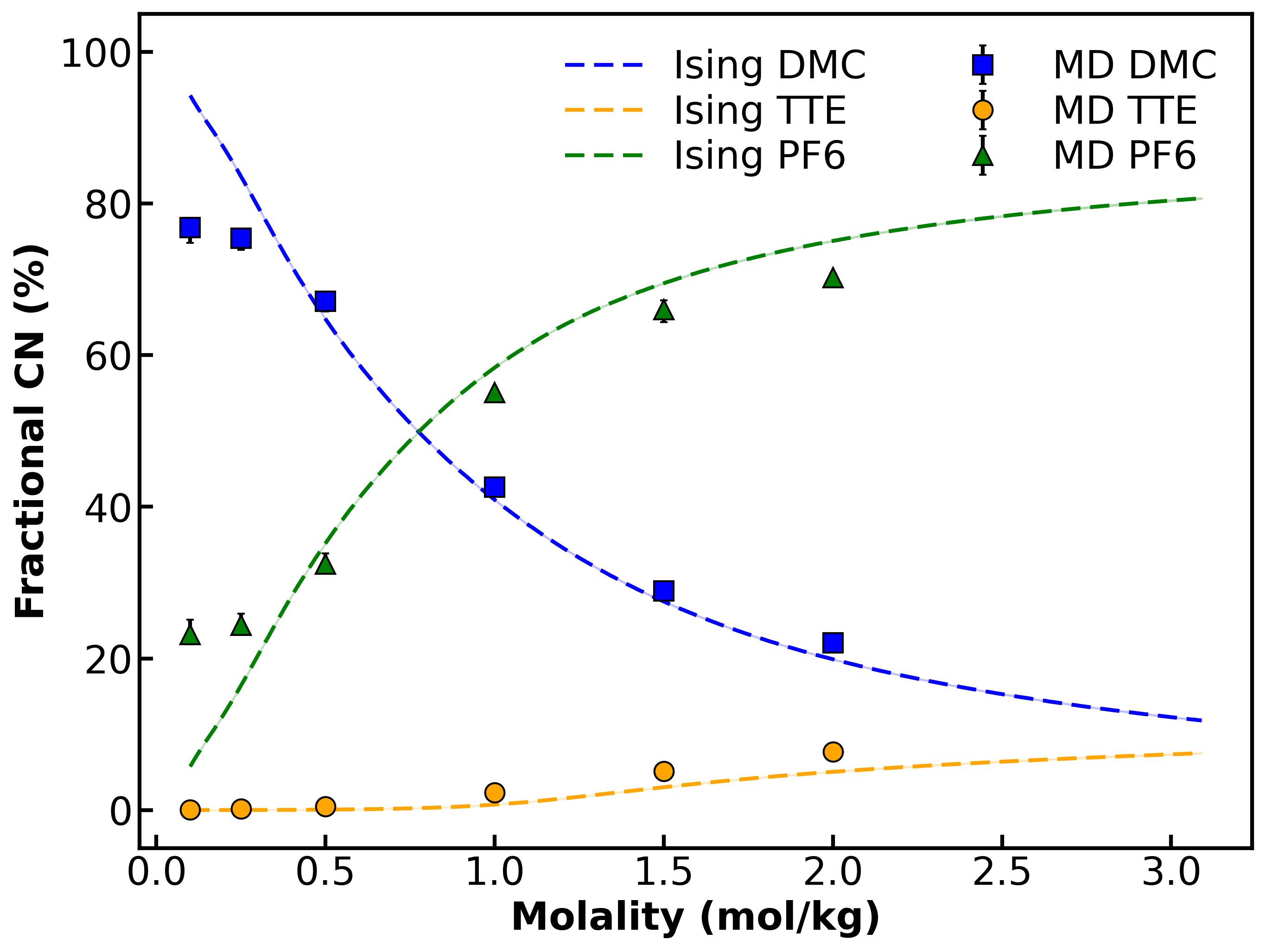}
        \label{fig:dmc_tte_lipf6}
    \end{subfigure}     
    \begin{subfigure}[b]{0.4\textwidth}
        \centering
        \caption{}
        \includegraphics[width=\linewidth]{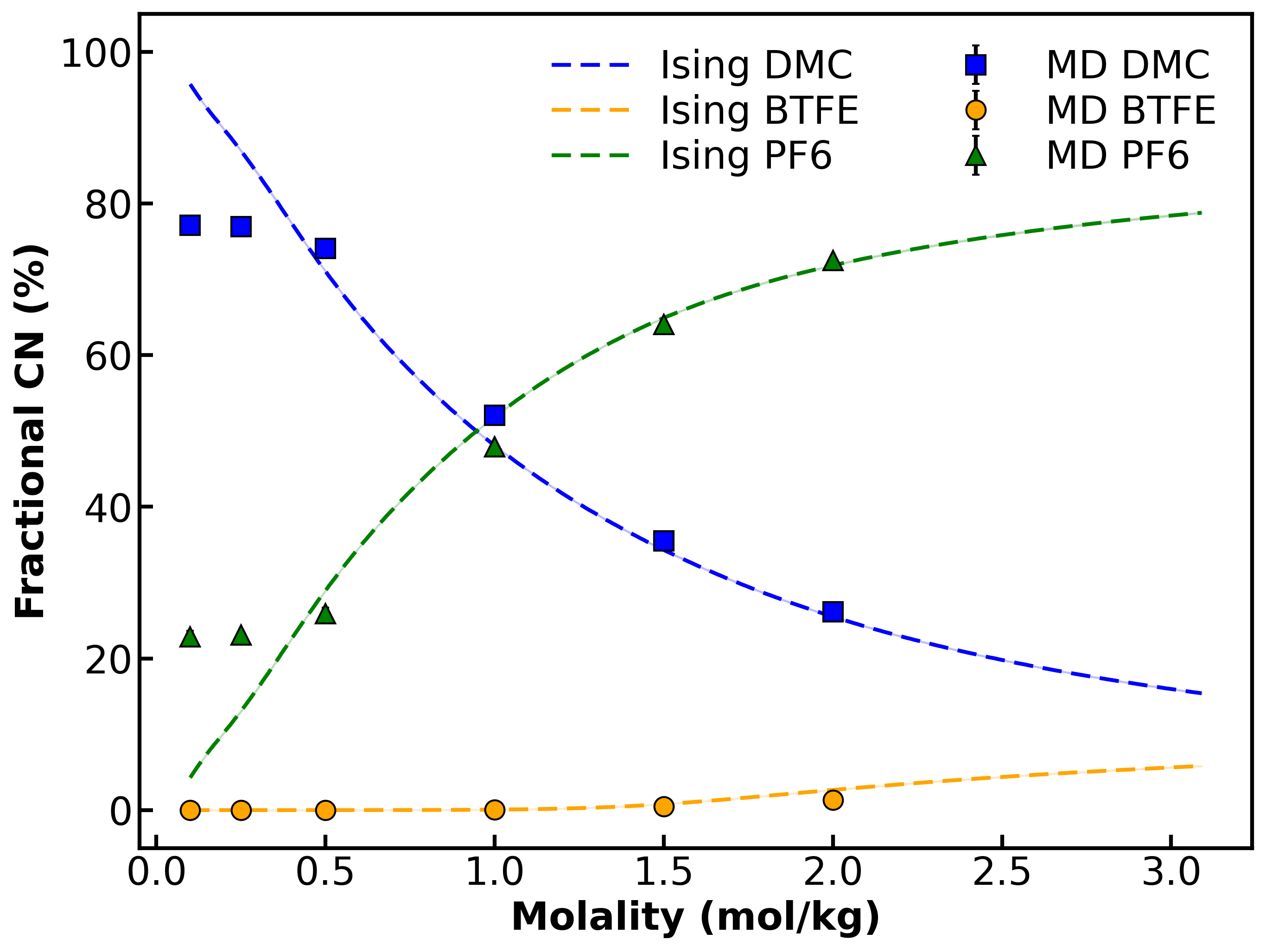}
        \label{fig:dme_btfe_lipf6}
    \end{subfigure}        
    \caption{Comparison between Ising model prediction and MD on commonly-used LHCE systems varying concentration: 
    (a) LiTFSI in DME:BTFE=1:2 (mol:mol), (b) LiPF$_6$ in DMC:TTE=1:2 (mol:mol), and (c) LiPF$_6$ in DMC:BTFE=1:2 (mol:mol).
    }
    \label{fig:real-system-result-si}
\end{figure}
\cref{fig:real-system-result-si} shows our model's performance on representative LHCE systems in addition to the LiTFSI DME/TTE system shown in main text. In all the systems, our model was able to quantitatively reproduce the solvation behavior of increasing anion-\lion \space coordination with increasing salt molality, with good match with MD-predicted values. In conjunction with MD data, our model predicts stronger dissociation trend of LiPF$_6$ compared with LiTFSI, stemming from PF$_6^-$'s lower DN. These results show that our model is capable of generalizing to different LHCE formulations.

We also try predicting the solvation shell composition of other classes of electrolytes that focus on solvation structure engineering. As mentioned in the main text, high-concentration-electrolytes (HCE) and high-entropy-electrolytes (HEE), another re two other novel electrolyte classes that highlights solvation shell composition designshow improved performance in LMBs. In addition, we test high concentration, high entropy electrolytes to explore out model's boundary. The formulations we test here are listed in \cref{tab:hce-formulations,tab:hee-formulations}.

We emphasize that \textbf{the model is not re-parameterized on any of these systems}. All predictions below are made with the same 5 models from the 5-fold cross validation of \cref{fig:parity_sep_mnl}, which are trained exclusively on LHCE formulations at solvent:diluent = 1:2 (mol:mol) and $\leq 2.5$m. Every formulation in this section is therefore a blind prediction, and each family lies outside the training distribution in at least one respect: HCE contains no diluent at all, and HEE contains four or five solvents rather than one solvent and one diluent. Agreement obtained under these conditions cannot be attributed to fitting, so the trends recovered below are evidence that our model encodes real transferable physics rather than being specific to the LHCE dataset.

\begin{table}[h]
\centering
\caption{HCE formulations. Each formulation is simulated at 7 molalities (1.0m to 4.0m in 0.5m increments) with LiTFSI as the salt and no diluent.}
\label{tab:hce-formulations}
    \begin{tabularx}{\linewidth}{X p{5.2cm} p{2.2cm}}
        \toprule
        Formulation & Solvent & DN (kcal$/$mol)\\
        \midrule
        LiTFSI/G4 & Tetraglyme & 16.6  \\
        LiTFSI/PC & Propylene carbonate & 15.1 \\
        LiTFSI/DMC & Dimethyl carbonate & 17.1 \\
        \bottomrule
    \end{tabularx}
\end{table}

\begin{table}[h]
\centering
\caption{HEE solvent mixtures. Solvents within each mixture are equimolar and grouped by DN so that no single solvent dominates the shell. Each mixture is paired with 4 salts (LiPF$_6$, LiBF$_4$, LiTFSI, LiNO$_3$). Abbreviations: DME, 1,2-dimethoxyethane; EA, ethyl acetate; MA, methyl acetate; THF, tetrahydrofuran; DMSO, dimethyl sulfoxide; DMF, N,N-dimethylformamide; DMA, N,N-dimethylacetamide; DEA, N,N-diethylacetamide; Py, pyridine; PC, propylene carbonate; ACN, acetonitrile; BZN, benzonitrile; SL, sulfolane; GBL, $\gamma$-butyrolactone; DMC, dimethyl carbonate.}
\label{tab:hee-formulations}
    \begin{tabularx}{\linewidth}{X p{4.0cm} p{2.0cm}}
        \toprule
        Solvent mixture & Solvent DN range (kcal$/$mol) & Molality \\
        \midrule
        DME/EA/MA/acetone/THF & 16.5 -- 20.0 & 1.0m, 3.0m\\
        DMSO/DMF/DMA/DEA/Py & 26.6 -- 33.1 & 1.0m, 3.0m \\
        PC/ACN/BZN/SL/GBL & 11.9 -- 18.0 & 1.0m, 3.0m \\
        DMF/DMA/DEA/Py & 26.6 -- 33.1 & 3.0m \\
        PC/BZN/SL/GBL/DMC & 11.9 -- 18.0 & 3.0m \\
        \bottomrule
    \end{tabularx}
\end{table}

\begin{figure}[h!]
    \centering
    \begin{subfigure}[b]{0.45\textwidth}
        \centering
        \caption{}
        \includegraphics[width=\linewidth]{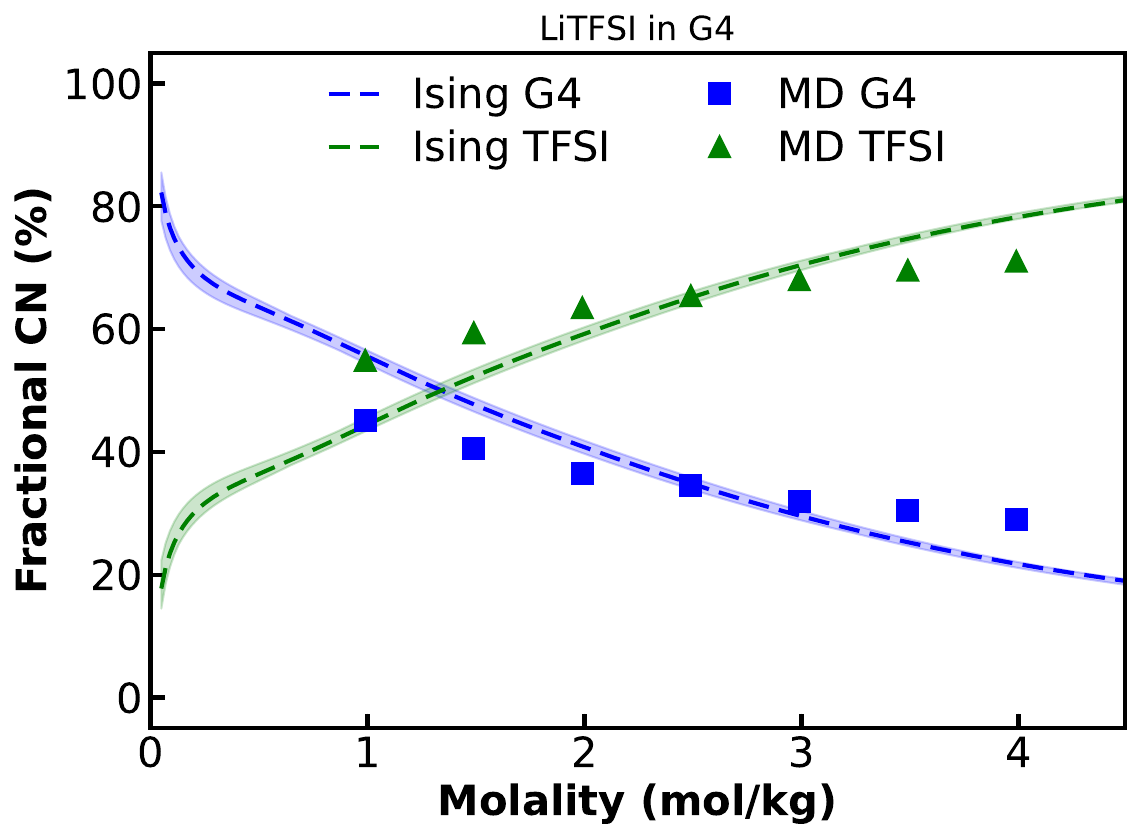}
        \label{fig:hce-g4}
    \end{subfigure}
    \begin{subfigure}[b]{0.45\textwidth}
        \centering
        \caption{}
        \includegraphics[width=\linewidth]{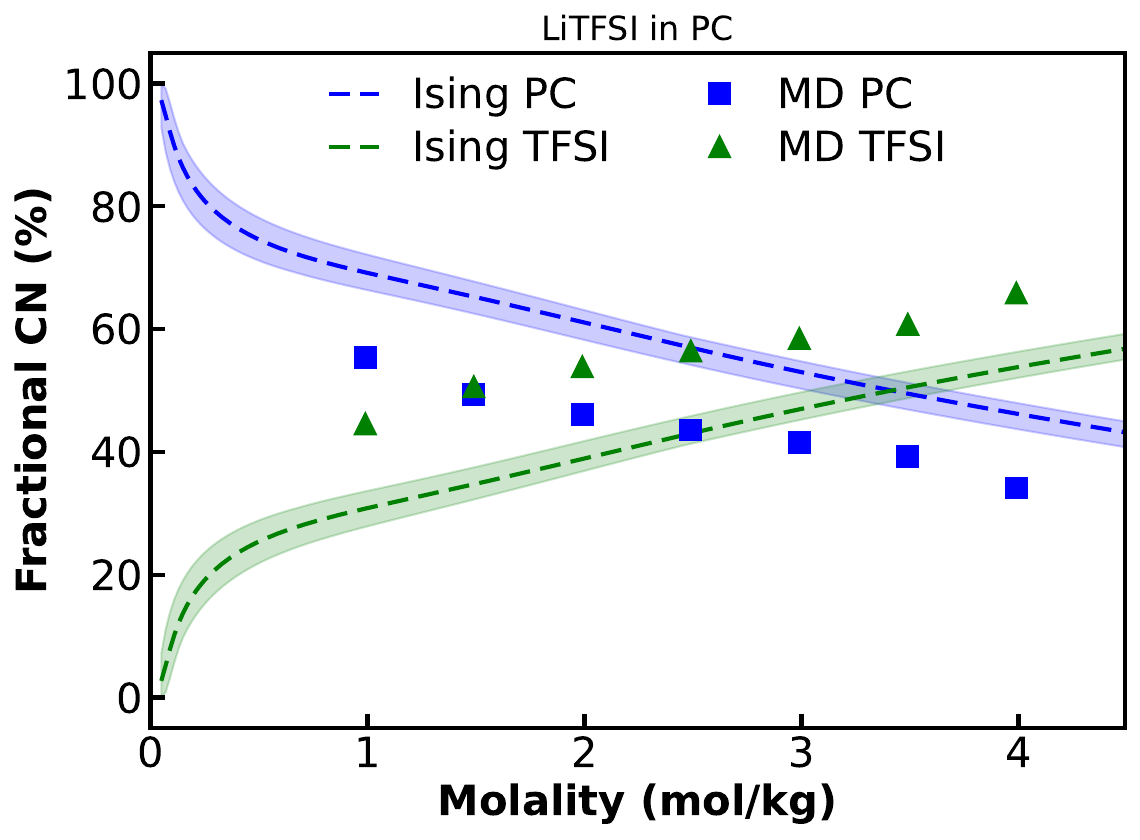}
        \label{fig:hce-pc}
    \end{subfigure}
    \begin{subfigure}[b]{0.45\textwidth}
        \centering
        \caption{}
        \includegraphics[width=\linewidth]{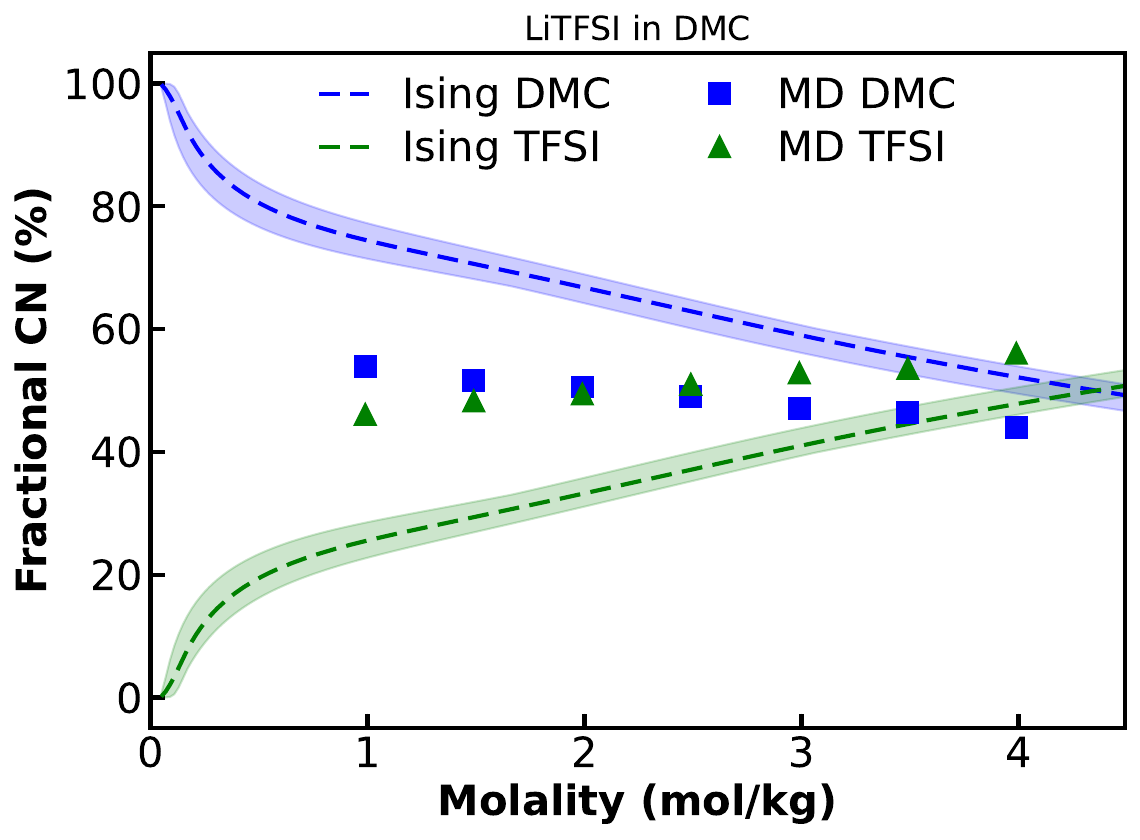}
        \label{fig:hce-dmc}
    \end{subfigure}
    \caption{Ising model predictions versus MD for HCE systems, as a function of LiTFSI molality: (a) DME, (b) G4, (c) PC, and (d) DMC. Shaded bands span the 5 cross-validation models. Agreement is best for the ether solvents (a, b); for the carbonates (c, d) the model underestimates anion occupation but preserves the monotonic rise with concentration.}
    \label{fig:hce-generalizability}
\end{figure}

\cref{fig:hce-generalizability} shows the HCE results. Our model achieves good agreement on the two ether systems, with mean absolute errors of $2.5\%$ for LiTFSI/DME and $5.3\%$ for LiTFSI/G4 in shell occupation. Agreement is poorer for the two carbonate systems, where the model systematically underestimates anion occupation by $13.1\%$ (LiTFSI/PC) and $14.1\%$ (LiTFSI/DMC). Importantly, the model preserves the correct physical trend in all four systems, with anion occupation rising monotonically with salt molality. The carbonate discrepancy is therefore an offset rather than a qualitative failure, and could be improved with more training data in the future.

\begin{figure}[h!]
    \centering
    \begin{subfigure}[b]{0.45\textwidth}
        \centering
        \caption{}
        \includegraphics[width=\linewidth]{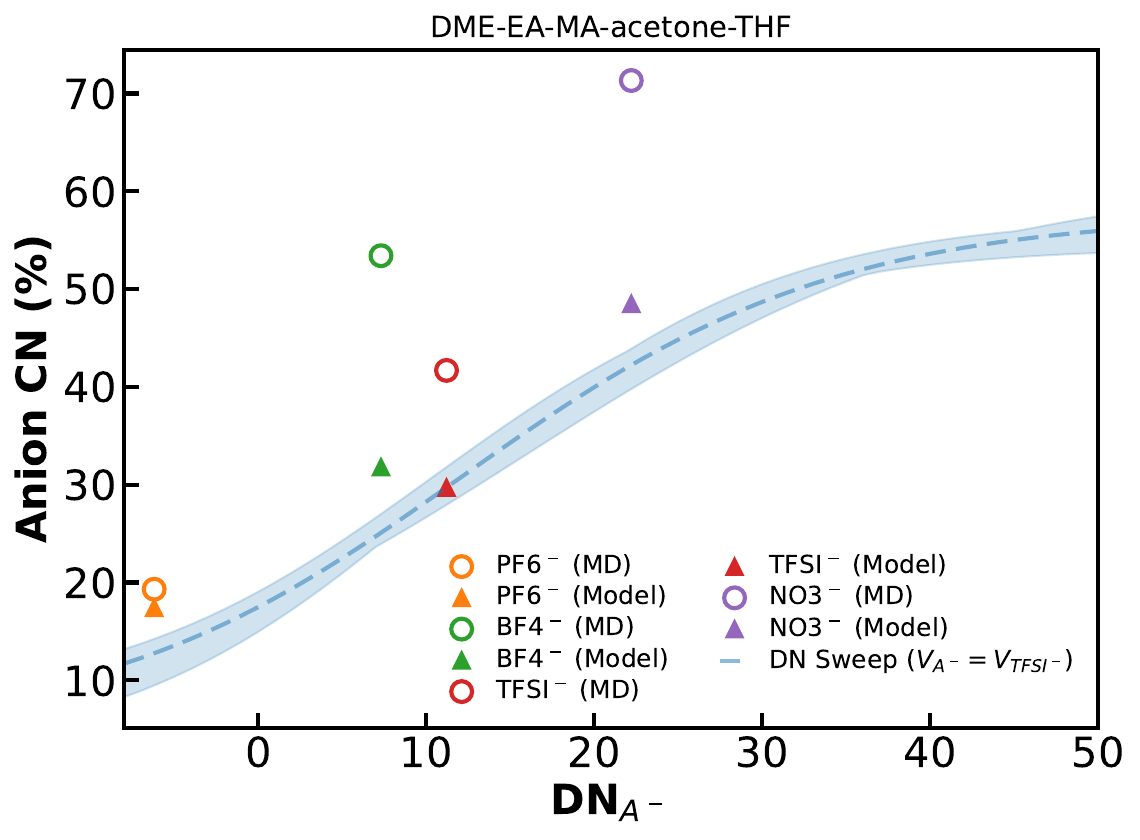}
        \label{fig:hee-dme}
    \end{subfigure}
    \begin{subfigure}[b]{0.45\textwidth}
        \centering
        \caption{}
        \includegraphics[width=\linewidth]{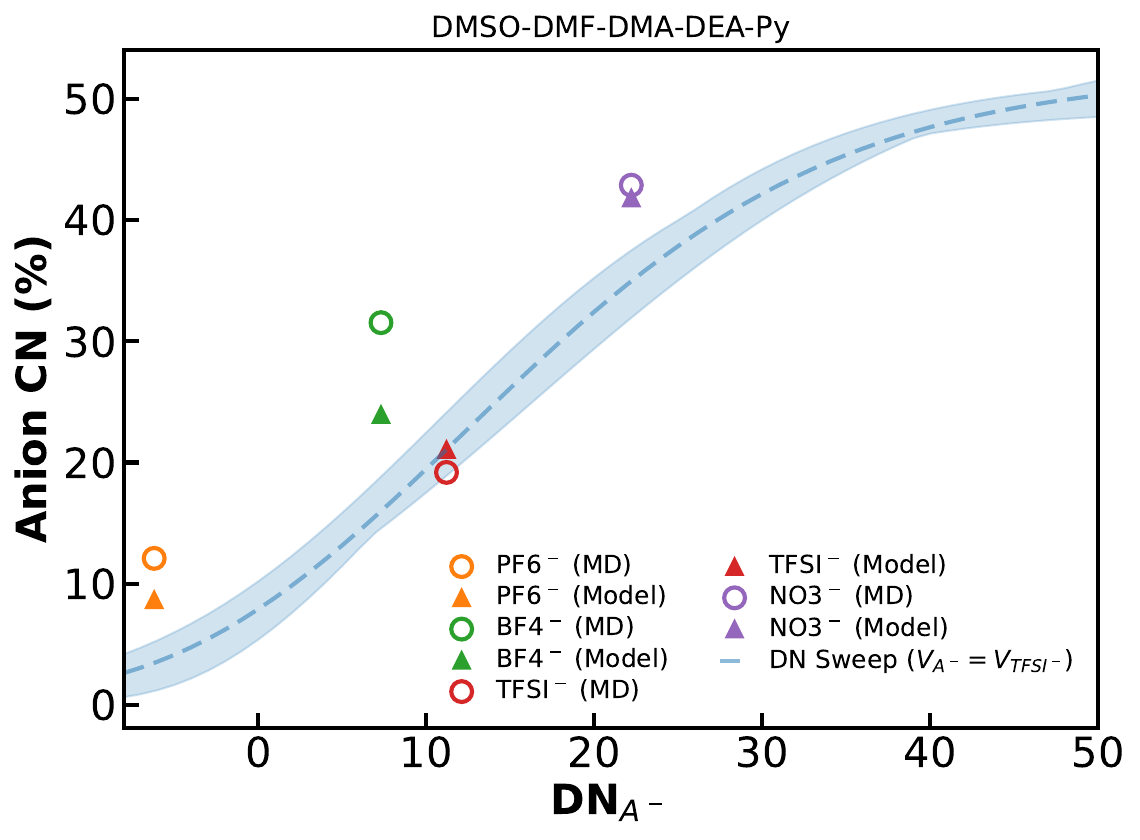}
        \label{fig:hee-dmso}
    \end{subfigure}
    \begin{subfigure}[b]{0.45\textwidth}
        \centering
        \caption{}
        \includegraphics[width=\linewidth]{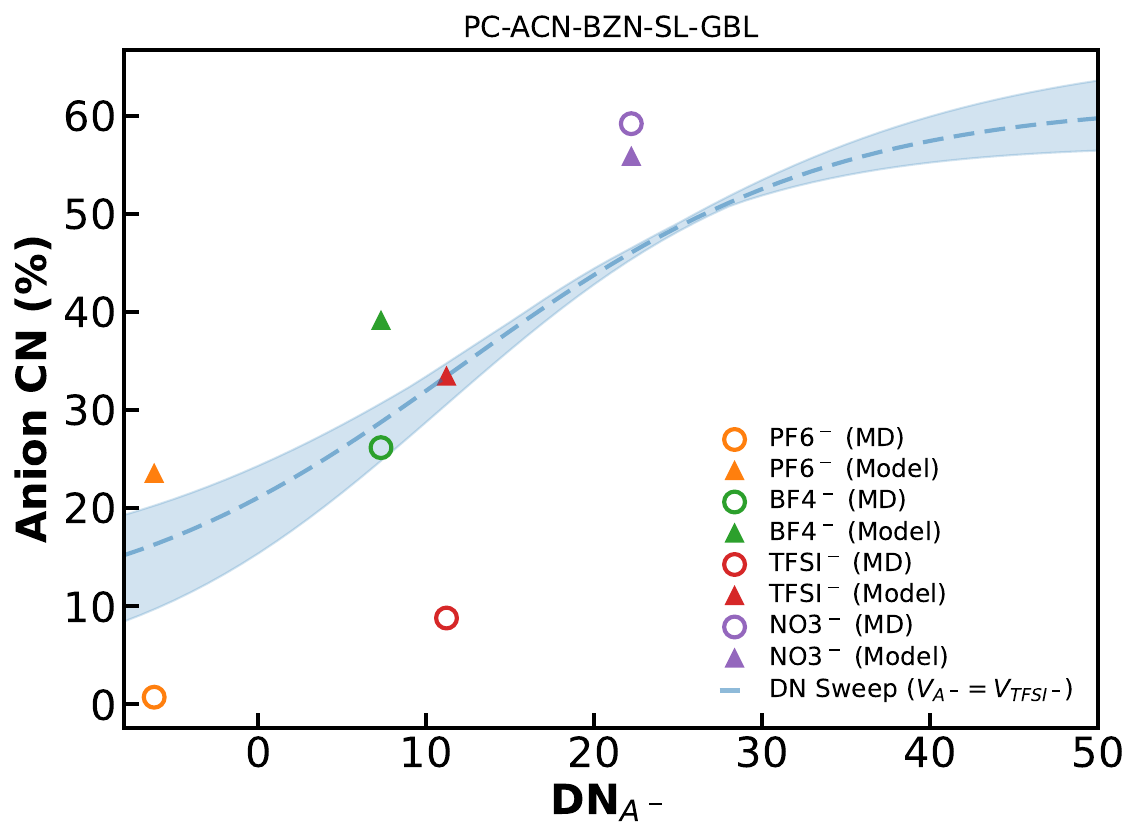}
        \label{fig:hee-pc}
    \end{subfigure}
    \caption{Anion occupation versus anion DN for HEE systems at 1.0m: (a) DME/EA/MA/acetone/THF, (b) DMSO/DMF/DMA/DEA/Py, and (c) PC/ACN/BZN/SL/GBL. Dashed curves sweep DN at fixed TFSI$^-$ volume and molar ratio; triangles are model predictions at each anion's own DN and volume, and open circles are MD. The model reproduces the MD ordering in all three mixtures, including the BF$_4^-$/TFSI$^-$ inversion driven by molecular size.}
    \label{fig:hee-generalizability}
\end{figure}

\cref{fig:hee-generalizability} shows the 1.0m HEE results, where we fix the solvent mixture and vary the anion. The dashed sweep curve varies DN alone while holding the anion volume at TFSI$^-$ value. Despite absolute errors reaching $16\%$ in anion occupation for some mixtures, the model reproduces the anion ordering exactly as the anion varies from low-DN PF$_6^-$ (DN$=-6.2$ kcal/mol) to high-DN NO$_3^-$ (DN$=22.2$ kcal/mol). Most notably, the model faithfully reproduces the reversal of BF$_4^-$ and TFSI$^-$: although BF$_4^-$ has a lower donor number (7.3 versus 11.2 kcal/mol), it occupies more of the shell in every mixture, because it is roughly a third the size of TFSI$^-$ ($72.9$ versus $209.8$ \AA$^3$). Our model recovers this inversion in all three mixtures, confirming that the molecular size term is doing real work and that the model is not merely ranking anions by DN. The largest errors occur for the PC/ACN/BZN/SL/GBL mixture, which has a mean \lion \space total coordination number of 4.10, well above our invariant assumption of 3.72, and very high free-solvent fraction ($54\%$ on average, reaching $97\%$ for LiPF$_6$); both are consistent with the limitations we reported earlier.

\begin{figure}[h!]
    \centering
    \begin{subfigure}[b]{0.45\textwidth}
        \centering
        \caption{}
        \includegraphics[width=\linewidth]{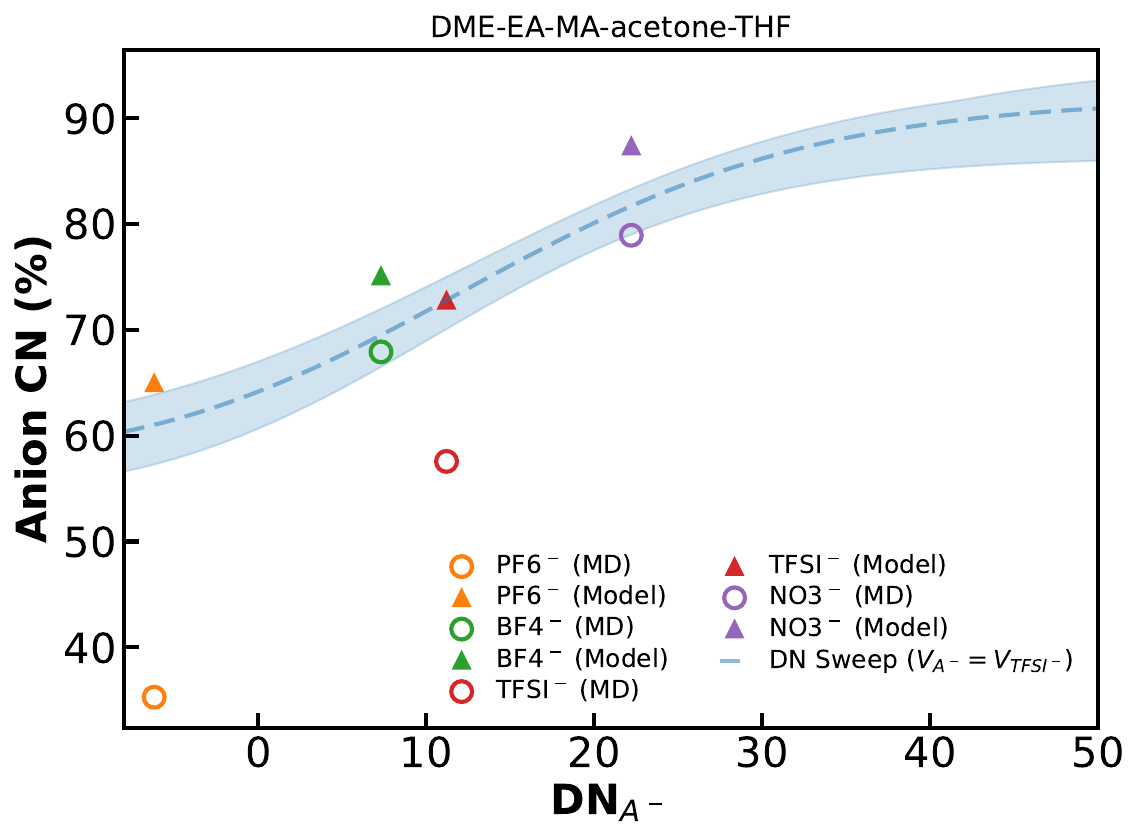}
        \label{fig:hcee-dme}
    \end{subfigure}
    \begin{subfigure}[b]{0.45\textwidth}
        \centering
        \caption{}
        \includegraphics[width=\linewidth]{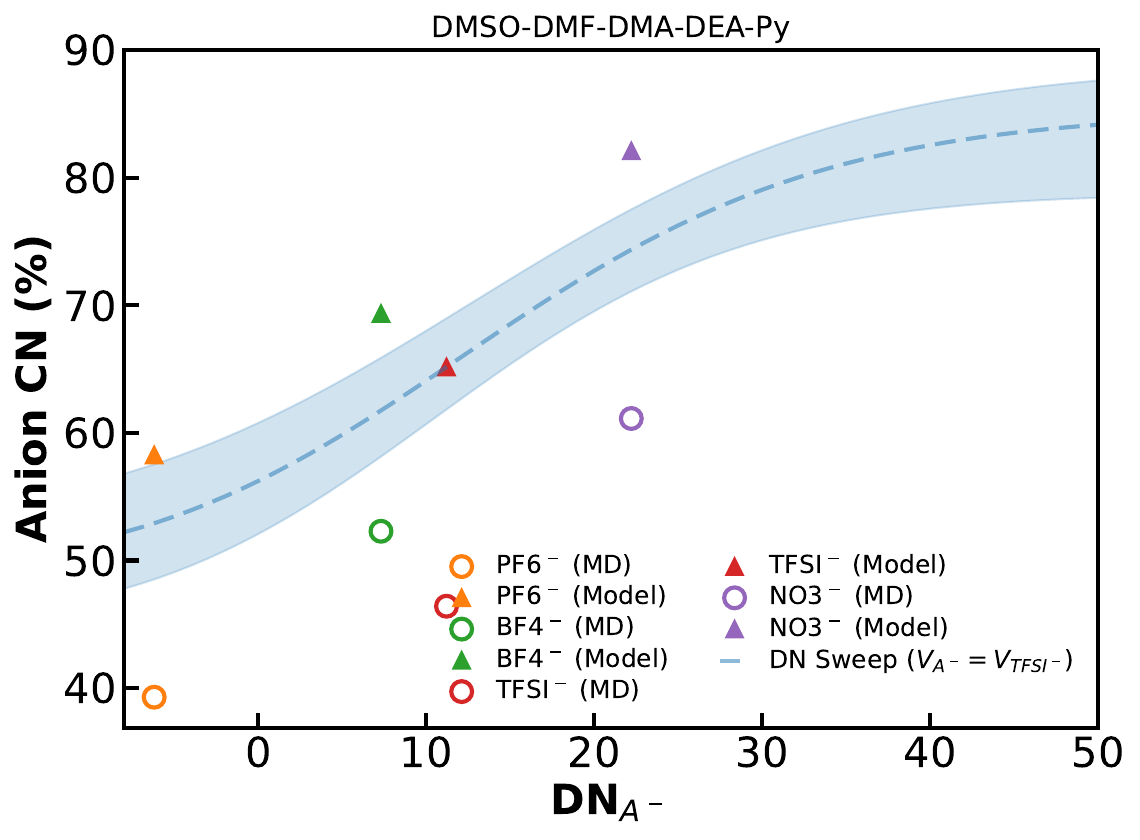}
        \label{fig:hcee-dmso}
    \end{subfigure}
    \begin{subfigure}[b]{0.45\textwidth}
        \centering
        \caption{}
        \includegraphics[width=\linewidth]{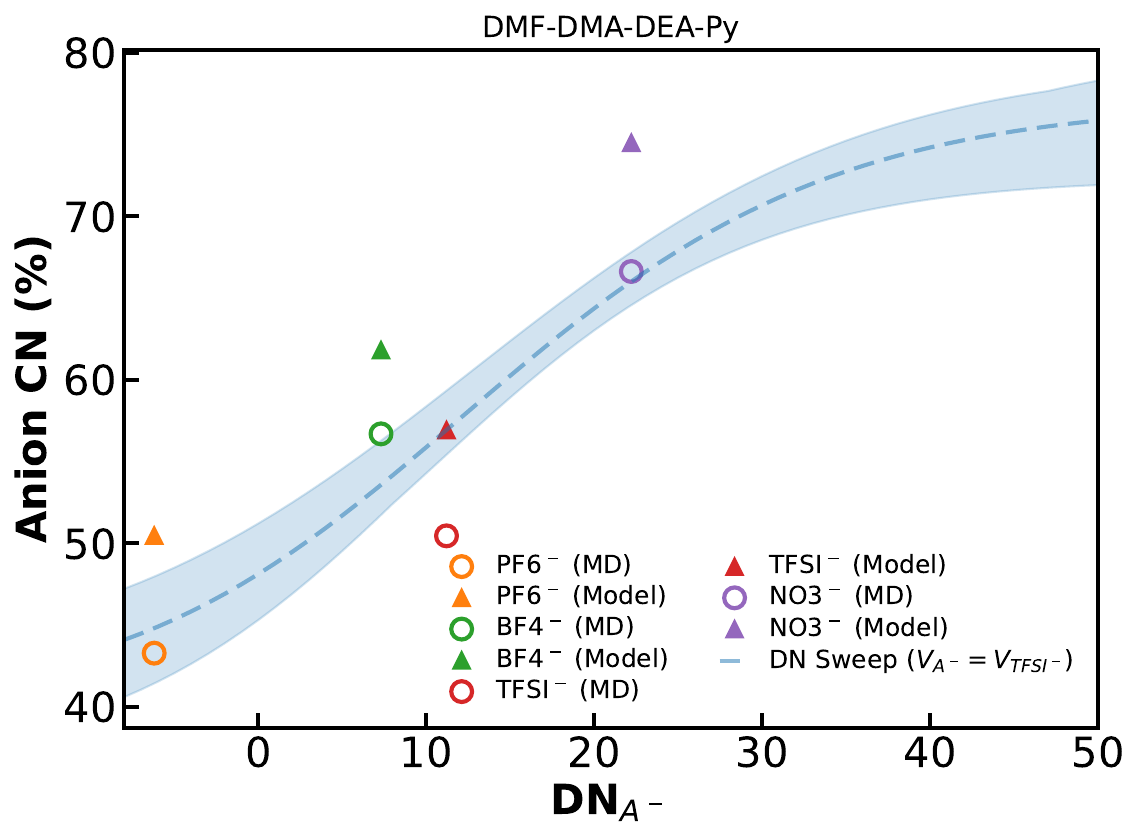}
        \label{fig:hcee-dmf}
    \end{subfigure}
    \begin{subfigure}[b]{0.45\textwidth}
        \centering
        \caption{}
        \includegraphics[width=\linewidth]{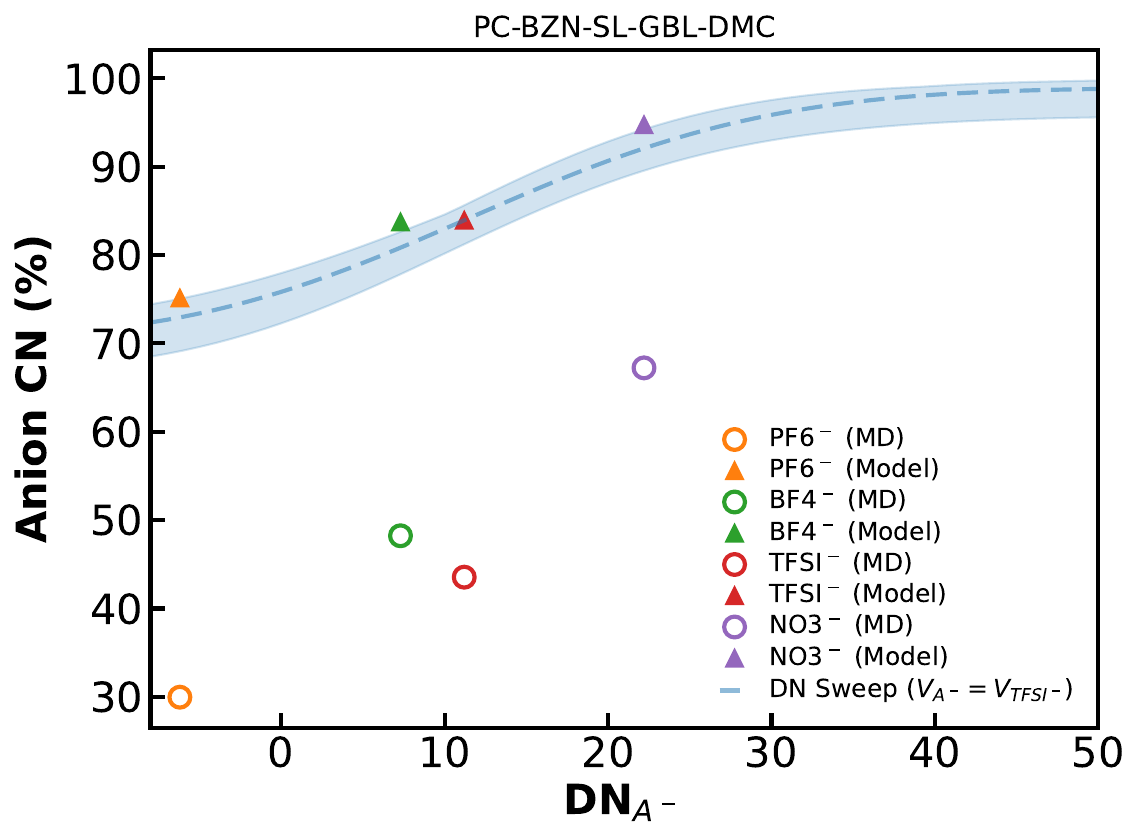}
        \label{fig:hcee-pc}
    \end{subfigure}
    \caption{Anion occupation versus anion DN for HEE systems at 3.0m: (a) DME/EA/MA/acetone/THF, (b) DMSO/DMF/DMA/DEA/Py, (c) DMF/DMA/DEA/Py, and (d) PC/BZN/SL/GBL/DMC. Markers follow \cref{fig:hee-generalizability}. The anion ordering is preserved, although the model overestimates anion occupation throughout, consistent with our findings in LHCE systems}
    \label{fig:hcee-generalizability}
\end{figure}

\cref{fig:hcee-generalizability} shows the same mixtures at 3.0m. The rank ordering remains robust, although quantitative agreement degrades relative to 1.0m: the mean absolute error rises from $11.3\%$ to $23.5\%$, with the model overestimating anion occupation in every mixture. This is the same failure mode documented above for the G4/LiTFSI system, where above 2m the model underestimates solvent occupation. 

Together, these tests indicate that our model shows decent transferability across multiple classes of electrolytes well beyond LHCE, and is able to capture some of the important physcial trends with increasing molality and DN. Despite more careful parameterization is needed for better prediction, these results show that our model can be readily extended to other types of novel electrolyte design if training data is provided.

\section{Visualization of Ising model interaction terms}
\begin{figure}[h]
    \begin{subfigure}[b]{0.4\textwidth}
        \centering
        \caption{}
        \includegraphics[width=\linewidth]{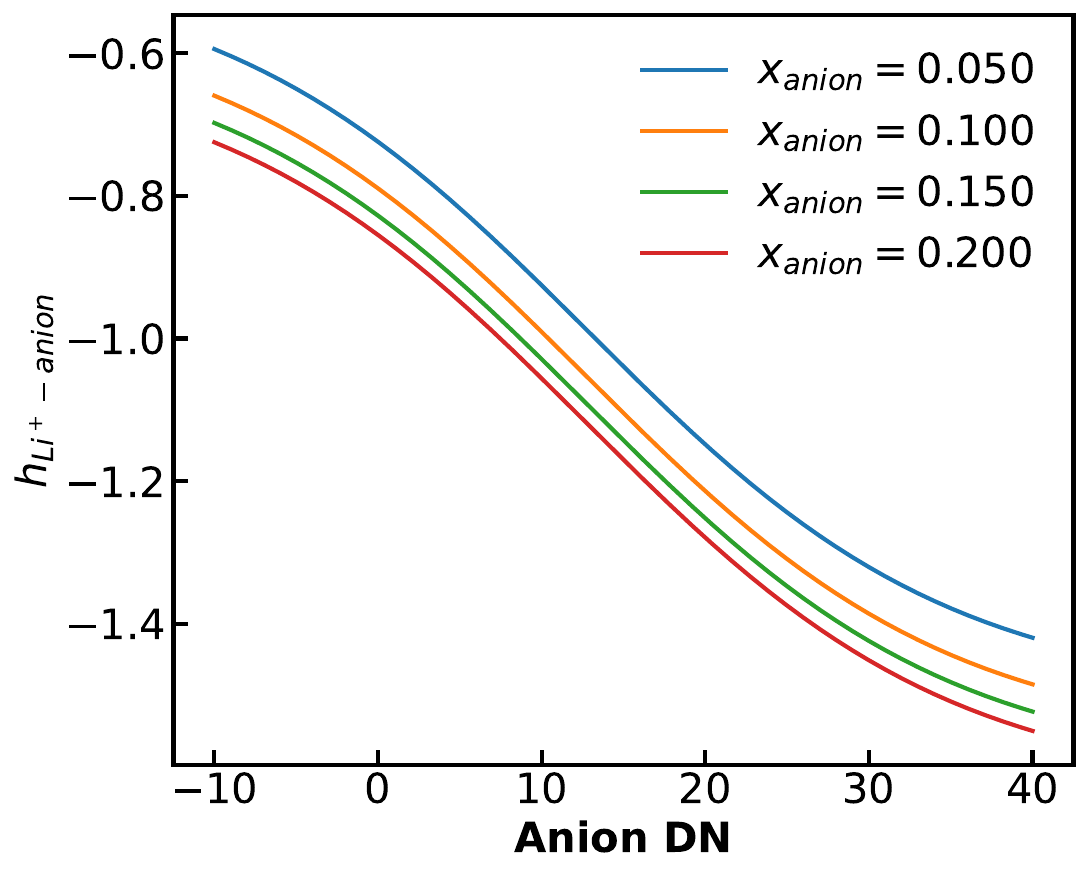}
        \label{fig:salt_dn_func}
    \end{subfigure}    
    \begin{subfigure}[b]{0.4\textwidth}
        \centering
        \caption{}
        \includegraphics[width=\linewidth]{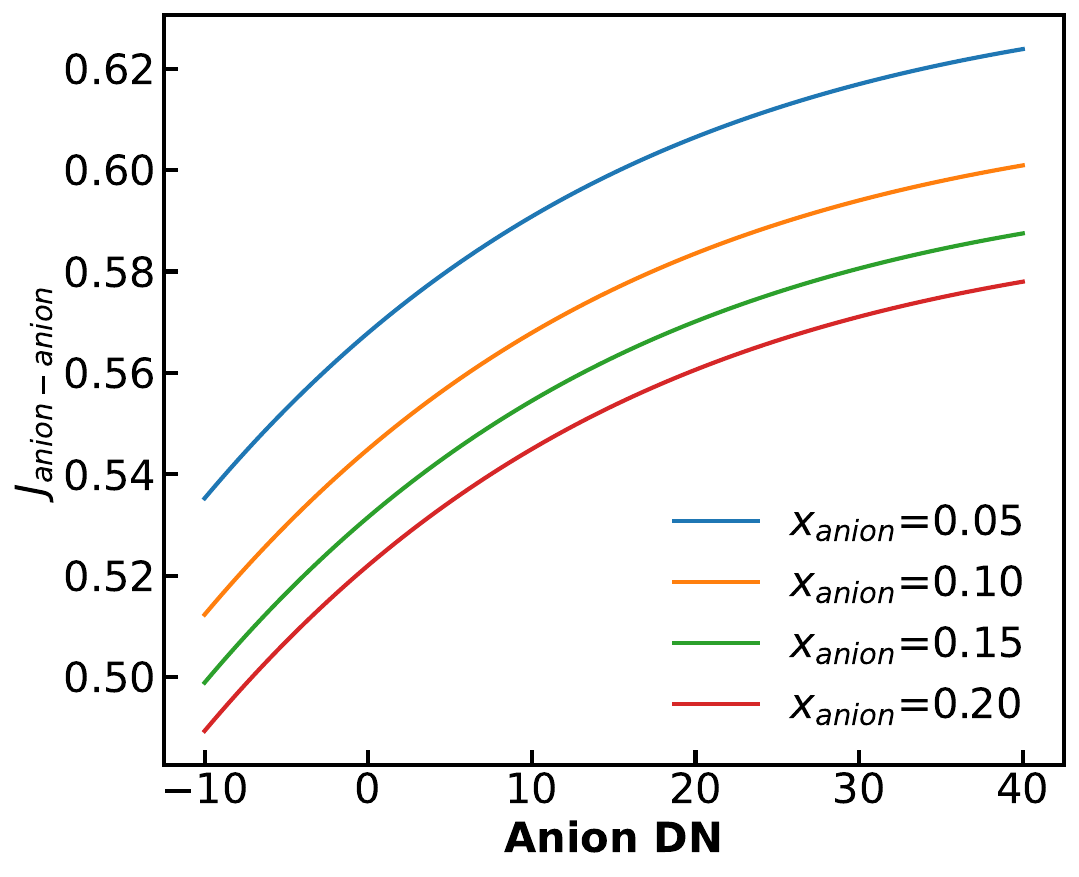}
        \label{fig:salt_salt}
    \end{subfigure}     
    \caption{Interactions within our model: (a) $h_{Li^+-anion}$, and (b) $J_{anion-anion}$, as functions of anion DN and molar ratio.}
\end{figure}

\begin{figure}[h]
    \centering
    \includegraphics[width=0.9\linewidth]{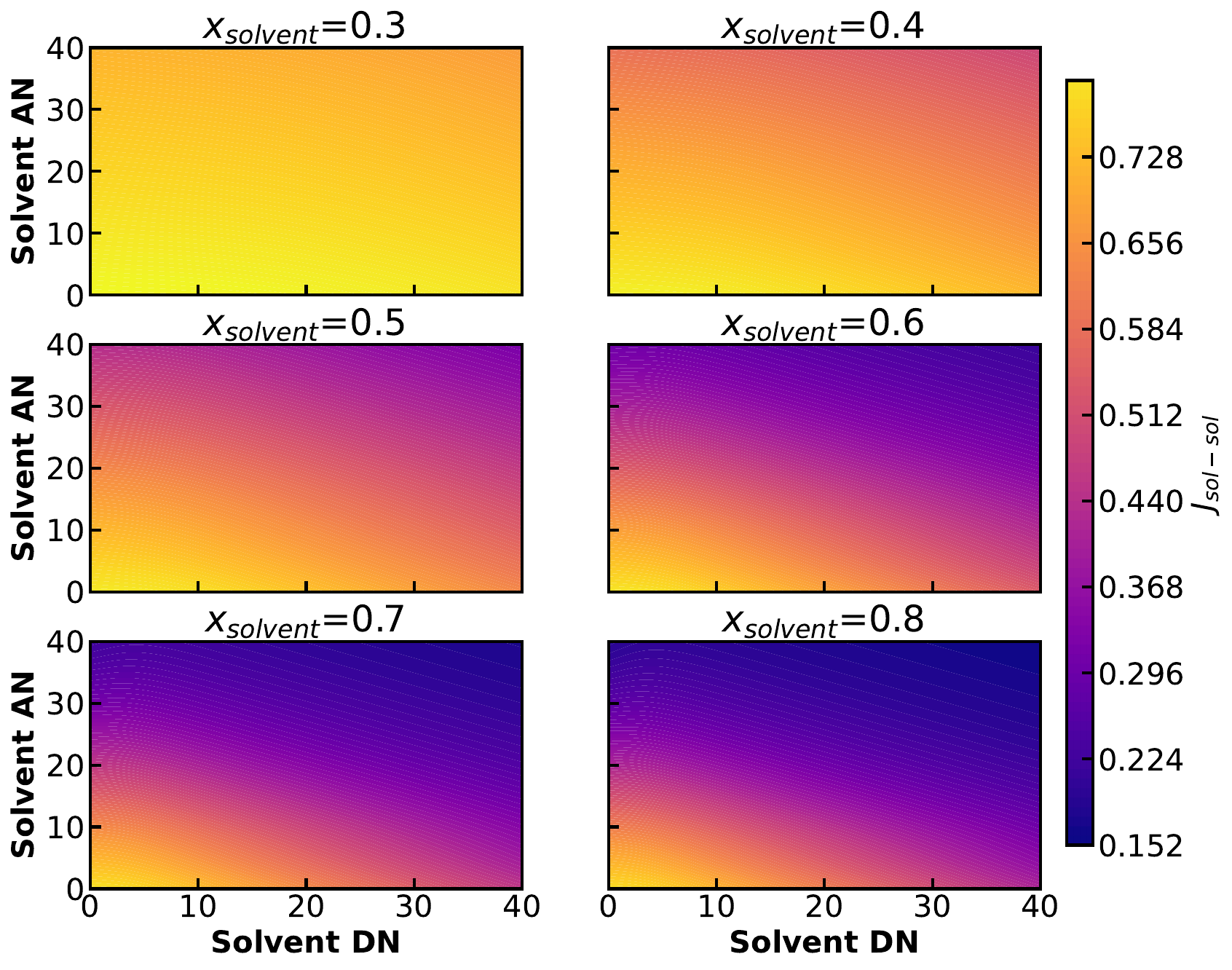}
    \caption{$J_{solvent-solvent}$ term of our model as functions of DN, AN and molar ratio. We assume a solvent molecular size equal to that of DME, and the anion is TFSI$^-$.}
    \label{fig:sol-sol}
\end{figure}

In \cref{fig:salt_dn_func,fig:salt_salt} we visualize the $h_{Li^+-anion}$ and $J_{anion-anion}$ terms. $h_{Li^+-anion}$ dependence on DN is similar to $h_{Li^+-sol}$. In \cref{fig:sol-sol} we visualize the solvent-solvent interaction terms as functions of DN, AN, and molar ratio.

\section{Discussion on physical monotonicity}
From a physical perspective, large DN is equivalent to greater enthalpic heat release from the exothermic coordination reaction between a Lewis acid and a Lewis base. Since \lion \space is a Lewis acid with a strong tendency of accepting electrons, this means that higher DN is equivalent to stronger and more energetically favorable \lion-solvent coordination in LHCE systems. The molar ratio effect contributes to both the entropic term in main text Eq.1, as well as enhanced effective DN in \cref{eq:dn_eff}, with the latter playing a more dominant role. At higher solvent molar ratio, the interaction between \lion and solvents becomes more favorable, resulting in a higher solvent content in the \lion solvation shell and more complete dissociation of \lion and salt anions, which is the case for conventional electrolytes. 

In the main text, we report a saturation behavior of our $h$ term describing \lion-solvent interactions. A natural concern is whether the high-DN saturation of $h_{Li^+-sol}$ (main text Fig.~3a) is an artifact of the sigmoid-type functional form of \cref{eq:h-term-si} rather than a genuine physical feature. We argue it is not. No constraints are placed on the range or slope of the sigmoid during parameterization, so a lower plateau over $20<DN<40$ is not guaranteed a priori. A counter-example appears in the same figure at low DN ($\sim 0$), where the characteristic \emph{upper} plateau of a sigmoid is absent. This upper plateau emerges below $DN\sim10$ only when \cref{eq:h-term-si} is fit directly to Li/\lion \space half-wave potential data (\cref{fig:sigmoid_fit_half_wave}), which we attribute to the scarcity of experimental data in that DN range. That neither the learned $h_{Li^+-sol}$ nor the underlying half-wave potential data exhibits an upper plateau indicates the saturation reflects the true DN dependence captured from the LHCE MD dataset, enabled by the end-to-end differentiable parameterization.

\section{Discussion on diluent DN boundary}
\begin{figure}[h]
    \centering
    \begin{subfigure}[b]{0.45\textwidth}
        \centering 
        \caption{}
        \includegraphics[width=\linewidth]{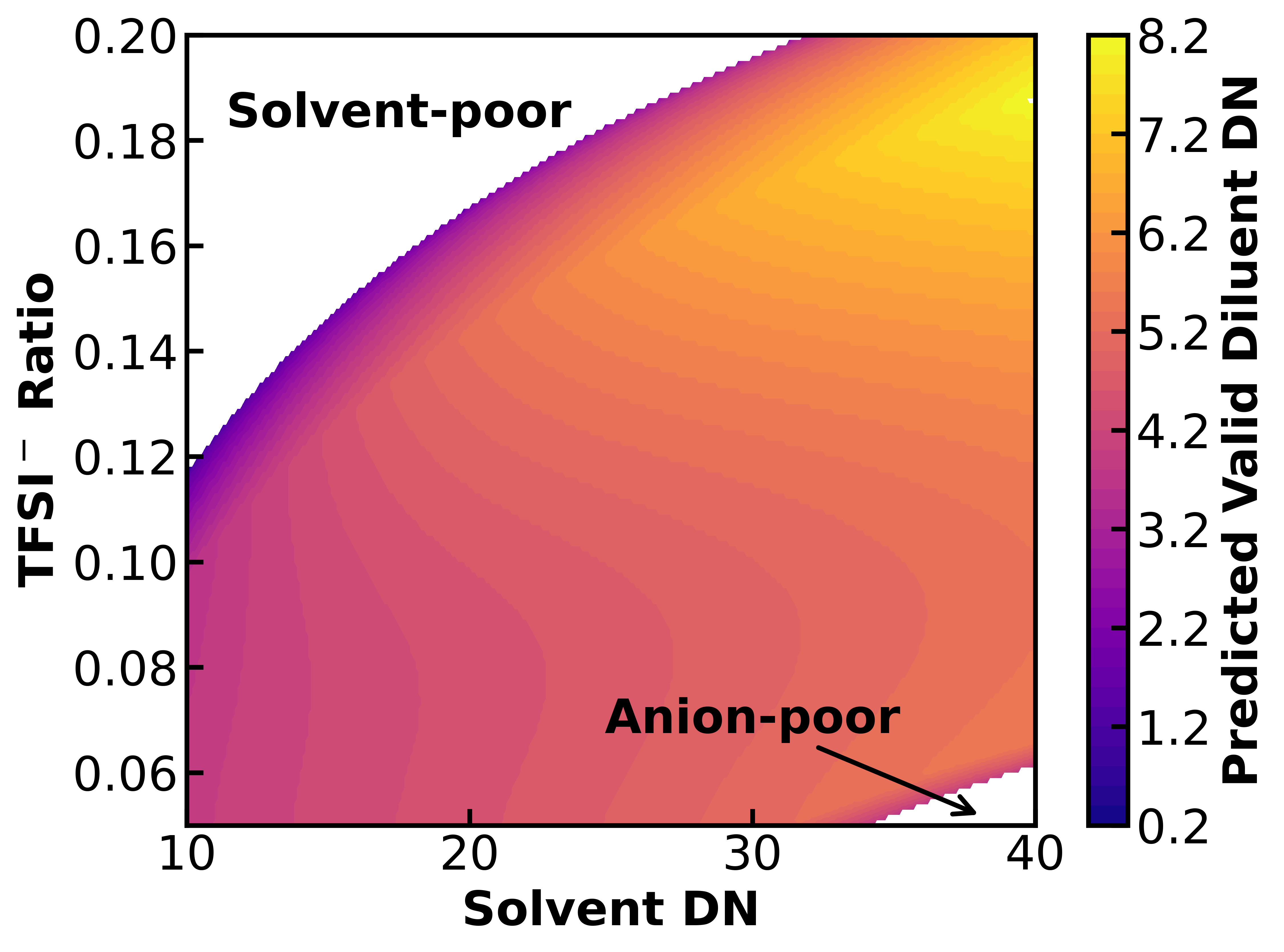}
         \centering
         \label{fig:dil_tfsi}
    \end{subfigure}            
    \begin{subfigure}[b]{0.45\textwidth}
        \centering 
        \caption{}
        \includegraphics[width=\linewidth]{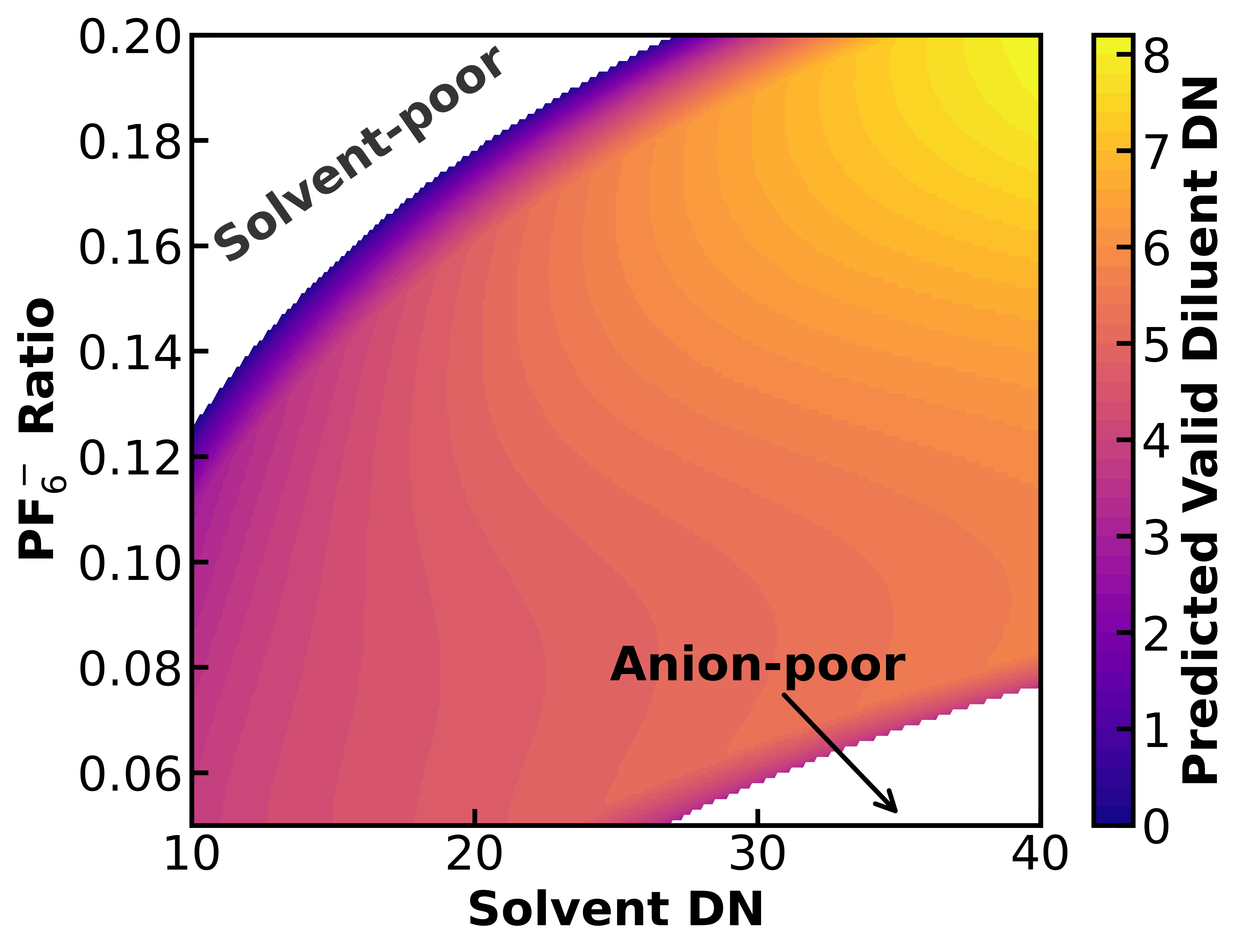}
         \centering
         \label{fig:dil_pf6}
    \end{subfigure}
    \caption{Diluent design rule for highest possible diluent DN when fixing anion molar ratio and solvent DN, with salt choice being (a) LiTFSI and (b) LiPF$_6$. This design rule can be used in the scenario where an HCE has already been made, and the goal is to find a diluent that does not affect HCE solvation structure. Molecules with DN lower than the contour value are valid diluents.}
    \label{fig:dil-design-rule}
\end{figure}

We show in \cref{fig:dil-design-rule} the predicted DN boundaries for diluents while taking solvent and anion occupations into account. The maximum diluent occupation is set to 10\%, while anion and solvent occupations should both be $\geq$30\%. The solvent-poor and anion-poor regions are labeled with text. At low solvent DN, the solvation shell becomes solvent-poor easily with increasing salt concentrations. At high solvent DN, due to excessive favorability of solvents coordinating with \lion, there is a high likelihood to form anion-poor solvation shells. \cref{fig:dil_tfsi} and \cref{fig:dil_pf6} compare the effect of using different anions, TFSI$^-$ vs PF$_6^-$. Due to PF$_6^-$ being more easily dissociated, it is more likely to form anion-poor solvation shells in LiPF$_6$-based electrolytes. Interestingly, the anion choice has very little impact on diluent DN boundary values in valid LHCE regions. This echoes previous literature findings that diluent choices are almost invariant to the salt choice \cite{chen_design_2023}.

We also note that the diluent DN boundary does not always increase monotonically with anion molar ratio, particularly at low solvent DN: near the solubility limit, the solvation shell becomes dominated by anions, and diluents compete primarily with anions rather than solvent for \lion \space coordination. Diluent occupations also rise slightly with increasing salt concentration, shifting the DN boundary toward lower values for a fixed maximum allowable diluent occupation.

\begin{figure}[h]
    \centering
    \includegraphics[width=0.9\linewidth]{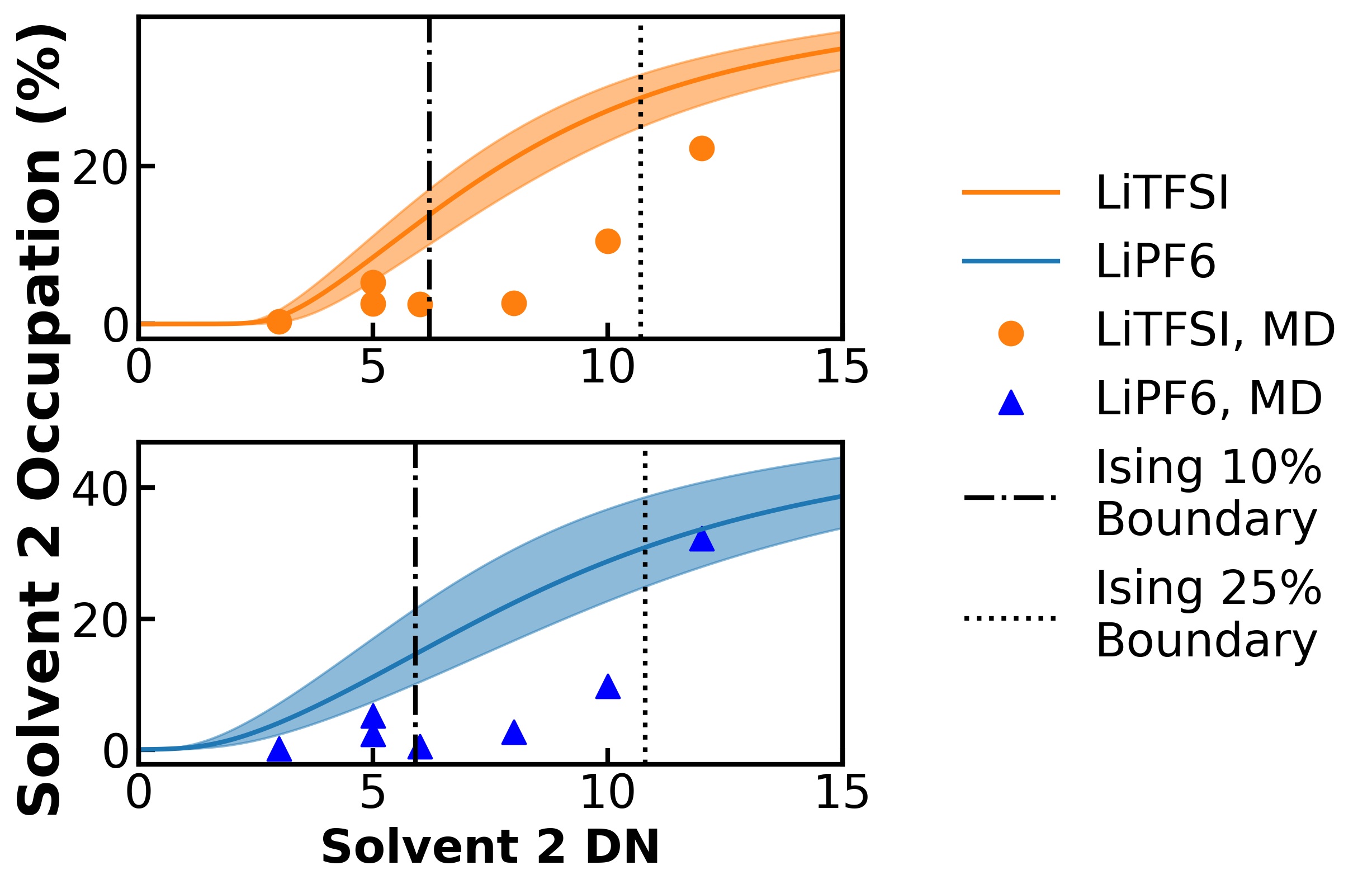}
    \caption{Comparison between Ising-predicted and MD-predicted diluent DN boundary. Our Ising model tend to underestimate diluent DN boundary, making our diluent selection criterion a sufficient but not necessary criterion.}
    \label{fig:diluent-choice-md-comparison}
\end{figure}

We also show a comparison between diluent DN boundaries predicted by our model and MD-simulated boundaries in \cref{fig:diluent-choice-md-comparison}. We use DME as the solvent and fix the anion molar ratio to be 0.15. The uncertainty is evaluated among the 5 models from the 5-fold cross validation. Our model prediction tends to underestimate diluent DN ranges in both LiTFSI and LiPF$_6$ electrolyte systems. If we set 10\% diluent occupation as the threshold, the diluent DN boundary predicted by our model is smaller by $\sim 5$ kcal/mol in DN values. Better agreement with MD can be achieved if we set 25\% as the threshold. The reason for high errors in the DN$=6-10$ range may originate from lack of training data in this range. 
Our model's underestimation tendency may leave out possible diluent candidates when used in high throughput screening, but the diluents that we identify are still valid diluents. In other words, the diluent selection criterion identified by our model is a sufficient but not necessary criterion.

\begin{figure}[h!]
    \centering
    \includegraphics[width=0.9\linewidth]{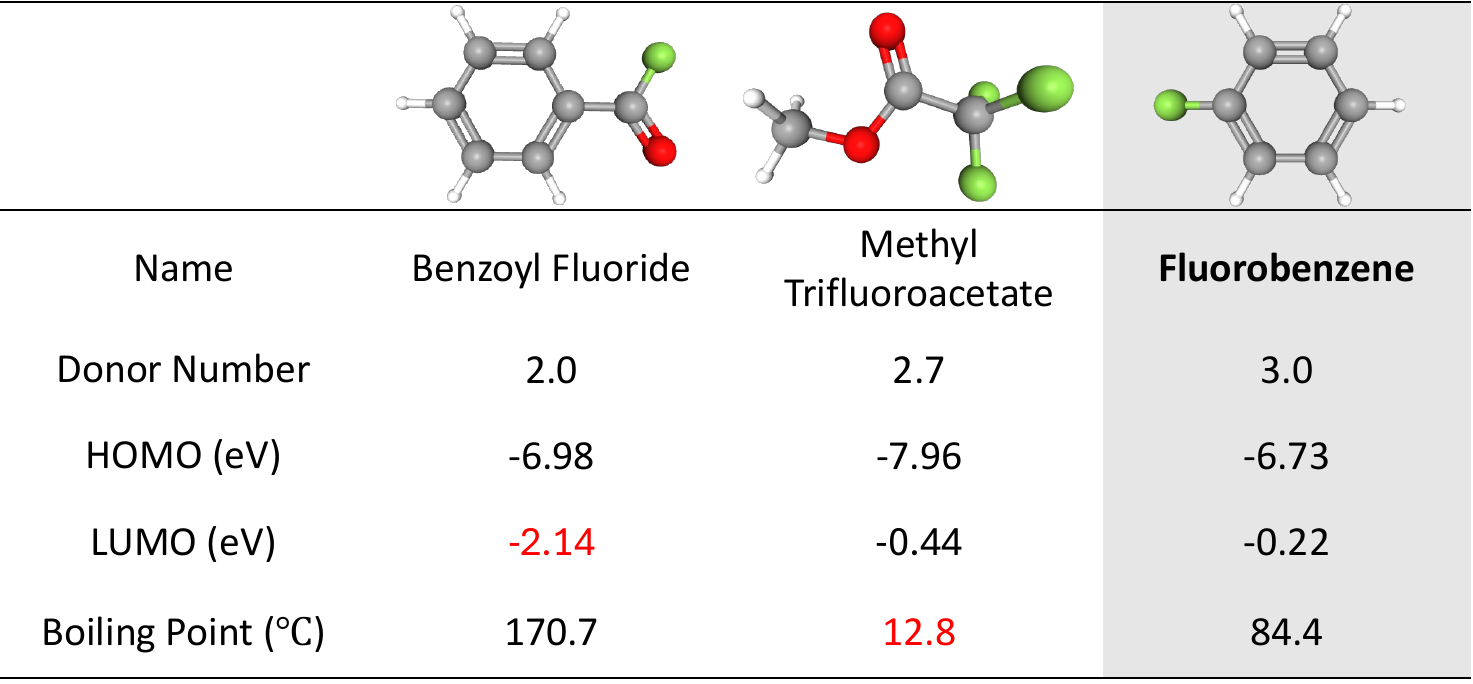}
    \caption{Comparison between 3 fluorine-containing diluent candidates, benzoyl fluoride, methyl trifluoroacetate, and fluorobenzene. HOMO, LUMO and boiling points are predicted by the MIST model \cite{wadell_foundation_2025}.}
    \label{fig:3-diluent-candidate-comparison}
\end{figure}

Promising candidates from the Stenutz database \cite{stenutz_gutmann_2026} that meet the diluent DN boundary criterion are shown in \cref{fig:3-diluent-candidate-comparison}. Among these candidates, fluorobenzene (FB) offers the best combination of reductive stability (highest LUMO energy), high boiling point, and low viscosity (0.551 cP at room temperature), motivating its selection as the diluent in the G4/LiTFSI design example (main text).

\section{Discussion on design workflow screening criteria}
In main text Fig.4, we define an anion-rich solvation shell using criteria based on fractional shell occupation: anion occupation should exceed 30\% and solvent occupation should exceed 20\%.
This criterion ensures that LHCE solvation shells possess a sufficiently high anion fraction (i.e., anion-rich), while avoiding excessive salt content that could risk salt precipitation and low ionic conductivity.
To the best of our knowledge, there is no universal quantitative definition in the literature for the minimum anion occupation required to define an anion-rich solvation structure. 
This has not been a critical issue, as the precise relationship between solvation shell composition and electrochemical performance remains unclear, and the latter is ultimately the more relevant metric. Nevertheless, the absence of a quantitative criterion introduces ambiguity when attempting to narrow concentration ranges for further testing.
In our example, we choose these selection criteria based on literature knowledge, our MD simulation results, and model errors. 

The criterion of diluent occupation $\langle l \rangle \leq 0.1$ is primarily based on model RMSE. As shown in \cref{fig:parity_sep_mnl} and main text Fig.2b, our model achieves a total RMSE among all components of $\sim 10\%$, while single-component RMSEs fall in the range of $4\%-8\%$. Taking a deeper look at diluent occupation  $\langle l \rangle$ prediction, 
we find that the standard deviation of $\langle l \rangle$ is $4.03\%$. The point of having diluents in LHCE is that they should not participate in the formation of the solvation shell, so the occupation of the diluent should ideally be $\sim 0\%$. Given that $>99\%$ of diluent occupation predictions are enclosed within $10\%$ error, we adopt $\langle l \rangle \leq 10\%$ as the selection threshold. 

The solvent and anion occupation criteria are less straightforward. Previous researchers often considered electrolytes with $>3$M or $>3$m salt concentrations to be high-concentration electrolytes \cite{yamada_advances_2019,chen_design_2023,cao_reviewlocalized_2021}, mentioning that above this concentration anions start to enter the solvation shell and free solvents begin to diminish. 
While this criterion holds in most electrolyte formulations of interest to LIB and LMB chemistries, it is not guaranteed to be valid. For example, in our MD simulations, we observe non-trivial anion occupations between 30\% and 40\% in the first solvation shell in 
$\sim 2.2$m LiTFSI in DMC, which is far below the solubility limit \cite{li_concentrated_2023}. This also explains why there is no universal rule for classifying HCE, as different salt and solvent choices would affect the actual solvation shell compositions and free solvent ratios under the same molarity or molality. 
From an atomistic standpoint, whether an electrolyte is high-concentration or not should depend on (a) whether or not it forms anion-rich or AGG-rich solvation structure, and (b) whether or not there is sufficiently low free solvent population. However, quantitative thresholds for sufficiently high and sufficiently low remain open questions. 
Here we take inspiration from formulations with available CN, either from literature or in-house MD simulations. \citeauthor{yang_advancing_2025} shows that in an HCE with 10M LiFSI dissolved in DME, the anion and solvent CN roughly follows a 1:1 ratio \cite{yang_advancing_2025}. \citeauthor{chen_high-voltage_2018} designed an LHCE with 1.2M LiFSI in DMC/BTFE (1:2 mol/mol) \cite{chen_high-voltage_2018}, with a salt-to-solvent ratio of 1:1.1, and our in-house MD simulations show that LHCEs under similar molar ratio have 68\% anion occupations. Based on these literature results, we require solvent occupation to be $>30\%$ to mitigate the risk of salt precipitation, and in our screening design we use a looser $>20\%$ criterion to account for the model's $\sim10\%$ RMSE. We set the minimum anion occupation at 30\% based on the model 3$\sigma$ error (\cref{fig:kfold_error_hist}). Both bounds are intentionally loose, chosen to avoid excluding formulations likely to form anion-rich solvation structures due to false negative predictions.

In addition to local solvation shell occupations, another key property is the free solvent population -- a distinct metric describing the fraction of solvent molecules in the bulk electrolyte that are not coordinated to \lion \space at all, as opposed to the solvent shell occupation $\langle m \rangle$ discussed above. We use a 20\% threshold to downselect formulations with low free solvent population; this value is numerically identical to the solvent shell occupation threshold above but reflects an unrelated criterion. Ideally the free solvent molar ratio should be 0\% to prevent side reactions between electrode surfaces and free solvents. We use 20\% here to set a loose bound in a similar fashion to solvation shell occupation thresholds to reduce the effect of leaving out false negative predictions.

\section{Discussion on G4/LiTFSI design validation}
Comparing our model's predictions with higher-fidelity MD for the G4/FB/LiTFSI system (main text Fig.~4), discrepancies are concentrated outside the final design window. At very high concentrations ($>2$m), our model underestimates solvent occupation, likely because solvent-in-salt electrolytes behave differently from those at moderate concentration. Free solvent ratio predictions show a similar pattern, with larger discrepancies at high concentration. Since these high-error regions fall outside the 0.46-1.8m design window, they do not affect the model's ability to identify a promising LHCE formulation range for this system.

\section{Additional LHCE design examples}
\begin{figure}[h!]
    \centering
    \begin{subfigure}[b]{0.45\textwidth}
        \centering 
        \caption{}
        \includegraphics[width=\linewidth]{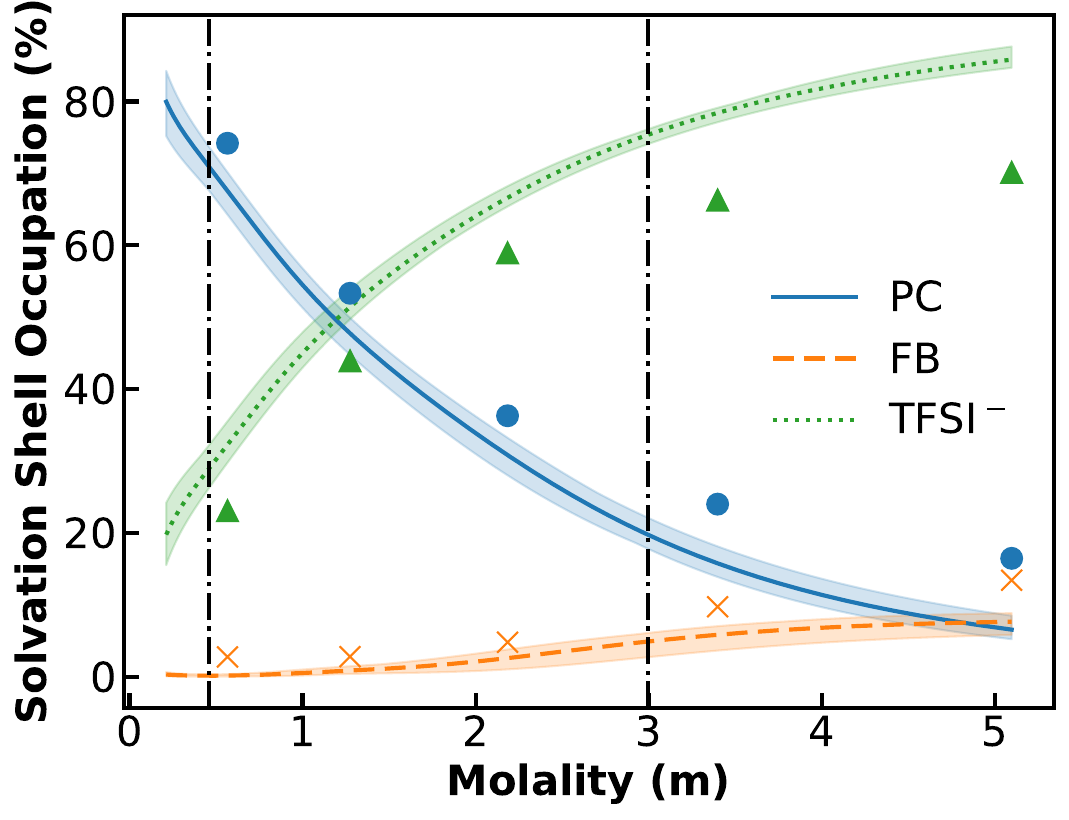}
         \centering
         \label{fig:pc_fb_tfsi}
    \end{subfigure}            
    \begin{subfigure}[b]{0.45\textwidth}
        \centering 
        \caption{}
        \includegraphics[width=\linewidth]{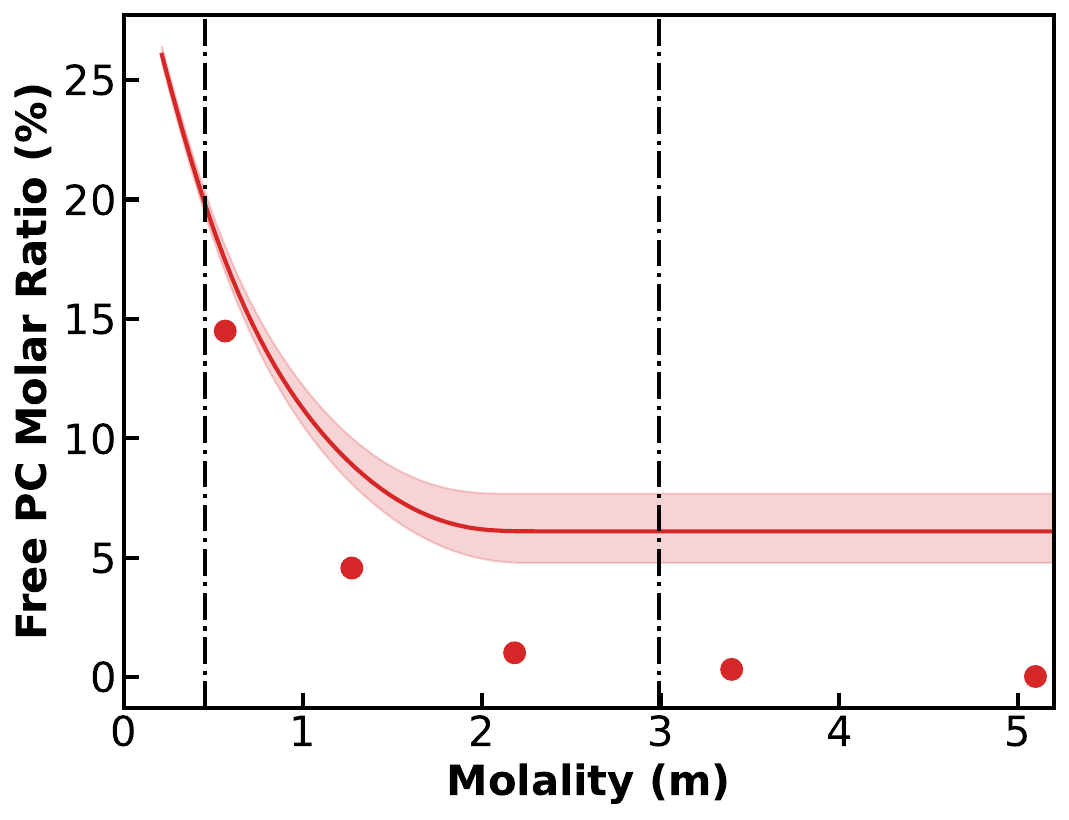}
         \centering
         \label{fig:pc_fb_tfsi_free}
    \end{subfigure}
    \caption{Example design of propylene carbonate (PC)/FB/LiTFSI system. (a) \lion \space fractional CN as a function of LiTFSI molality. (b) Free PC molar ratio as a function of LiTFSI molality. Our model predicts 0.46m-3.00m LiTFSI in PC:FB=1:2(mol:mol) as promising formulations for further LHCE design.}
    \label{fig:pc-fb-design}
\end{figure}

\begin{figure}[h!]
    \centering
    \begin{subfigure}[b]{0.45\textwidth}
        \centering 
        \caption{}
        \includegraphics[width=\linewidth]{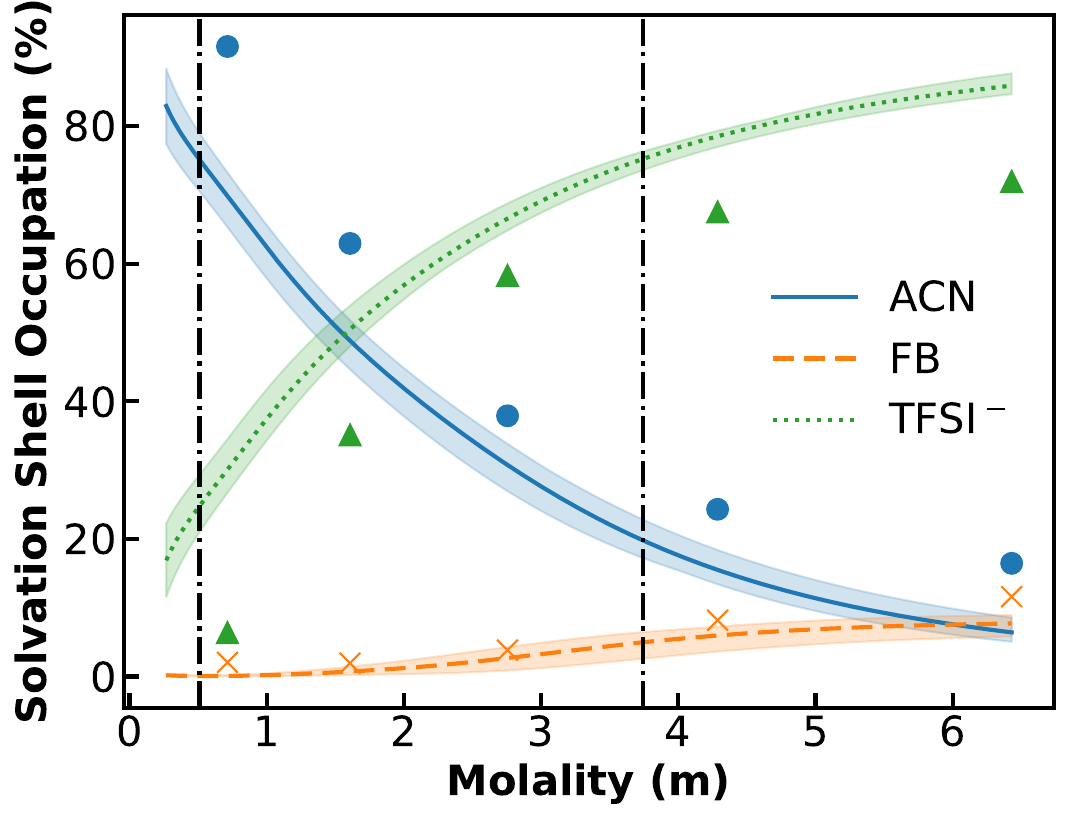}
         \centering
         \label{fig:acn_fb_tfsi}
    \end{subfigure}            
    \begin{subfigure}[b]{0.45\textwidth}
        \centering 
        \caption{}
        \includegraphics[width=\linewidth]{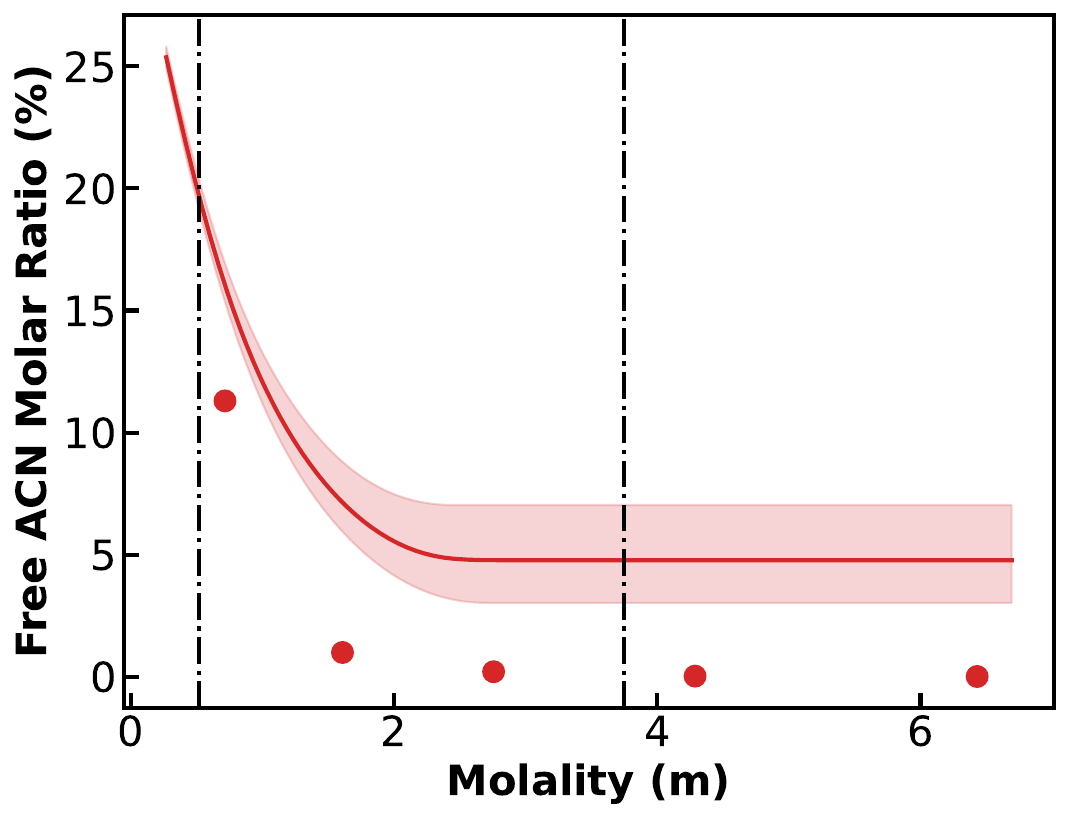}
         \centering
         \label{fig:acn_fb_tfsi_free}
    \end{subfigure}
    \caption{Example design of acetonitrile(ACN)/FB/LiTFSI system. (a) \lion \space fractional CN as a function of LiTFSI molality. (b) Free ACN molar ratio as a function of LiTFSI molality. Our model predicts 0.51m-3.74m LiTFSI in ACN:FB=1:2(mol:mol) as promising formulations for further LHCE design.}
    \label{fig:acn-fb-design}
\end{figure}

In the main text we demonstrate the applicability of our model on ether solvents. Here we show additional examples of applying the model to other classes of solvents commonly used in LIB and LMB electrolyte designs, including carbonates (\cref{fig:pc-fb-design}) and nitriles (\cref{fig:acn-fb-design}). We use force field parameters from the work of \citeauthor{you_dielectric_2016} for propylene carbonate (PC) MD simulations. Our model predicts that 0.46m-3.00m LiTFSI in PC:FB=1:2(mol:mol) are promising formulations for further selection, in good agreement with MD predictions. For acetonitrile (ACN) electrolytes, the model predicts that 0.51m-3.74m LiTFSI in ACN:FB = 1:2(mol:mol) are promising formulations for further design. Despite slightly underestimating \lion-solvent fractional CN, our model predictions reproduce the trend in MD results and serve as a good first-order approximation. 

\begin{figure}[h!]
    \centering
    \begin{subfigure}[b]{0.45\textwidth}
        \centering 
        \caption{}
        \includegraphics[width=\linewidth]{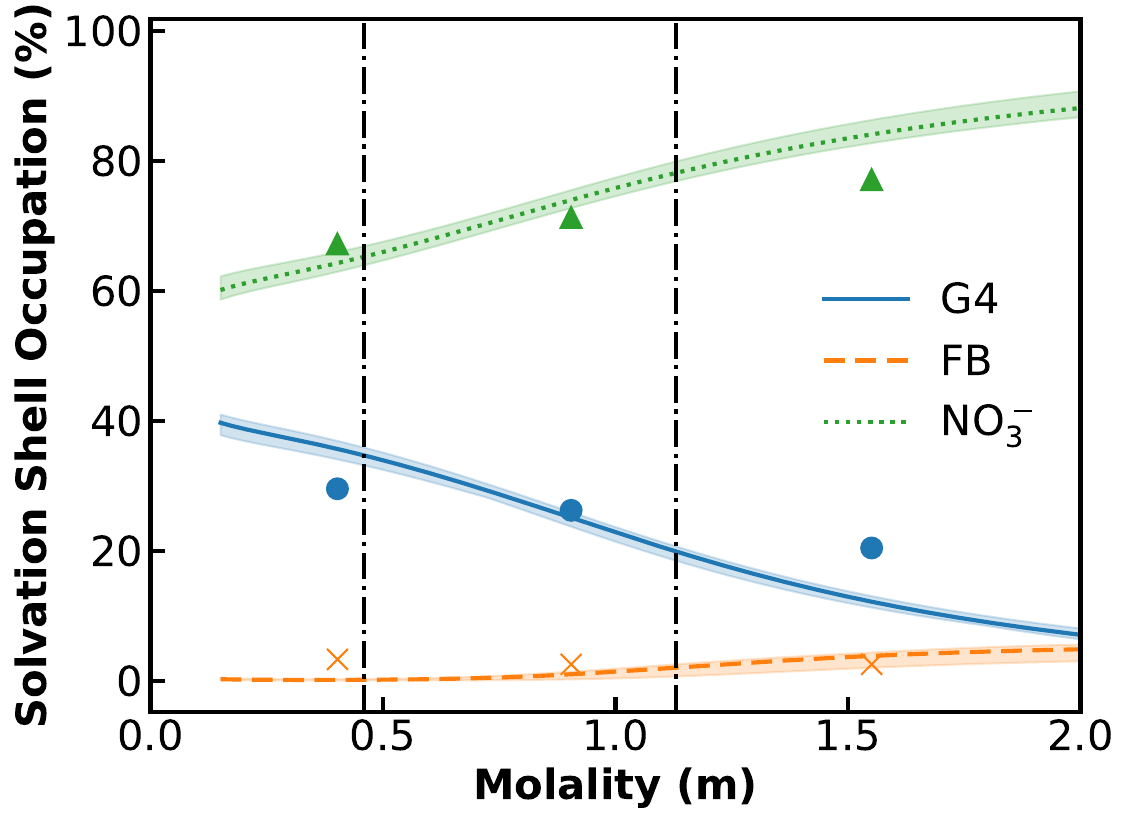}
         \centering
         \label{fig:g4_fb_no3}
    \end{subfigure}            
    \begin{subfigure}[b]{0.45\textwidth}
        \centering 
        \caption{}
        \includegraphics[width=\linewidth]{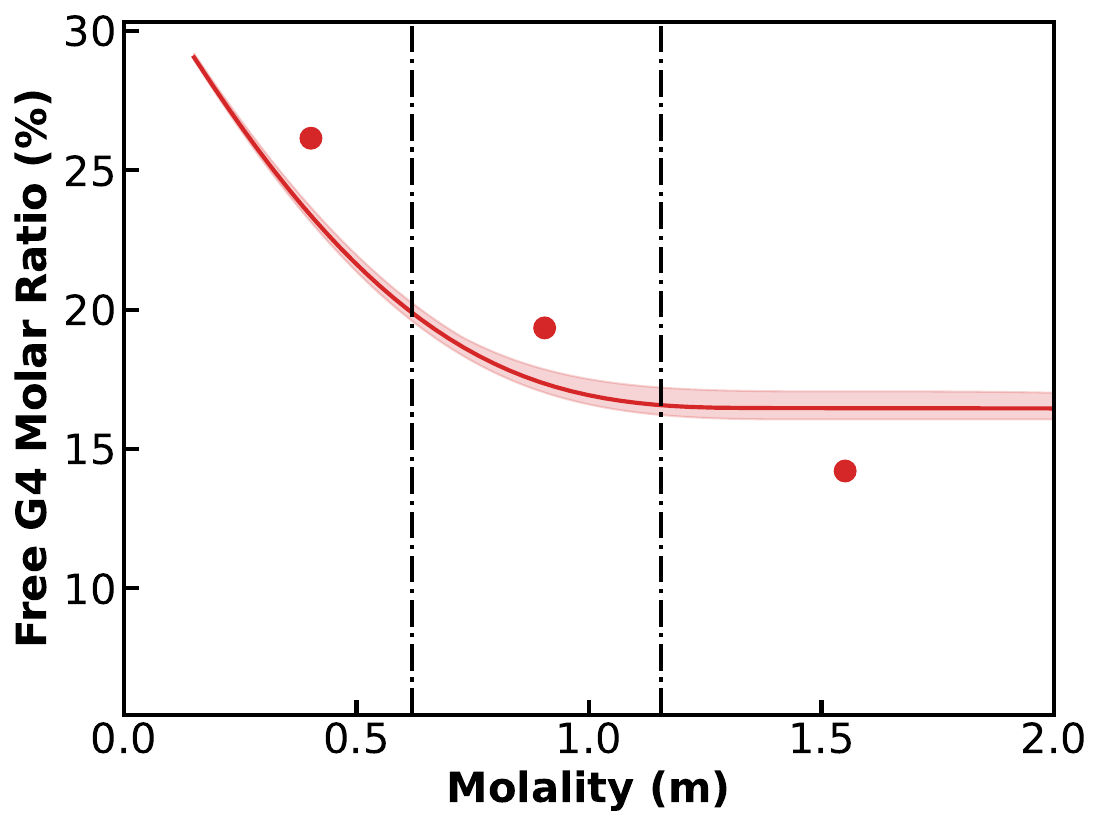}
         \centering
         \label{fig:g4_fb_no3_free}
    \end{subfigure}
    \caption{Example design of G4/FB/LiNO$_3$ system. (a) \lion \space fractional CN as a function of LiNO$_3$ molality. (b) Free G4 molar ratio as a function of LiNO$_3$ molality. Because NO$_3^-$ has very high donicity (DN=22.2 kcal/mol), even at low concentrations NO$_3^-$ are extremely favorable to coordinate with \lion.}
    \label{fig:g4-fb-no3-design}
\end{figure}

We also demonstrate the applicability of our model for electrolyte systems containing other anions of interest. \cref{fig:g4-fb-no3-design} shows how \lion \space fractional CN changes with varying salt molality in a LiNO$_3$/G4/FB system. NO$_3^-$ anions have very strong donating tendencies (DN=22.2 kcal/mol), so our model predicts that even at very low concentrations, NO$_3^-$ would favorably coordinate with \lion. This indicates that LiNO$_3$ would not be the best salt for designing LHCE, but they can be used to tune \lion \space solvation shell compositions and subsequently SEI/CEI compositions as additives.

\section{Discussion on model scope and future extensions}
The current model's applicability is bounded by available DN and AN data: only 43 molecules have both DN and AN measurements, and relaxing the criterion to molecules with available DN alone (or its counterpart, BF$_3$ affinity) yields only 347 candidates, many of which are unsuitable for LMB electrolytes. Expanding high-quality DN and AN datasets would significantly broaden the model's applicability. Additional molecular descriptors, such as the dielectric constant ($\varepsilon$), could also be incorporated into the current framework's $h$ and $J$ interactions to extend it toward conventional and dilute electrolyte regimes, where solvation behavior is less exclusively governed by DN and AN.

\newpage

\end{document}